\documentclass[journal]{IEEEtran}
\ifCLASSINFOpdf

\else

\fi
\usepackage[cmex10]{amsmath}
\usepackage{amssymb}
\usepackage{cite}
\usepackage{multirow}
\usepackage{makecell}
\usepackage{graphicx}
\usepackage{array,color}
\usepackage{amsmath}
\usepackage{stfloats}
\usepackage{graphicx}
\usepackage{subfigure}
\usepackage{tabularx}
\usepackage{epsfig,epsf,color,balance,cite}
\usepackage{algorithmic}
\usepackage{algorithm}
\usepackage{bm}
\usepackage{textcomp}
\allowdisplaybreaks[4]

\begin{document}

\title{ An Overview of Signal Processing Techniques for RIS/IRS-aided  Wireless Systems}
\author{ Cunhua Pan, Gui Zhou, Kangda Zhi, Sheng Hong, Tuo Wu, Yijin Pan, Hong Ren, Marco Di Renzo, \IEEEmembership{Fellow, IEEE}, A. Lee Swindlehurst, \IEEEmembership{Fellow, IEEE}, Rui Zhang, \IEEEmembership{Fellow, IEEE}, and Angela Yingjun Zhang,  \IEEEmembership{Fellow, IEEE}

\thanks{C. Pan is  with the National Mobile Communications Research Laboratory, Southeast University, Nanjing 210096, China. He was with the School of
Electronic Engineering and Computer Science at Queen Mary University of London, London E1 4NS, U.K. (Email: cunhuapan21@gmail.com).
G. Zhou, K. Zhi and T. Wu are with the School of Electronic Engineering and Computer Science at Queen Mary University of London, London E1 4NS, U.K. (Email: \{g.zhou, k.zhi, tuo.wu\}@qmul.ac.uk). S. Hong is with Information Engineering School of Nanchang University, Nanchang 330031, China. (Email: shenghong@ncu.edu.cn). Y. Pan and H. Ren are the National Mobile Communications Research Laboratory,
Southeast University, Nanjing 210096, China. (Email: \{panyj, hren\}@seu.edu.cn). M. Di Renzo is with Universit\'e Paris-Saclay, CNRS, CentraleSup\'elec, Laboratoire des Signaux et Syst\`emes, 3 Rue Joliot-Curie, 91192 Gif-sur-Yvette, France. (Email: marco.direnzo@centralesupelec.fr). A. L. Swindlehurst is with the Center for Pervasive Communications and Computing, University of California, Irvine, CA 92697, USA. (Email: swindle@uci.edu). R. Zhang is with the Department of Electrical and Computer Engineering, National University of Singapore, Singapore 117583 (Email: elezhang@nus.edu.sg). Y. J. Zhang is with the Department of Information Engineering, The Chinese University of Hong Kong, Shatin, Hong Kong (Email: yjzhang@ie.cuhk.edu.hk).  }
}

\maketitle

\begin{abstract}
In the past as well as present  wireless communication systems,  the wireless propagation environment is regarded as an uncontrollable black box that impairs the received signal quality, and its negative impacts are compensated for by relying on the design of various sophisticated transmission/reception schemes. However, the improvements through applying such schemes operating at two endpoints (i.e., transmitter and receiver) only are limited even after five generations of wireless systems.  Reconfigurable intelligent surface  (RIS) or intelligent reflecting surface (IRS) have emerged as a new and revolutionary technology that can configure the wireless environment in a favorable manner by properly tuning the phase shifts of a large number of quasi  passive and low-cost reflecting elements, thus standing out as a promising candidate technology for the next-/sixth-generation (6G) wireless system. However, to reap the performance benefits promised by  RIS/IRS, efficient signal processing techniques are crucial, for a variety of purposes such as channel estimation, transmission design, radio localization, and so on. In this paper, we provide a comprehensive overview of recent advances on RIS/IRS-aided wireless systems from the signal processing perspective. We also highlight promising research directions that are worthy of investigation in the future.
\end{abstract}

\IEEEpeerreviewmaketitle
\section{Introduction}
While the fifth-generation (5G) wireless communication system is under deployment worldwide,  research interest has  shifted to the future sixth-generation (6G) wireless system \cite{Saad2020,you2021towards,Zhengquanzhang}, which targets  supporting not only  cutting-edge applications like multisensory augmented/virtual reality applications, wireless brain computer interactions, and fully autonomous systems, but also the wireless evolution from ``connected things'' to ``connected intelligence''. The required key performance indicators (KPIs), including data rates, reliability, latency, spectrum/energy efficiency, and connection density,  will be superior to those for  5G. For example, the energy and spectrum efficiency for 6G are expected to be 10-100 times and 5 times better than those of 5G, respectively. These KPIs, however, cannot be fully achieved by the existing three-pillar 5G physical layer techniques \cite{Andrews2014}, which include massive multiple-input multiple-output (MIMO), millimeter wave (mmWave) communications, and ultra-dense heterogeneous networks. In particular, a large number of antennas along with active radio frequency (RF) chains are needed for massive MIMO to achieve high spectrum efficiency, which leads to high energy consumption and hardware cost. Moreover, moving to the mmWave frequency band renders the electromagnetic waves more susceptible to blockage by obstacles such as furniture and walls in indoor scenarios. In addition, more costly RF chains and sophisticated hybrid precoding are necessary for mmWave communication systems. The dense deployment of small base stations (BSs) also incurs high maintenance cost, network energy consumption, and hardware cost due to high-speed backhaul links. Furthermore, sophisticated interference management techniques are necessary in ultra-dense networks.

Conventionally, the wireless environment is perceived as a randomly varying entity that impairs the signal quality due to uncontrolled reflections, refractions and unexpected interference. Although a plethora of physical layer techniques such as advanced modulation/demodulation and precoding/decoding schemes have been developed at the endpoints of communication links to compensate for these negative impacts, it is undeniable that a certain level of saturation has been reached in terms of achievable data rate and performance reliability. Huge performance gains are expected   when regarding the wireless environment as an additional variable to optimize. This is made possible by exploiting the new and revolutionary idea of reconfigurable intelligent surfaces (RISs) or intelligent reflecting surfaces (IRSs) \cite{di2019smart,qingqingwu2020,cunhuamagazine,xiaojun2021,Chongwewc}, which are capable of reconfiguring the wireless propagation environment into a  transmission medium with more desirable characteristics.

An RIS/IRS is a planar surface composed of a large number of quasi passive and low-cost reflecting elements, each of which can impose an independent phase shift/amplitude on the impinging electromagnetic signals in a fully customized way. Thanks to recent advances in metamaterials \cite{cui2014coding}, the phase shifts imposed on the incident electromagnetic signals can be adjusted in real-time in reaction to the rapid variations in the wireless propagation environment. By judiciously tuning the phase shifts of the RIS/IRS, the signals reradiated from the RIS/IRS can be  added constructively with the signals from other paths to enhance the received signal power at the desired users, or can be combined  destructively to mitigate the undesired signals at unintended users such as multiuser interference and signal leakage at the eavesdroppers. Due to these functionalities,  RIS/IRS can be used to extend the coverage area, improve the channel rank, mitigate the interference,  enhance the reliability, and improve the positioning accuracy. Unlike  conventional relaying techniques,  an RIS/IRS is free from  RF chains and amplifiers, and thus entails much reduced power consumption and hardware cost. Furthermore, due to their quasi passive nature, RIS/IRS can be fabricated with a low profile, light weight, and  limited  thickness, which enables them to be readily layered on surfaces  available in the environment, including building facades, ceilings, street lamps, and so on.

The appealing advantages of RIS/IRS have led to extensive research focused on its fundamental performance limits \cite{Ertugrul2019,taoqincl,liangyang2020,zhiguoding,ren2021intelligent,liangyang,jiayizhangifs,pengxu2021,dongli,shaoqingzhou}, channel modeling \cite{wankai,bjoson2020near,Najafi2021,Yildirim2021,Yingzhuo}, and prototype design \cite{Linglongdaiaccess,wankaijsac,Xilongpei,zhang2021wireless}. It was shown in \cite{Xilongpei} that, in  an indoor realistic propagation environment, 26 dB power gain  can be achieved by an  RIS/IRS prototype consisting of 1100 controllable elements operating in the  5.8 GHz band. In addition, several tutorial/overview papers have  summarized recent developments in this area, including the technical challenges \cite{qingqingtcomtutorial},  operation principles \cite{yuanweiliu2021}, transmission design and applications \cite{Shimin}, electromagnetic modeling \cite{Renzojsac}, practical design issues \cite{zheng2021survey} and channel estimation \cite{Lee}.

In this paper, we provide a systematic overview of existing works on IRS/RIS mainly from the signal processing point of view, by focusing on channel estimation, transmission design and radio localization issues. Specifically, in Section \ref{channelestimation}, we overview existing contributions on channel estimation  based on the structure of the channels, including  unstructured channels that model  low-frequency rich-scattering scenarios and  structured channels that are appropriate for high-frequency sparse channels. In Section \ref{transmissiondesign}, we overview existing works on transmission design from two aspects, namely, optimization techniques and the  availability of channel state information (CSI).  The existing contributions on transmission design are classified into three cases, based on fully instantaneous CSI, two-timescale CSI and fully long-term CSI, respectively. Then, in Section \ref{radiolocalization}, we introduce  RIS/IRS-aided localization techniques by differentiating between far-field  and near-field channel models. Promising research directions for future work are highlighted in Section \ref{futuredirection} and conclusions are drawn in Section \ref{hrforh}.

\textbf{Notations:} ${{\bf{1}}_M}$ and ${{\bf{0}}_M}$ are  column vectors of all ones and all zeros, respectively. The Hadamard, Kronecker and Khatri-Rao products between two matrices ${\bf A}$ and ${\bf B}$ are denoted by ${\bf {A}}\odot{\bf {B}}$, $\mathbf{A}\otimes\mathbf{B}$ and $\mathbf{A}\diamond\mathbf{B}$, respectively.  ${\left\Vert {\bf {A}}\right\Vert _{2}}$ denote the    2-norm of matrix ${\bf {A}}$. $\mathbf{A}_{(:,n)}$ and $\mathbf{A}_{(m,:)}$ denote the $n$-th column and the $m$-th row of matrix $\mathbf{A}$, respectively. ${\left( \cdot \right)^{*}}$, ${\left( \cdot \right)^{\rm{T}}}$ and ${\left(  \cdot \right)^{\rm{H}}}$ denote the conjugate, transpose and Hermitian operators, respectively.  Given  two matrices ${\bf{X}}_1$ and ${\bf{X}}_2$, we define $\left\langle {{{\bf{X}}_1},{{\bf{X}}_2}} \right\rangle  \buildrel \Delta \over = {\rm Re}\left\{ {{\rm{tr}}\left( {{\bf{X}}_1^{\rm{H}}{{\bf{X}}_2}} \right)} \right\}$, where ${\rm Re}\left\{ \cdot \right\}$ is the real-value operator. ${\rm Im}\left\{ \cdot \right\}$ is the imaginary-value operator. The symbol $\mathbb{C}$ denotes the complex field, $\mathbb{R}$ represents the real field, and $j\triangleq\sqrt{-1}$ is the imaginary unit.

\section{Channel Estimation}\label{channelestimation}
\begin{figure}
\centering
\includegraphics[width=2.2in]{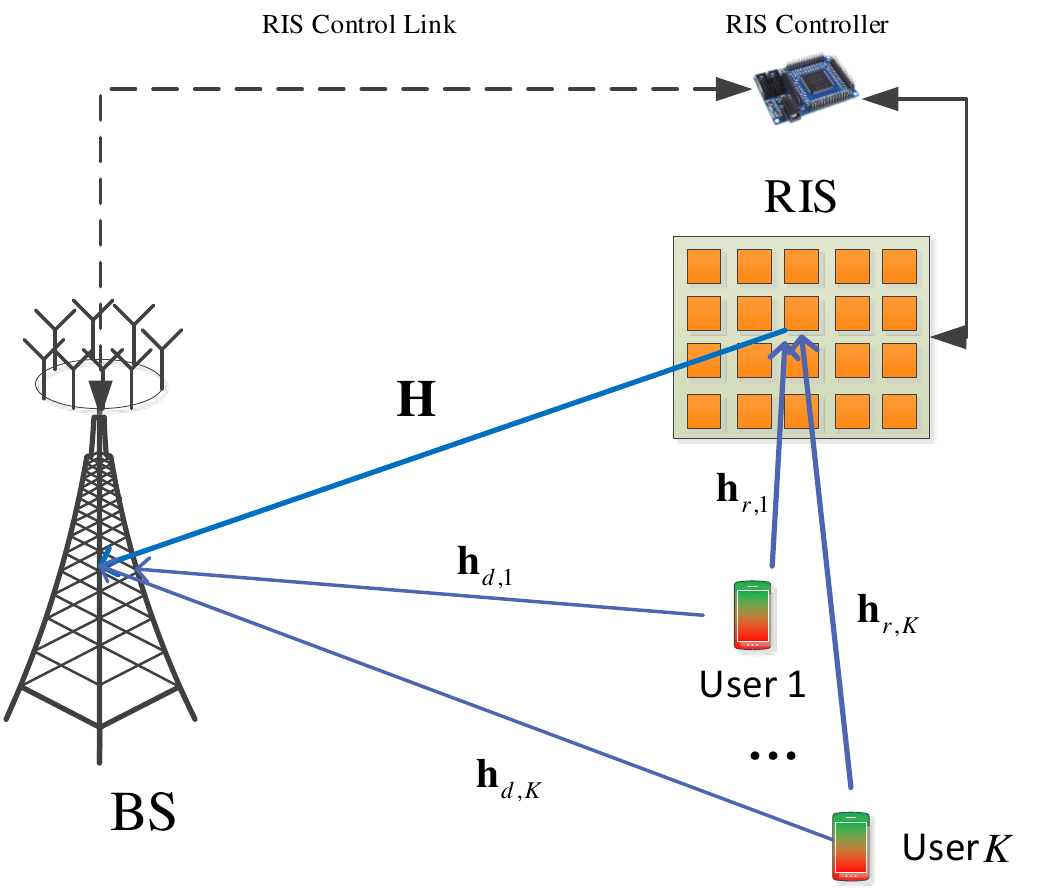}
\caption{Illustration of an RIS-aided uplink communication system.}
\vspace{-0.3cm}
\label{systemodel}
\end{figure}

In this section, we overview existing methods for channel estimation. For illustrative purposes, we focus our attention on the uplink, but similar considerations can be made for the downlink.
The uplink channel estimation scenario for a narrow-band RIS-aided\footnote{In the remaining parts, we use RIS to represent RIS/IRS for notational simplicity. } multi-user communication system is  shown in Fig.~\ref{systemodel}. The system model consists of one BS with $N$ antennas, one RIS with $M$ reflecting elements, and $K$ single-antenna users. Let  ${\bf{H}}\in {\mathbb{C}}^{N \times M}$ denote the RIS-BS channel, ${\bf{h}}_{r,k}\in {\mathbb{C}^{M \times 1}}$  the user-RIS channel of user $k$, and ${\bf{h}}_{d,k}\in {\mathbb{C}}^{N \times 1}$  the user-BS channel of user $k$. We assume that all the channels are subject to quasi-static fading and hence  the channel coefficients remain constant within one channel coherence interval.

At time slot $t$, the received baseband signal at the BS is given by
\begin{equation}\label{jofieio}
 {\bf{y}}_t = \sum\nolimits_{k = 1}^K \sqrt {{P_k}} {\left( {{{\bf{h}}_{d,k}} + {\bf{H }}{\bf{\Theta}}_t{{\bf{h}}_{r,k}}} \right)} {x_{k,t}} + {\mathbf{n}}_{t},
\end{equation}
where $x_{k,t}\in{\mathbb{C}}$ is the pilot signal transmitted by user $k$ satisfying the constraint $\left| {{x_{k,t}}} \right| = 1$, $P_k$ is the transmit power, ${\bf{n}}_t \sim {\cal CN}\left( {0,{\sigma ^2}{{\bf{I}}_N}} \right)$ is the additive white Guassian noise (AWGN), and ${\bf{\Theta }}_t$ is the reflection coefficient matrix of  the RIS. In general, ${\bf{\Theta }}_t$ is a diagonal matrix given by ${\bf{\Theta }}_t = {\rm{diag}}\left( {\bm{\theta }}_t \right)$, where ${\bm{\theta }}_t = {\left[ {{\theta_{t,1}}, \cdots ,{\theta_{t,M}}} \right]^{\rm{T}}}$ is the corresponding reflection coefficients at the RIS with $\theta_{t,m}={\alpha_m e^{j{\varphi _{t,m}}}}$ being the reflection coefficient corresponding to the $m$-th reflecting element. Here,  $\alpha_m$ and  ${\varphi _{t,m}} \in [0,2\pi ) $ are the amplitude and the phase shift  of the $m$-th element of the RIS, respectively. For simplicity, the reflection amplitude is assumed to be $\alpha_m=1, \forall m$.\footnote{Reflection amplitude variation can be exploited for further enhancing the multiuser communication performance, especially when the CSI is not perfectly estimated in practice\cite{MMzhao2021Amplitude}.}

A key  property for  channel estimation in RIS-aided systems is that there is a scaling ambiguity issue that prevents the RIS-BS channel and the user-RIS channel from being individually identifiable. Specifically, for any invertible  $M\times M$ diagonal matrix ${\bf{\Lambda }}$, we have
\begin{equation}\label{ederf}
  {\bf{H }}{\bf{\Theta}}_t{{\bf{h}}_{r,k}} = {\bf{H\Lambda  }}{\bf{\Theta}}_t{{\bf{\Lambda }}^{ - 1}}{{\bf{h}}_{r,k}} = {\bf{\tilde H }}{\bf{\Theta}}_t{{{\bf{\tilde h}}}_{r,k}},
\end{equation}
where ${\bf{\tilde H}}={\bf{H\Lambda }}$ and ${{{\bf{\tilde h}}}_{r,k}}={{\bf{\Lambda }}^{ - 1}}{{\bf{h}}_{r,k}}$. Hence, even if ${\bf{H }}{\bf{\Theta}}_t{{\bf{h}}_{r,k}}$ can be estimated  based on (\ref{jofieio}), one is still not able to extract the individual channels ${\bf{H}}$ and ${{\bf{h}}_{r,k}}$. Fortunately, there is generally no need to address this ambiguity issue when designing the phase shift matrix at the RIS for data transmission without loss of optimality.

Denote the cascaded channel for user $k$ as ${{\bf{G}}_k} = {\bf{H}}{\rm{diag}}\left( {{{\bf{h}}_{r,k}}} \right)\in \mathbb{C}^{N \times M}$. Then, the received signal in (\ref{jofieio}) can be rewritten as
\begin{equation}\label{jofswqswqio}
{\bf{y}}_t = \sum\nolimits_{k = 1}^K \sqrt {{P_k}}  {\left( {{{\bf{h}}_{d,k}} + {{\bf{G}}_k}{\bm{\theta}}_t} \right)} {x_{k,t}} + {\bf{n}}_t.
\end{equation}
Hence, as will be seen in Section III,  the cascaded channel ${\bf{G}}_k$ and the direct channel ${\bf{h}}_{d,k}$ are sufficient for  designing RIS-aided communications. As a result, most of the existing contributions have focused on designing algorithms to estimate the cascaded channels and the direct channel ${\bf{h}}_{d,k}$ separately\footnote{Due to the page limit, we do not discuss methods where the individual channels ${\bf{H }}$ and ${{\bf{h}}_{r,k}}$ are  estimated individually. Interested readers can refer to \cite{xiaojun,hangliu,Gilderlan,weili,huchen,Hu2021semiPassive} for more details.}.

In the following, we review the state-of-art techniques based on the structure of the channel models, namely, unstructured channel models and structured channel models.

\subsection{Unstructured Channel Models}

In this subsection, we consider the case when the channels are characterized by rich scattering. This is often the case in sub-6 GHz communication systems, where the propagation environment cannot be efficiently parameterized. In the following, we first consider the simple single-user case, and then address the multiuser case.

\subsubsection{\textbf{Single-user Case}} Since a single user is considered, the user index is omitted in the following derivations and the received signal model in (\ref{jofswqswqio}) becomes
\begin{equation}\label{jscdsio}
{\bf{y}}_t = \sqrt {{P}} {\left( {{{\bf{h}}_{d}} + {{\bf{G}}}{\bm{\theta}}_t} \right)} {x_{t}} + {\bf{n}}_t.
\end{equation}
The $m$-th column of ${\bf{G}}$ is denoted by ${\bf{g}}_m$. Define ${\bf{H}} = \left[ {{{\bf{h}}_1}, \cdots ,{{\bf{h}}_M}} \right]$ and ${{\bf{h}}_{r}} = \left[ {h_{r}^1; \cdots ;h_{r}^M} \right]$. Then, ${{\bf{g}}_m} = h_{r}^m{{\bf{h}}_m}$.  By defining the overall channel as ${\bf{c}} = {\left[ {{\bf{h}}_d^{\rm{T}},{\bf{g}}_1^{\rm{T}}, \cdots ,{\bf{g}}_M^{\rm{T}}} \right]^{\rm{T}}}$, (\ref{jscdsio}) can be rewritten as
\begin{equation}\label{dwdwdsio}
{{\bf{y}}_t} =\sqrt {{P}} \left({x_t} {\left[ {1,{\bm{\theta }}_t^{\rm{T}}} \right] \otimes {{\bf{I}}_N}} \right){\bf{c}} + {{\bf{n}}_t}.
\end{equation}
Assume that $T$ time slots are used for channel training, i.e, $T$ time slots are reserved for estimating the end-to-end channel, and define
\begin{equation}\label{dwwed}
{\bm{\Phi}}  = \left[ \begin{array}{c}
1,{\bm{\theta }}_1^{\rm{T}}\\
 \cdots \\
1,{\bm{\theta }}_T^{\rm{T}}
\end{array} \right] \in {{\mathbb{C}}^{T \times (M + 1)}},{\bm{\Xi}}  = {\bm{\Phi}}  \otimes {{\bf{I}}_N}.
\end{equation}
Stacking the $T$ training time slots together,  the overall received signal vector can be expressed as
\begin{align}
  \left[ {\begin{array}{*{20}{l}}
{{{\bf{y}}_1}}\\
 \vdots \\
{{{\bf{y}}_T}}
\end{array}} \right] &= \sqrt P \left[ {\begin{array}{*{20}{c}}
{{x_1}\left[ {1,{\bm{\theta }}_1^{\rm{T}}} \right] \otimes {{\bf{I}}_N}}\\
 \vdots \\
{{x_T}\left[ {1,{\bm{\theta }}_T^{\rm{T}}} \right] \otimes {{\bf{I}}_N}}
\end{array}} \right]{\bf{c}} + \left[ {\begin{array}{*{20}{l}}
{{{\bf{n}}_1}}\\
 \vdots \\
{{{\bf{n}}_T}}
\end{array}} \right]\nonumber\\
    &=\sqrt P {\bf{X}}{\bm \Xi} {\bf{c}} + {\bf{n}}, \label{yhdt}
\end{align}
where we have defined ${\bf{X}} = {\rm{diag}}\left( {\left[ {{x_1}{{\bf{1}}_N}; \cdots ;{x_T}{{\bf{1}}_N}} \right]} \right)$ and ${\bf{n}} = {\left[ {{\bf{n}}_1^{\rm{T}}, \cdots ,{\bf{n}}_T^{\rm{T}}} \right]^{\rm{T}}}$. By defining ${\bf{y}} = {\left[ {{\bf{y}}_1^{\rm{T}}, \cdots ,{\bf{y}}_T^{\rm{T}}} \right]^{\rm{T}}}$ and ${\bf{Z}}={\bf{X}}{\bm{\Xi}}$, the overall received signal vector in (\ref{yhdt}) can be written as
\begin{equation}\label{hnjikgujhy}
 {\bf{y}} =\sqrt {{P}} {\bf{Zc}} + {\bf{n}}.
\end{equation}
Our aim is to estimate $\bf{c}$ based on the pilot matrix $\bf{Z}$ that is assumed to be known and depends on the pilot sequence being used for channel estimation. To ensure that $\bf{c}$ can be uniquely estimated, $\bf{Z}$ must be a full rank matrix. Hence, the  number of time slots for channel training need to satisfy the condition $T\ge (M+1)$.

In general, based on (\ref{hnjikgujhy}), there are two common methods for estimating ${\bf{c}}$.

\textbf{\emph{Method I: Least Squares (LS) Estimator }}

The simplest method to estimate ${\bf{c}}$ is the LS estimator, which is formulated as
\begin{equation}\label{dggtr}
{\bf{\hat c}} = \mathop {\arg \min }\limits_{\bf{c}} {\left\| {{\bf{y}} - \sqrt P {\bf{Zc}}} \right\|^2},
\end{equation}
for which the  solution is
\begin{equation}\label{grg}
  {\bf{\hat c}} = \frac{1}{{\sqrt P }}{\left( {{{\bf{Z}}^{\rm{H}}}{\bf{Z}}} \right)^{ - 1}}{{\bf{Z}}^{\rm{H}}}{\bf{y}}.
\end{equation}
Assume that the noise vectors are uncorrelated, that is ${\bf{n}} \sim {\cal C}{\cal N}\left( {0,{\sigma ^2}{{\bf{I}}_{NT}}} \right)$. Then, the LS channel estimate is equivalent to the maximum likehood (ML) estimate, and it is an unbiased estimator, e.g., ${\mathbb{E}}\left\{ {{\bf{\hat c}}} \right\} = {\bf{c}}$. The error covariance matrix of the estimated channel is equal to the Cram${\rm {\acute{e}}}$r-Rao bound (CRB):
\begin{align}
	{\bf R}_{e} & \!=\!\mathbb{E}\left\{ \left({\bf c}-{\bf \hat{c}}\right)\left({\bf c}-{\bf \hat{c}}\right)^{\rm H}\right\} \nonumber\\
	&\!=\!\frac{1}{P}\mathbb{E}\left\{ \left({\bf Z}^{{\rm H}}{\bf Z}\right)^{-1}{\bf Z}^{{\rm H}}{\bf n}{\bf n}^{{\rm H}}{\bf Z}\left({\bf Z}^{{\rm H}}{\bf Z}\right)^{-1}\right\} \nonumber \\
	& \!=\!\frac{\sigma^{2}}{P}\left({\bf Z}^{{\rm H}}{\bf Z}\right)^{-1}=\frac{\sigma^{2}}{P}\left(\bm{\Xi}^{{\rm H}}\bm{\Xi}\right)^{-1}=\frac{\sigma^{2}}{P}\left(\bm{\Phi}^{\rm H}\bm{\Phi}\right)^{-1}\otimes{\bf I}_{N}.\label{fhoihtg}
\end{align}

Note that the error covariance matrix in (\ref{fhoihtg}) is not related to the training signals sent by the users, e.g., $\bf{X}$. Therefore, the optimization of the channel estimation boils down to the design of the training phase shift matrix ${\bm{\Phi}}$. In the following, two commonly used schemes are introduced.
 \begin{itemize}
   \item \textbf{\emph{On-off Scheme \cite{Mishraonoff,yifeiyang}:}} The simplest scheme is the on-off scheme adopted in \cite{Mishraonoff,yifeiyang}. The main idea is to switch each reflecting element on and off \cite{Mishraonoff}, or to switch groups of reflecting elements on and off \cite{yifeiyang} to reduce the training overhead. Specifically, in the first time slot, all the reflecting elements are switched off, and the BS estimates the direct channel ${\bf{h}}_d$. Then, in the remaining time slots, only one element (group) is switched on while keeping the others off. In this scheme, the number of time slots for channel training is equal to $T=M+1$. Note that ${\theta_{t,m}}={0,1}$ corresponds to the case when the reflecting element is off and on, respectively. Hence, the training phase shift matrix is given by
       \begin{equation}\label{jyfhy}
         {\bm{\Phi}}  =\left[\begin{array}{cc}
         	1 & {\bf 0}_{M}^{{\rm T}}\\
         	{\bf 1}_{M} & {\bf I}_{M}
         \end{array}\right].
       \end{equation}
       Then, based on (\ref{fhoihtg}), the error covariance matrix is equal to
      \begin{align}
      	{\bf {R}}_{e} & =\frac{{\sigma^{2}}}{P}{\left({{{\bf {\Xi}}^{{\rm {H}}}}{\bf {\Xi}}}\right)^{-1}}=\frac{{\sigma^{2}}}{P}{\left({{{\bf {\Phi}}^{\rm H}}{\bf {\Phi}}}\right)^{-1}}\otimes{{\bf {I}}_{N}}\nonumber \\
      	& =\frac{{\sigma^{2}}}{P}\left[\begin{array}{cc}
      		1 & -{\bf 1}_{M}^{{\rm T}}\\
      		-{\bf 1}_{M} & {\bf E}_{M}
      	\end{array}\right]\otimes{{\bf {I}}_{N}},
      \end{align}
       where ${\bf{E}}_M={{\bf{1}}_M}{\bf{1}}_M^{\rm{T}} + {{\bf{I}}_M}$. Hence, the error variance of the channels is obtained as
       \begin{equation}\label{ythytd}
        {\mathop{\rm var}} \left( {{{{\bf{\hat h}}}_d}} \right) = \frac{{{\sigma ^2}}}{P}{{\bf{I}}_N},{\mathop{\rm var}} \left( {{{{\bf{\hat g}}}_m}} \right) = \frac{{2{\sigma ^2}}}{P}{{\bf{I}}_N},m = 1, \cdots ,M.
       \end{equation}
       The factor  2 in the numerator of ${\mathop{\rm var}} \left( {{{{\bf{\hat g}}}_m}} \right)$
        is due to the error propagation of the estimation error of ${{\bf{h}}}_d$.  The main drawback of the on-off scheme is that switching off all the reflecting elements except one would reduce the reflected power, which degrades the received signal-to-noise ratio (SNR).
   \item \textbf{\emph{Discrete Fourier Transform (DFT) Scheme \cite{Beixiong,Jensen,zhengyi,zhengbx2021ofdm}:}} To enhance the reflected signal power, the discrete DFT training scheme was proposed in \cite{Beixiong}. Specifically, the training phase shifts at the RIS are optimized to minimize the mean squared error, and it is  demonstrated that the DFT training scheme achieved the optimal performance. Based on the DFT method, the training phase shift matrix ${\bm{\Phi}}$  is equal to the first $M+1$ columns of a $T\times T$ DFT matrix, which is given by
    \begin{equation*}
    {\left[ {\bm{\Phi}}  \right]_{t,m}} = {e^{ - j\frac{{2\pi (t - 1)(m - 1)}}{T}}},t = 1, \cdots ,T,m = 1, \cdots ,M.
    \end{equation*}
    Then, the error covariance matrix is equal to
    \begin{align}
    	{\bf {R}}_{e} & =\frac{{\sigma^{2}}}{P}{\left({{{\bf {\Xi}}^{{\rm {H}}}}{\bf {\Xi}}}\right)^{-1}}=\frac{{\sigma^{2}}}{P}{\left({{{\bf {\Phi}}^{\rm H}}{\bf {\Phi}}}\right)^{-1}}\otimes{{\bf {I}}_{N}}\nonumber \\
    	& =\frac{{\sigma^{2}}}{TP}{{\bf {I}}_{N(M+1)}}.\label{eq:hwsihduw}
    \end{align}
       Hence, the error variance of the channels is given by
       \begin{equation}\label{dxscs}
        {\mathop{\rm var}} \left( {{{{\bf{\hat h}}}_d}} \right) \!=\! {\mathop{\rm var}} \left( {{{{\bf{\hat g}}}_m}} \right) \!=\! \frac{{{\sigma ^2}}}{TP}{{\bf{I}}_N},m \!=\! 1, \cdots ,M+1.
       \end{equation}

   By comparing  (\ref{dxscs}) with (\ref{ythytd}), the DFT training scheme  reduces the channel error by a factor equal to $1/T$ for the direct channel ${{\bf{\hat h}}}_d$ and by a factor equal to $1/(2T)$ for the RIS-reflected channels ${{\bf{\hat g}}}_m, \forall m$. Recently, the DFT training scheme has been extended to the multiple-input multiple-output (MIMO) case in \cite{zhengyi} and to the orthogonal frequency division multiplexing (OFDM) system in \cite{zhengbx2021ofdm}.

   \item \emph{\textbf{Hadamard Matrix \cite{Youcs2020jsac,zhengyi}:}} The training phase shift matrix can also be designed using the first $M+1$ columns of  the $T
   \times T$ Hadamard  matrix \cite{zhengyi}, where $T=2$ or $T$ is a multiple of 4.  Specifically, one can construct a $T$-dimensional Hadamard matrix as follows
    \begin{equation*}
      {{\bf{D}}_{T}} = \left[ {\begin{array}{*{20}{c}}
{{{\bf{D}}_{T/2}}}&{{{\bf{D}}_{T/2}}}\\
{{{\bf{D}}_{T/2}}}&{ - {{\bf{D}}_{T/2}}}
\end{array}} \right],{{\bf{D}}_2} = \left[ {\begin{array}{*{20}{c}}
1&1\\
1&{ - 1}
\end{array}} \right],
    \end{equation*} with $T=2^n,n=1,2,\cdots$.
    It can be readily verified that ${\bf{D}}_{{2^B}}^{\rm{H}}{{\bf{D}}_{{2^B}}} = T {\bf{I}}$. Then,  the training phase shift matrix ${\bm{\Phi}}$  can be set equal to the first $M+1$ columns of  the matrix ${{\bf{D}}_{{2^B}}}$. Then, the error covariance matrix is calculated as  in (\ref{dxscs}), which means that setting the training phase shift matrix as the Hadamard matrix results in the minimum MSE of the estimator. Compared to the DFT matrix, the main advantage is that only two discrete phase shifts, namely $\{0,\pi\}$, are required for channel training, which can reduce the hardware complexity, and is thus an appealing solution from the implementation standpoint.

 \end{itemize}

\textbf{\emph{Method II: Linear Minimum Mean-Squared-Error (LMMSE)}}

The LS estimators do not exploit prior knowledge of the channel distributions. When such information is available, the optimal estimate of $\bf{c}$ that minimizes the MSE, which is defined as ${\mathbb{E}}\left[ {{{\left\| {{\bf{c}} - {\bf{\hat c}}} \right\|}^2}} \right]$, is the MMSE estimate, which is given by
\begin{equation}\label{yhfht}
 {{{\bf{\hat c}}}_{{\rm{mmse}}}} = {\mathbb{E}}\left[ {\left. {\bf{c}} \right|{\bf{y}}} \right].
\end{equation}
Since $\bf{c}$ depends on the cascaded channel that is the product of the user-RIS channel and the RIS-BS channel, the unknown vector $\bf{c}$ is, in general, not Gaussian distributed. This means that the posterior distribution $p(\left. {\bf{c}} \right|{\bf{y}})$ is difficult to obtain, and thus ${{{\bf{\hat c}}}_{{\rm{mmse}}}}$ cannot be readily calculated in a closed form expression. To address this issue, the LMMSE method was proposed in \cite{Kundu}.

Before introducing the LMMSE method, we first provide the distributions of various channels. The most common approach is to assume that the channel coefficients are correlated. Specifically, the three channels in the RIS-aided system model in Fig.~\ref{systemodel} can be written as
\begin{equation}\label{htydhgrs}
{\bf{H}} = {\bf{R}}_{{\rm{HB}}}^{\frac{1}{2}}{\bf{\tilde HR}}_{{\rm{HR}}}^{\frac{1}{2}},{{\bf{h}}_r} = {\bf{R}}_{{{\rm{h}}_{\rm{r}}}{\rm{R}}}^{\frac{1}{2}}{{{\bf{\tilde h}}}_r},{{\bf{h}}_d} = {\bf{R}}_{{{\rm{h}}_{\rm{d}}}{\rm{B}}}^{\frac{{\rm{1}}}{{\rm{2}}}}{{{\bf{\tilde h}}}_d},
\end{equation}
where  ${\bf{R}}_{{\rm{HB}}}$ and ${\bf{R}}_{{{\rm{h}}_{\rm{d}}}{\rm{B}}}$ are the spatial correlation matrices with unit diagonal elements at the BS for channel  $\bf{H}$ and ${{\bf{h}}_r}$, respectively, ${\bf{R}}_{{{\rm{h}}_{\rm{r}}}{\rm{R}}}$ and ${\bf{R}}_{{\rm{HR}}}$ are the spatial correlation matrices with unit diagonal elements at the RIS for channel ${{\bf{h}}_r}$ and ${{\bf{H}}}$, respectively. Then,  ${{\bf{\tilde H}}}$, ${{{\bf{\tilde h}}}_r}$, and ${{{\bf{\tilde h}}}_d}$ have probability distributions equal to ${\bf{\tilde H}} \sim {\cal C}{\cal N}\left( {{\bf{0}},{{\bf{I}}_M} \otimes {{\bf{I}}_N}} \right)$, ${{{\bf{\tilde h}}}_r} \sim {\cal C}{\cal N}\left( {{\bf{0}},{{\bf{I}}_M}} \right)$ and ${{{\bf{\tilde h}}}_d} \sim {\cal C}{\cal N}\left( {{\bf{0}},{{\bf{I}}_N}} \right)$, respectively.

Based on the above definitions, using the linear model in (\ref{hnjikgujhy}), the LMMSE of $\bf{c}$ is given by \cite{Kundu}
\begin{equation}\label{sgdrtg}
 {\bf{\hat c}} \!=\! {\mathbb{E}}\left[ {\bf{c}} \right] \!+\!\! \sqrt P {{\bf{C}}_{{\bf{cc}}}}{{\bf{Z}}^{\rm{H}}}{\left( {P{\bf{Z}}{{\bf{C}}_{{\bf{cc}}}}{{\bf{Z}}^{\rm{H}}} \!+\! {\sigma ^2}{{\bf{I}}_{TN}}} \right)^{\! - 1}}\!\left( {{\bf{y}}\! -\! {\bf{\bar y}}} \right),
\end{equation}
where ${{\bf{C}}_{{\bf{cc}}}} = {\mathbb{E}}\left[ {{\bf{c}}{{\bf{c}}^{\rm{H}}}} \right]$ and ${\bf{\bar y}}={\mathbb{E}}\{\bf{y}\}$. ${\bf{\bar y}}$ can be readily  shown to be the zero vector. Since $\bf{H}$ and ${{\bf{h}}_r}$ are independent and have zero mean, we  have ${\mathbb{E}}\left[ {\bf{c}} \right] =\bf{0}$. Hence, the matrix ${{\bf{C}}_{{\bf{cc}}}} $ can be formulated as
 \begin{equation}\label{rgdtrg}
 {{\bf{C}}_{{\bf{cc}}}} = \left[ {\begin{array}{*{20}{c}}
{{{\bf{R}}_{{{\rm{h}}_{\rm{d}}}{\rm{B}}}}}&{{{\bf{0}}_{N \times MN}}}\\
{{{\bf{0}}_{MN \times N}}}&{\left( {{{\bf{R}}_{{{\rm{h}}_{\rm{r}}}{\rm{R}}}} \odot {{\bf{R}}_{{\rm{HR}}}}} \right) \otimes {{\bf{R}}_{{\rm{HB}}}}}
\end{array}} \right].
 \end{equation}
Then, the error covariance matrix for the channel estimate is given by \cite{Kundu}
\begin{equation}\label{xdwacfefe}
  {{\bf{R}}_e} \!\!=\! {\left(\!\! {{\bf{C}}_{{\bf{cc}}}^{ - 1} \!+\! \frac{P}{{{\sigma ^2}}}{{\bf{Z}}^{\rm{H}}}{\bf{Z}}} \right)^{ - 1}}\!\!\!\!=\! {\left( {{\bf{C}}_{{\bf{cc}}}^{ - 1} \!+\! \frac{P}{{{\sigma ^2}}}{{\bm{\Phi}} ^{\rm{H}}}{\bm{\Phi}} \otimes {{\bf{I}}_N}} \right)^{\! -\! 1}}.
\end{equation}
To minimize the MSE, the majorization-minimization (MM) algorithm was proposed in \cite{Kundu} to optimize the training phase shift matrix ${\bm{\Phi}}$. However, the simulation results in \cite{Kundu} show that the performance of the optimized phase shift matrix is similar to that of the DFT matrix.  For this reason, we assume the DFT-based training phase shift matrix, and the corresponding error covariance matrix of the LMMSE channel estimator is given by
\begin{equation}\label{xdwwdwedwe}
  {{\bf{R}}_e} = {\left( {{\bf{C}}_{{\bf{cc}}}^{ - 1} + \frac{{TP}}{{{\sigma ^2}}}{{\bf{I}}_{N(M + 1)}}} \right)^{ - 1}}.
\end{equation}
As for the special case when all the channels undergo uncorrelated Rayleigh fading, we obtain ${{\bf{C}}_{{\bf{cc}}}} = {{\bf{I}}_{N(M + 1)}}$ and
\begin{equation}\label{ygrsr}
  {{\bf{R}}_e} = \frac{{{\sigma ^2}}}{{{\sigma ^2} + TP}}{{\bf{I}}_{N(M + 1)}}.
\end{equation}
Hence, the error variance of the channels is given by
       \begin{equation}\label{cwedwcs}
        {\mathop{\rm var}} \left( {{{{\bf{\hat h}}}_d}} \right) = {\mathop{\rm var}} \left( {{{{\bf{\hat g}}}_m}} \right) = \frac{{{\sigma ^2}}}{{{\sigma ^2} + TP}}{{\bf{I}}_N},m = 1, \cdots ,M.
       \end{equation}
By comparing (\ref{cwedwcs}) with (\ref{dxscs}), we  observe that the LMMSE has a smaller estimation error as compared to the LS method.

\textbf{Training Overhead Analysis:} To ensure that $\bf{c}$ can be uniquely estimated using the LS or the LMMSE estimators,   the  number of time slots for channel training needs to fulfill the condition $T\ge (M+1)$. In typical setups, this results in an excessive training overhead, e.g.,when the number of reflecting elements $M$ is large. In this case, the remaining time slots for data transmission are, in fact, significantly reduced.

To reduce the pilot overhead, the authors of \cite{yifeiyang} and \cite{Beixiong} proposed the element grouping (EG) method. In practical RIS-aided communication systems, the channel associated with adjacent RIS reflecting elements may be highly correlated. In such cases, the EG method groups the adjacent elements and assigns the same reflection pattern to them. This scheme is effective when the reflecting elements are installed closely together. Assume that the group size is $J$. Thus, the number of groups is $M'=M/J$, which is assumed to be an integer. Then, define ${{\bm{\theta}}_t} = {{\bm{\theta}} '_t} \otimes {{\bf{1}}_J}$, where  ${\bm{\theta}} '_t\in {\mathbb{C}}^{M' \times 1}$, so that
\begin{equation}\label{hsrgrf}
 {\bf{G}}{{\bm{\theta}} _t} = {\bf{G}}\left( {{{{\bm{\theta}} '}_t} \otimes {{\bf{1}}_J}} \right) = {\bf{G}}\left( {{{\bf{I}}_{M'}} \otimes {{\bf{1}}_J}} \right){\bm{\theta}_t'}  = {\bf{G'}}{\bm{\theta}_t'} ,
\end{equation}
where $ {\bf{G'}}={\bf{G}}\left( {{{\bf{I}}_{M'}} \otimes {{\bf{1}}_J}} \right)\in {\mathbb{C}}^{N\times M'}$, and each column of ${\bf{G'}}$ is the unit-coefficient combination of the columns of ${\bf{G}}$ corresponding to the group of reflecting elements. Then, the model in (\ref{hsrgrf}) has the same form as the original models. Hence,  the above mentioned channel estimation methods can be directly applied. In this case, to estimate the direct channel ${\bf{h}}_d$ and the reflected channel ${\bf{G'}}$, the minimum training pilot overhead is $J+1$, which is lower than $M+1$ for the original model. When ${\bf{G'}}$ is estimated, the original channel ${\bf{G}}$ can be approximately recovered as ${\bf{G}} \approx {{\left( {{\bf{G'}} \otimes {\bf{1}}_J^{\rm{T}}} \right)} \mathord{\left/
 {\vphantom {{\left( {{\bf{G'}} \otimes {\bf{1}}_J^{\rm{T}}} \right)} J}} \right.
 \kern-\nulldelimiterspace} J}$.

\begin{figure}
	\centering
	\includegraphics[width=3.3in]{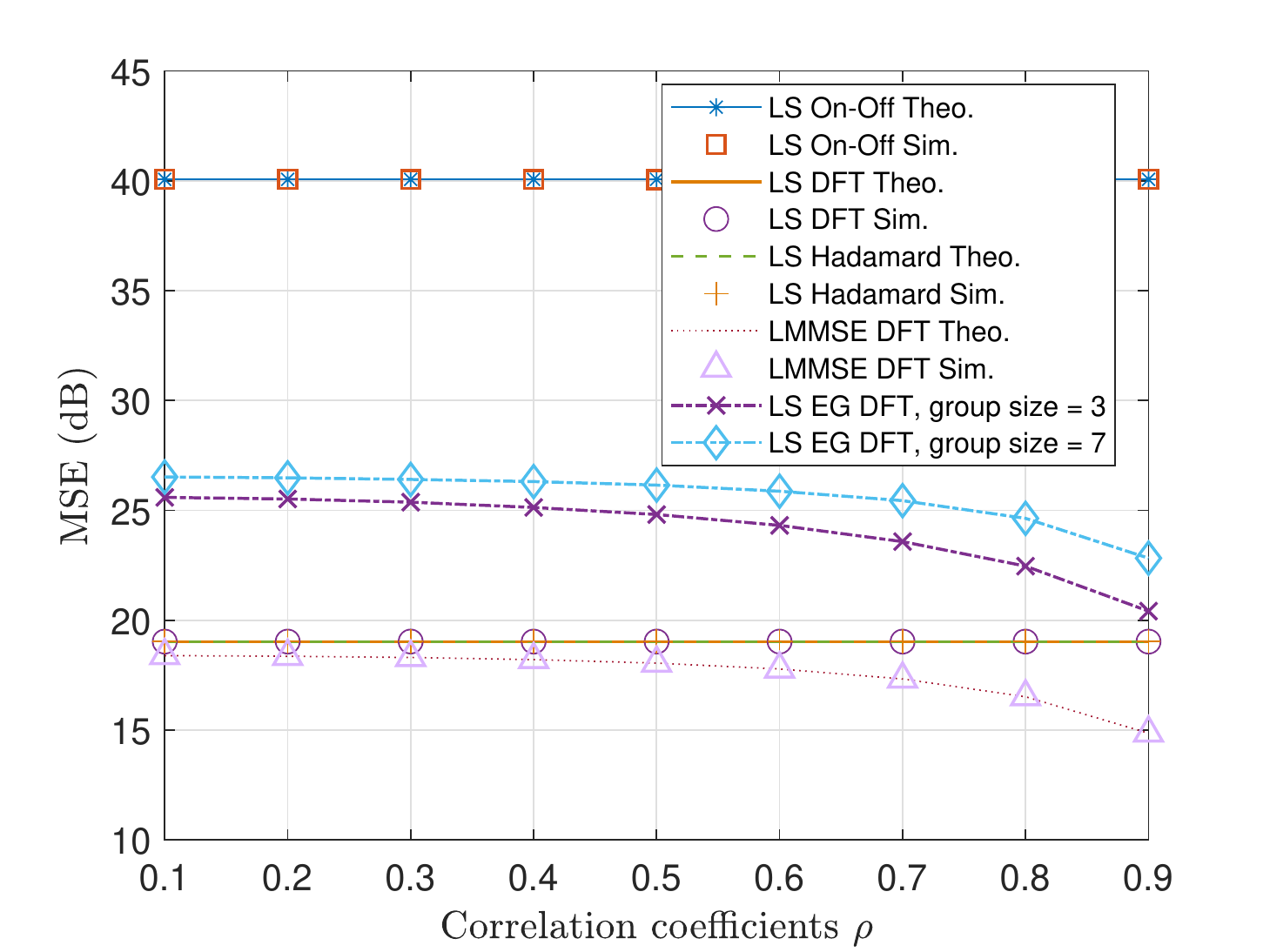}
	\caption{MSE comparison of different channel estimation schemes under the single-user setup, where $\rho=\rho_1=\rho_2=\rho_3=\rho_4$, $T=M+1=64$, $\gamma =  - 10$ dB.}
	\vspace{-0.4cm}
	\label{cdshhc}
\end{figure}
\textbf{Simulation Results: }For illustration, we consider a single-user scenario and compare the MSE performance of various algorithms in Fig.~\ref{cdshhc}. For simplicity, the exponential spatial correlation model is adopted, namely ${\left[ {{{\bf{R}}_{{\rm{HB}}}}} \right]_{i,j}} = \rho _1^{\left| {i - j} \right|},{\left[ {{{\bf{R}}_{{\rm{HR}}}}} \right]_{i,j}} = \rho _2^{\left| {i - j} \right|},\left[{{\bf{R}}_{{{\rm{h}}_{\rm{r}}}{\rm{R}}}}\right]_{i,j} = \rho _3^{\left| {i - j} \right|},\left[{{\bf{R}}_{{{\rm{h}}_{\rm{d}}}{\rm{B}}}}\right]_{i,j} = \rho _4^{\left| {i - j} \right|}$. The number of antennas and reflecting elements are  $N=8$ and $M=63$, respectively. The SNR is defined as $\gamma  = {P \mathord{\left/
 {\vphantom {P {{\sigma ^2}}}} \right.
 \kern-\nulldelimiterspace} {{\sigma ^2}}}$. The simulation results confirm the theoretical results. The LS-on-off scheme provides the worst MSE performance, while the LMMSE DFT approach offers the best estimation accuracy. Note that except for the LS EG DFT approach, the MSE performance of the other LS methods does not depend on the correlation coefficients since they do not exploit the prior knowledge  of the spatial correlation. In contrast, the MSE performance of the LMMSE DFT scheme decreases as the correlation increases. It is interesting to observe that the MSE of   the LS EG DFT  method decreases by increasing the spatial correlation and approaches the MSE of the LS DFT method without grouping, which demonstrates that the EG scheme is effective when the reflecting elements are strongly correlated.

\subsubsection{\textbf{Multi-user Case}}

Now, we consider the multi-user case, and briefly review several existing methods for channel estimation.

\textbf{\emph{Method I: Direct Channel Estimation Method \cite{Nadeem}}}

   In \cite{Nadeem}, the authors  proposed a channel estimation algorithm that is a direct extension of the single-user case. Specifically, the channel estimation period is divided into $T$ sub-phases, and the RIS applies the phase shift vector ${{\bm{\theta}}_t}$ in the $t$-th sub-phase. In each sub-phase, the users transmit orthogonal pilot sequences with length $K$, and the $k$-th user's pilot sequence is defined as ${\bf{x}}_k = {\left[ {{x_{k,1}}, \cdots ,{x_{k,K}}} \right]^{\rm{T}}}\in {\mathbb{C}}^{K\times 1}$, such that ${\bf{x}}_k^{\rm{H}}{{\bf{x}}_l} = 0$ for $k\ne l, \forall l,k$, and ${\bf{x}}_k^{\rm{H}}{{\bf{x}}_k} = K$. In different sub-phases, the users adopt the same set of pilot sequences while the RIS uses  different phase shift vectors. In the $t$-th sub-phase, the received training signal ${{\bf{Y}}_t}\in {\mathbb{C}}^{N\times K}$ at the BS is given by
      \begin{equation}\label{gdjgott}
        {{\bf{Y}}_t} \!\!=\! \sum\nolimits_{k = 1}^K {\!\!\sqrt {{P_k}} \left( {{{\bf{h}}_{d,k}} + {{\bf{G}}_k}{{\bm{\theta }_t}}} \right)} {\bf{x}}_k^{\rm{T}} + {\bf{N}}_t, t=1,\cdots, T,
      \end{equation}
      where ${\bf{N}}_t\in {\mathbb{C}}^{N\times K}$ is the noise matrix at the BS, whose columns are independent and have probability distribution  $ {\cal C}{\cal N}\left( {{\bf{0}},{\sigma ^2}{{\bf{I}}_N}} \right)$. By right multiplying both sides of (\ref{gdjgott}) with ${\bf{x}}_k^*$ and by taking into account that the users' pilot sequences are orthonormal, we have
      \begin{equation}\label{grse}
        {{\bf{y}}_{t,k}} =  {\sqrt {{P_k}} \left( {{{\bf{h}}_{d,k}} + {{\bf{G}}_k}{{\bm{\theta}} _t}} \right)}  + {{\bf{n}}_{t,k}},t = 1, \cdots ,T,
      \end{equation}
      where ${{\bf{y}}_{t,k}} = {{\bf{Y}}_t}{\bf{x}}_k^*$ and ${{\bf{n}}_{t,k}} = {{\bf{N}}_t}{\bf{x}}_k^*$. The obtained expression in (\ref{grse}) is similar to the single-user case in (\ref{jscdsio}), and the already described single-user channel estimation algorithms can be directly applied. For this scheme, the minimum number of sub-phases is equal to $T=M+1$ \cite{Nadeem}. Hence, the pilot overhead is   equal to $K(M+1)$.

\textbf{\emph{Method II: Exploiting the Common RIS-BS Channel \cite{Zhaorui}}}

The above  channel estimation scheme fails to exploit the inherent structure of RIS-aided communication systems, and requires a large amount of pilot training overhead, especially when $K$ and/or $M$ are large. Specifically, all users' cascaded channels share the same RIS-BS channel and, therefore, it is possible to reduce the channel training overhead since the number of independent complex variables to be estimated is $KN+NM+MK$ rather than $(M+1)NK$.

In \cite{Zhaorui}, the authors exploited the common RIS-BS channel and proposed a novel three-stage channel estimation algorithm.   In the first stage, the RIS is switched off, and the direct user-BS channels can be readily estimated. In the second stage, one reference user is selected and only this user can transmit its pilot signals. For ease of notation, let us denote this selected user as user 1. Since the direct channel ${\bf{h}}_{d,1}$ has been estimated in stage one, its impact when estimating the cascaded channel of user 1, i.e., ${\bf{G}}_1$, can be subtracted from the received signal at the BS, and the cascaded channel ${\bf{G}}_1$ can be estimated  using the same channel estimation methods as for the single-user case.

Now, we focus on the third stage. For simplicity, we assume that the channels estimated in the first two stages are perfect. By substituting ${{\bf{G}}_1} = {\bf{H}}{\rm{diag}}\left( {{{\bf{h}}_{r,1}}} \right)$ into (\ref{jofieio}), the received signal can be rewritten as
      \begin{equation}\label{grsrgt}
        \begin{array}{l}
{{\bf{y}}_t} = \sum\nolimits_{k = 1}^K {\sqrt {{P_k}} } {{\bf{h}}_{d,k}}{x_{k,t}} + \sqrt {{P_1}} {{\bf{G}}_1}{{\bm\theta} _t}{x_{1,t}}\\
\qquad + \sum\nolimits_{k = 2}^K {\sqrt {{P_k}} } {{\bf{G}}_1}{\rm{diag}}\left( {{{\bm\theta} _t}} \right){{{\bf{\tilde h}}}_k}{x_{k,t}} + {{\bf{n}}_t},
\end{array}
      \end{equation}
      where ${{{\bf{\tilde h}}}_k} = {\rm{diag}}{\left( {{{\bf{h}}_{r,1}}} \right)^{ - 1}}{{\bf{h}}_{r,k}},k = 2, \cdots ,K$. If ${{\bf{h}}_{r,k}}$ can be estimated, the cascaded channel ${{\bf{G}}_k}$ for $k=2, \cdots, K$ can  be calculated as ${{\bf{G}}_k} = {{\bf{G}}_1}{\rm{diag}}
      ( {{{\bf{\tilde h}}}_k}
      ) $. Hence, the remaining task is to estimate ${{\bf{h}}_{r,k}},k = 2, \cdots ,K$. In \cite{Zhaorui}, the authors assume that one user sends its pilot signals $x_{k,t}=1, P_k=1$ at each time and all the other users keep silent. Thus, we have $x_{j,t}=0$ for all $j\ne k$.  Then, the received signal is given by
      \begin{equation}\label{feafre}
        {{\bf{y}}_t} = {{\bf{h}}_{d,k}} + {{\bf{G}}_1}{\rm{diag}}\left({{{\bm{\theta}} _t}} \right){{{\bf{\tilde h}}}_k} + {{\bf{n}}_t}.
      \end{equation}
      In the following, we discuss the two cases where $M$ is less and greater than $N$, respectively.

      \emph{\textbf{Case I: $M\le N$}.}
   By setting ${{\bm{\theta}} _t} = {\bf{1}}$,  ${{{\bf{\tilde h}}}_k}$ can be directly estimated as
      \begin{equation}\label{dythtgr}
        {{{\bf{\tilde h}}}_k} = {\bf{G}}_1^{ - 1}\left( {{{\bf{y}}_t} - {{\bf{h}}_{d,k}}} \right).
      \end{equation}
     Accordingly, a single time slot is required to estimate $ {{{\bf{\tilde h}}}_k}$ in this case.

     \emph{\textbf{Case II: $M> N$}.}
  In this case,  ${{{\bf{\tilde h}}}_k}$ cannot be estimated using (\ref{dythtgr}) since the system of equations in (\ref{feafre})   is under-determined. To address this issue, a larger number of time slots for training is required. In the first training time slot,  the first $N$ reflecting elements are switched on and the reflection coefficients are set equal to  $\left[ {{\theta _{t,1}},{\theta _{t,2}} \cdots ,{\theta _{t,N}}} \right]^{\rm{T}} = {\bf{1}}$. Then, the first $N$ coefficients of ${{{\bf{\tilde h}}}_k}$ can be estimated as follows
     \begin{equation}\label{vgrsfde}
       \left[ \begin{array}{l}
{{\tilde h}_{k,1}}\\
{{\tilde h}_{k,2}}\\
 \vdots \\
{{\tilde h}_{k,N}}
\end{array} \right] \!\!=\!\! {\left[ {\begin{array}{*{20}{c}}
{{G_{1,11}}}& \cdots &{{G_{1,1N}}}\\
 \vdots & \ddots & \vdots \\
{{G_{1,N1}}}& \cdots &{{G_{1,NN}}}
\end{array}} \right]^{ - 1}}\left( {{{\bf{y}}_t} - {{\bf{h}}_{d,k}}} \right),
     \end{equation}
     where ${\tilde h}_{k,i}$ is the $i$-th element of ${{{\bf{\tilde h}}}_k}$ and ${{G_{1,ij}}}$ is the $(i,j)$-th entry of the matrix ${{\bf{G}}_1}$.
     In the second time slot, the next $N$  reflecting elements are switched on, i.e., $\left[ {{\theta _{t,N + 1}},{\theta _{t,2}} \cdots ,{\theta _{t,2N}}} \right]^{\rm{T}} = {\bf{1}}$, while  the others are switched off. The corresponding $N$ coefficients in ${{{\bf{\tilde h}}}_k}$ can be estimated  using again (\ref{vgrsfde}). This procedure is repeated until all the coefficients in ${{{\bf{\tilde h}}}_k}$ are estimated. For each user,  at least $\left\lceil {\frac{M}{N}} \right\rceil $ time slots are required to estimate ${{{\bf{\tilde h}}}_k}$.

      By combining Case I with Case II, the minimum number of required pilots is equal to $K + M + \max \left( {K - 1,(K - 1)\left\lceil {M/N} \right\rceil } \right)$, which is much less than the number of pilots that is needed for the  channel estimation algorithm in \cite{Nadeem}. The idea of this approach has been extended to multiuser OFDMA in \cite{beixiongtwcchan}.

The main issue of the method introduced in  \cite{Zhaorui} is the error propagation. The authors of \cite{wei2021channel} and \cite{guo2021cascaded} have proposed  channel estimation protocols to address this issue, where the direct channel and the reflected channels are estimated at the same time. The amount of pilot overhead required  by the solutions proposed in \cite{wei2021channel} and \cite{guo2021cascaded} is the same as that of the method proposed in \cite{Zhaorui}.

\subsection{Structured Channel Models}

\label{structured}

A geometric channel model or the Saleh-Valenzuela (SV) channel model
\cite{SV-MODEL} is usually used to characterize the channels in the mmWave
and THz frequency bands, where multipath scattering is sparse and the number of channel
parameters is small. Assuming a uniform linear array (ULA) at the BS and that the RIS elements are arranged in  uniform planar array (UPA), the SV channel model of the RIS-BS link in the presence of ${L_{\mathrm{BR}}}$ spatial paths is given by \cite{SV-MODEL}
\begin{align}
	\mathbf{H} & =\sum\nolimits_{l=1}^{L_{\mathrm{BR}}}\alpha_{l}\mathbf{a}_{\mathrm{B}}\left(\omega_{\mathrm{BH},l}\right)\mathbf{a}_{\mathrm{R}}^{\mathrm{H}}\left(\boldsymbol{\omega}_{\mathrm{RH},l}\right)\nonumber \\
	& =\mathbf{A}_{\mathrm{B}}(\boldsymbol{\omega}_{\mathrm{BH}})\boldsymbol{\Lambda}_{\mathrm{BR}}\mathbf{A}_{\mathrm{R}}^{\mathrm{H}}(\boldsymbol{\omega}_{\mathrm{RH}}),
\end{align}
where the diagonal matrix $\boldsymbol{\Lambda}_{\mathrm{BR}}=\mathrm{diag}(\alpha_{1},\ldots,\alpha_{L_{\mathrm{BR}}})$
contains the complex path gains $\boldsymbol{\alpha}=[\alpha_{1},\ldots,\alpha_{L_{\mathrm{BR}}}]^{\mathrm{T}}$.
The columns of ${{\bf{A}}_{\rm{R}}}({{\bf{\omega }}_{{\rm{RH}}}}){\rm{ }} = \left[ {{{\bf{a}}_{\rm{R}}}\left( {{{\bf{\omega }}_{{\rm{RH}},1}}} \right), \ldots ,{{\bf{a}}_{\rm{R}}}\left( {{{\bf{\omega }}_{{\rm{RH}},{L_{{\rm{BR}}}}}}} \right)} \right]$
denote the steering vectors of an $M_{x}\times M_{z}$ UPA, namely the RIS, on the
$xz$-plane, which are
expressed as
\begin{align}
	\mathbf{a}_{\mathrm{R}}\left(\boldsymbol{\omega}_{\mathrm{RH},l}\right) & =\mathbf{a}_{x}(\omega_{x,l})\otimes\mathbf{a}_{z}(\omega_{z,l}),\forall l,\label{eq:res-UPA}
\end{align}
where $\mathbf{a}_{x}(\omega_{x,l})=[1,e^{j\omega_{x,l}},\ldots,e^{j(M_{x}-1)\omega_{x,l}}]^{\mathrm{T}}$ is the horizontal array response with spatial frequency $\omega_{x,l}=2\pi\frac{d}{\lambda_{c}}\sin(\phi_{el,l})\cos(\phi_{az,l})$, and $\mathbf{a}_{z}(\omega_{z,l})=[1,e^{j\omega_{z,l}},\ldots,e^{j(M_{z}-1)\omega_{z,l}}]^{\mathrm{T}}$ is the vertical array response with spatial frequency $\omega_{z,l}=2\pi\frac{d}{\lambda_{c}}\cos(\phi_{el,l})$.
Here, $d$ denotes the antenna spacing, $\lambda_{c}$ is the
carrier wavelength, and $\phi_{el,l}$ and $\phi_{az,l}$ denote the elevation and the azimuth, respectively, angles of departures (AoDs) from the RIS. The two-dimensional (2D) spatial frequency vector
$\boldsymbol{\omega}_{\mathrm{RH},l}=[\omega_{x,l},\omega_{z,l}]^{\mathrm{T}}$
corresponds to the 2D spatial angles $\{\phi_{el,l},\phi_{az,l}\}$.

The steering vectors of an $N$-element ULA  ${{\bf{A}}_{\rm{B}}}({{\bf{\omega }}_{{\rm{BH}}}}){\rm{ }} = \left[ {{{\bf{a}}_{\rm{B}}}\left( {{\omega _{{\rm{BH}},1}}} \right), \ldots ,{{\bf{a}}_{\rm{B}}}\left( {{\omega _{{\rm{BH}},{L_{{\rm{BR}}}}}}} \right)} \right]$
 are the same as (\ref{eq:res-UPA}), but the angles of elevation are
$\phi_{el,l}=\frac{\pi}{2},\forall l$. Hence, $\mathbf{a}_{\mathrm{B}}(\omega_{\mathrm{BH},l})$ is
\begin{align}
	\mathbf{a}_{\mathrm{B}}(\omega_{\mathrm{BH},l}) & =[1,e^{j\omega_{\mathrm{BH},l}},\ldots,e^{j(N-1)\omega_{\mathrm{BH},l}}]^{\mathrm{T}},\forall, l,\label{eq:res-ULA}
\end{align}
where $\boldsymbol{\omega}_{\mathrm{BH}}$ is the angle of arrival (AoA) spatial frequency vector at the BS, which is given by $\boldsymbol{\omega}_{\mathrm{BH}}=[\omega_{\mathrm{BH},1},\ldots,\omega_{\mathrm{BH},L_{\mathrm{BR}}}]$.

The SV channel model for  the user-RIS
link in the presence of ${L_{\mathrm{RU}}}$ spatial paths has a similar structure. Precisely, denoting
by $\mathbf{h}_{r,k}\in\mathbb{C}^{M\times 1}$ the channel from user
$k$ to the RIS, it can be formulated as
\begin{align}
	\mathbf{h}_{r,k} & =\sum\nolimits_{l=1}^{L_{\mathrm{RU}}}\beta_{k,l}\mathbf{a}_{\mathrm{R}}\left(\boldsymbol{\omega}_{\mathrm{RH}_{r},k,l}\right)\nonumber \\
	& =\mathbf{A}_{\mathrm{R}}(\boldsymbol{\omega}_{\mathrm{RH}_{r},k})\boldsymbol{\beta}_{k},
\end{align}
where  $\boldsymbol{\beta}_{k}=[\beta_{k,l},\ldots,\beta_{k,L_{\mathrm{RU}}}]^{\mathrm{T}}$
is the vector of complex path gains, and the matrix $\mathbf{A}_{\mathrm{R}}(\boldsymbol{\omega}_{\mathrm{RH}_{r},k})=[\mathbf{a}_{\mathrm{R}}\left(\boldsymbol{\omega}_{\mathrm{RH}_{r},k,1}\right),\ldots,\mathbf{a}_{\mathrm{R}}\left(\boldsymbol{\omega}_{\mathrm{RH}_{r},k,L_{\mathrm{RU}}}\right)]$
collects the steering vectors of the propagation paths whose AoA spatial
frequency vector is  $\boldsymbol{\omega}_{\mathrm{RH}_{r},k}=[\boldsymbol{\omega}_{\mathrm{RH}_{r},k,1}^{\mathrm{T}},\ldots,\boldsymbol{\omega}_{\mathrm{RH}_{r},k,L_{\mathrm{RU}}}^{\mathrm{T}}]^{\mathrm{T}}$.

Accordingly, the received baseband signal in (\ref{jofieio}) is rewritten
as\footnote{For simplicity, we ignore the direct BS-user channels  since they are likely
	blocked by obstacles at high frequency bands.}
\begin{align}
	{\bf {y}}_{t} & =\sum\nolimits _{k=1}^{K}\sqrt{{P_{k}}}{\bf {H}}{\bf {\Theta}}_{t}\mathbf{h}_{r,k}{x_{k,t}}+{\mathbf{n}}_{t}\nonumber \\
	& =\sum\nolimits _{k=1}^{K}\sqrt{P_{k}}{\bf G}_{k}\bm{\theta}_{t}x_{k,t}+{\mathbf{n}}_{t}.\label{eq:FE}
\end{align}

\subsubsection{Single-user case\label{structured-single}}

Let us start with the analysis of the single-user case. By omitting the user index for simplicity,
(\ref{eq:FE}) reduces to
\begin{equation}
	{\bf y}_{t}=\sqrt{P}{\bf G}\bm{\theta}_{t}x_{t}+\mathbf{n}_{t}.\label{jofieio-2}
\end{equation}
The overarching  idea of estimating a channel with an inherent geometric structure  is to reconstruct
the channel by estimating a small number  of angles and gains instead
of directly estimating all the channel coefficients. We overview several existing channel estimation methods based on this approach.

\textbf{\emph{Method I: One-Stage Channel Estimation}} \cite{Peilan,hengliu}

In this method, the estimation of the cascaded channel is reformulated in terms of  an AoA estimation problem
\begin{equation}
	{\bf y}_{t}=\mathbf{Z}\mathbf{A}(\boldsymbol{\omega})\boldsymbol{\gamma}+\mathbf{n}_{t},\label{eq:aoa-prp}
\end{equation}
where  $\mathbf{Z}$ is the known sensing matrix that contains the training pilots, the matrix $\mathbf{A}(\boldsymbol{\omega})$
is the array response that depends on the spatial frequency
vector $\boldsymbol{\omega}$ and  $\boldsymbol{\gamma}$ is the vector with the unknown channel gains. The detailed definitions of $\mathbf{Z}$,  $\mathbf{A}(\boldsymbol{\omega})$ and  $\boldsymbol{\gamma}$ will be given later.

Accordingly, any known AoA estimation algorithms, including
the multiple signal classification (MUSIC) \cite{Music}, the estimation of signal parameters via rational invariance techniques (ESPRIT) \cite{ESPRIT} and compressed sensing (CS) based
techniques, can be directly applied. Specifically, by utilizing the identity
$\mathbf{A}\mathbf{B}\diamond\mathbf{C}\mathbf{D}=(\mathbf{A}\otimes\mathbf{C})(\mathbf{B}\diamond\mathbf{D})$,
the cascaded channel ${\bf G}$ can be formulated as
\begin{align}
	\!\!\!{\bf G} & \!=\!{\bf H}\mathrm{diag}(\mathbf{h}_{r})={\bf H}\diamond\mathbf{h}_{r}^{\mathrm{T}}\nonumber \\
	& \!=\!\mathbf{A}_{\mathrm{B}}(\boldsymbol{\omega}_{\mathrm{BH}})\!\left(\boldsymbol{\Lambda}_{\mathrm{BR}}\otimes\boldsymbol{\beta}^{\mathrm{T}}\right)\!\left(\mathbf{A}_{\mathrm{R}}^{\mathrm{H}}(\boldsymbol{\omega}_{\mathrm{RH}})\diamond\mathbf{A}_{\mathrm{R}}^{\mathrm{T}}(\boldsymbol{\omega}_{\mathrm{RH}_{r}})\right).\label{eq:G}
\end{align}

Denoting by $\boldsymbol{\omega}=[\boldsymbol{\omega}_{\mathrm{BH}}^{\mathrm{T}},\boldsymbol{\omega}_{\mathrm{RH}}^{\mathrm{T}},\boldsymbol{\omega}_{\mathrm{RH}_{r}}^{\mathrm{T}}]^{\mathrm{T}}$, using the identity $\mathrm{vec}(\mathbf{A}\left(\mathrm{diag}(\mathbf{b})\otimes\mathbf{d}^{\mathrm{T}}\right)\mathbf{C})=(\mathbf{C}^{\mathrm{T}}\diamond\mathbf{A})(\mathbf{b}\otimes\mathbf{d})$
and vectorizing both sides of (\ref{eq:G}), we arrive at
\begin{align}
	\mathrm{vec}(\mathbf{G}) & =\mathbf{A}(\boldsymbol{\omega})\boldsymbol{\gamma},\label{eq:fe}
\end{align}
where
\begin{align}
	\mathbf{A}(\boldsymbol{\omega}) & =\left(\mathbf{A}_{\mathrm{R}}^{\mathrm{T}}(\boldsymbol{\omega}_{\mathrm{RH}_{r}})\diamond\mathbf{A}_{\mathrm{R}}^{\mathrm{H}}(\boldsymbol{\omega}_{\mathrm{RH}})\right)^{\mathrm{T}}\diamond\mathbf{A}_{\mathrm{B}}(\boldsymbol{\omega}_{\mathrm{BH}}),\label{eq:re}\\
	\boldsymbol{\gamma} & =\boldsymbol{\alpha}\otimes\boldsymbol{\beta}.\label{eq:ew}
\end{align}
The matrix $\left(\mathbf{A}_{\mathrm{R}}^{\mathrm{T}}(\boldsymbol{\omega}_{\mathrm{RH}_{r}})\diamond\mathbf{A}_{\mathrm{R}}^{\mathrm{H}}(\boldsymbol{\omega}_{\mathrm{RH}})\right)^{\mathrm{T}}$
is  the cascaded array response at the RIS \cite{Jiechen},
whose $k$-th column is
\begin{align}
	\left[\left(\mathbf{A}_{\mathrm{R}}^{\mathrm{T}}(\boldsymbol{\omega}_{\mathrm{RH}_{r}})\diamond\mathbf{A}_{\mathrm{R}}^{\mathrm{H}}(\boldsymbol{\omega}_{\mathrm{RH}})\right)_{k,:}\right]^{\mathrm{T}} & =\left[\mathbf{a}_{\mathrm{R}}^{\mathrm{T}}(\boldsymbol{\omega}_{\mathrm{RH}_{r},p})\diamond\mathbf{a}_{\mathrm{R}}^{\mathrm{H}}(\boldsymbol{\omega}_{\mathrm{RH},q})\right]^{\mathrm{T}}\nonumber \\
	& =\mathbf{a}_{\mathrm{R}}(\boldsymbol{\omega}_{\mathrm{RH}_{r},p}-\boldsymbol{\omega}_{\mathrm{RH},q}),\label{eq:ji-1}
\end{align}
where $p=\left\lfloor k/L_{\mathrm{BR}}\right\rfloor $ and $q=\mathrm{mod}_{L_{\mathrm{BR}}}(k)$.
The vector $\boldsymbol{\gamma}$ depends on the complex path gains of the
 cascaded channel.

By combining (\ref{jofieio-2}) with (\ref{eq:fe}),  (\ref{jofieio-2})
can be reformulated in a compact form as
\begin{align}
	{\bf y}_{t} & =\sqrt{P}\left(x_{t}\bm{\theta}_{t}^{\mathrm{T}}\otimes {\bf{I}}_{N}\right)\mathrm{vec}(\mathbf{G})+\mathbf{n}_{t}\nonumber \\
	& =\sqrt{P}\left(x_{t}\bm{\theta}_{t}^{\mathrm{T}}\otimes {\bf{I}}_{N}\right)\mathbf{A}(\boldsymbol{\omega})\boldsymbol{\gamma}+\mathbf{n}_{t}.\label{eq:ko}
\end{align}

Similar to (\ref{yhdt}), the measurement (pilot) signals received at the BS
over $T$ training time slots are stacked into a vector, as
\begin{align}
	{\bf y} & =\sqrt{P}\mathbf{Z}\mathbf{A}(\boldsymbol{\omega})\boldsymbol{\gamma}+\mathbf{n}\in\mathbb{C}^{NT\times1},\label{eq:hi}
\end{align}
where
\[
\mathbf{Z}=\left[\begin{array}{c}
	x_{1}\bm{\theta}_{1}^{\mathrm{T}}\otimes {\bf{I}}_{N}\\
	\vdots\\
	x_{T}\bm{\theta}_{T}^{\mathrm{T}}\otimes {\bf{I}}_{N}
\end{array}\right].
\]

By inspection of (\ref{eq:hi}), the obtained signal model corresponds to a standard  AoA estimation problem as formulated in (\ref{eq:aoa-prp}). Therefore, the deterministic
maximum likehood  (DML) criterion can be adopted to estimate $\boldsymbol{\omega}$
 \cite{Lee}
\[
\hat{\boldsymbol{\omega}}=\mathrm{arg}\min_{\boldsymbol{\omega}}\thinspace{\bf y}^{\mathrm{H}}\mathbf{P}_{\mathbf{Z}\mathbf{A}(\boldsymbol{\omega})}^{\perp}{\bf y},
\]
where $\mathbf{P}_{\mathbf{Z}\mathbf{A}(\boldsymbol{\omega})}^{\perp}$
is the projection orthogonal to the columns
of  $\mathbf{Z}\mathbf{A}(\boldsymbol{\omega})$. Given the estimate, the vector of channel gains is given by $\hat{\boldsymbol{\gamma}}=(\mathbf{Z}\mathbf{A}(\hat{\boldsymbol{\omega}}))^{\dagger}{\bf y}/\sqrt{P}$.

However, the DML criterion ignores the structural properties of the signal in (\ref{eq:hi}). More precisely,
the cascaded gain vector $\boldsymbol{\gamma}$ in (\ref{eq:ew})
is a nonlinear function of $\boldsymbol{\alpha}$ and $\boldsymbol{\beta}$,
and the elements of $\boldsymbol{\gamma}$ are not independent. In addition, the spatial frequency
of each cascaded steering vector in (\ref{eq:ji-1}) is different. In order to exploit the inherent relationship
between the columns of the cascaded array response matrix, the authors of  \cite{Peilan} and \cite{hengliu}
adopted the CS technique, according to which each array response matrix  $\mathbf{A}(\boldsymbol{\omega})$
is replaced by a dictionary matrix composed of potential steering
vectors. Specifically, (\ref{eq:fe}) is approximated  using the virtual
angular domain (VAD) representation, i.e., \cite{Peilan}
\begin{align}
	\mathrm{vec}(\mathbf{G}) & \approx \mathbf{A}_{D}\boldsymbol{\gamma}_{D},\label{eq:fe-1}
\end{align}
where $\boldsymbol{\gamma}_{D}$ is an $L_{\mathrm{BR}}L_{\mathrm{RU}}$-sparse
vector with $L_{\mathrm{BR}}L_{\mathrm{RU}}$ non-zero elements corresponding
to the cascaded channel gains. The matrix $\mathbf{A}_{D}$ is the composite dictionary
of $\mathbf{A}(\boldsymbol{\omega})$ and is defined as $\mathbf{A}_{D}=\left(\mathbf{A}_{\mathrm{R}D}^{\mathrm{T}}\diamond\mathbf{A}_{\mathrm{R}D}^{\mathrm{H}}\right)^{\mathrm{T}}\otimes\mathbf{A}_{\mathrm{B}D}$.
The columns of the overcomplete dictionary matrix $\mathbf{A}_{\mathrm{B}D}\in{\mathbb{C}}^{N\times G_{\mathrm{B}}}$
($N\ll G_{\mathrm{B}}$) are the BS steering vectors whose sample spatial
frequencies are chosen from pre-discretized grids. However, the overcomplete
dictionary matrix $\mathbf{A}_{\mathrm{R}D}\in{\mathbb{C}}^{M\times G_{\mathrm{R}}}$
($M\ll G_{\mathrm{R}}$) is specified by a two-dimensional grid and
its columns can be written as $\mathbf{a}_{x}(\omega_{x,q})\otimes\mathbf{a}_{z}(\omega_{z,p}),1\leq q,p\leq\sqrt{G_{\mathrm{R}}}$
with $\omega_{x,q}$ and $\omega_{z,p}$ being independently selected from pre-discretized
grids. By capitalizing on the sparse approximation in (\ref{eq:fe-1}),
the estimation of $\mathbf{G}$ can be transformed into a sparse signal
recovery problem as
\begin{subequations}
	\label{Problem-4}
	\begin{align}
		\min_{\boldsymbol{\gamma}_{D}} & \thinspace\thinspace||\boldsymbol{\gamma}_{D}||_{1}\label{omiga}\\
		\mathrm{s.t.} & \thinspace\thinspace||\mathbf{y}-\sqrt{P}\mathbf{Z}\mathbf{A}_{D}\boldsymbol{\gamma}_{D}||_{2}\leq\xi,\label{alpha}
	\end{align}
\end{subequations}
where the 1-norm of $\boldsymbol{\gamma}_{D}$ enforces the sparse
structure and $\xi\geq0$ is an error tolerance parameter that depends on the noise power. Problem (\ref{Problem-4}) can be solved by using several classical CS algorithms,
such as the orthogonal matching pursuit (OMP) \cite{Peilan} and the alternating
direction method of multipliers (ADMM) \cite{hengliu}. After recovering
$\boldsymbol{\gamma}_{D}$, the  cascaded channel $\mathbf{G}$
can be retrieved from (\ref{eq:fe-1}). However, the main issue with the
CS approach used for solving Problem (\ref{Problem-4}) is the large size of
$\mathbf{A}_{D}$, which has $G_{\mathrm{B}}G_{\mathrm{R}}^{2}$ columns.
If   each spatial frequency dimension has 100 grids, this matrix has $10^{10}$ columns. Therefore, a more tractable and practical
method is needed. According to \cite{Candes}, $l\log(\frac{m}{l})$
measurements are needed for successful recovering an  $l$-sparse vector
of dimension $m\times1$. Thus, the required pilot overhead for
solving Problem (\ref{Problem-4}) is $T\geq\frac{L_{\mathrm{BR}}L_{\mathrm{RU}}}{N}\log(\frac{G_{\mathrm{B}}G_{\mathrm{R}}^{2}}{L_{\mathrm{BR}}L_{\mathrm{RU}}})$.

\textbf{\emph{Method II: Two-Stage Channel Estimation \cite{Khaled,TianLin,jiguan}}}

A different approach for channel estimation consists of splitting the cascaded channel
into two stages. In the first stage, the AoAs at the BS are estimated.
After eliminating these estimated angles from the variables to the estimated, only the
 cascaded spatial frequencies and the channel gains need to be
determined. The estimation of both stages can be formulated in terms of several several
AoA estimation problems, thus any AoA estimation algorithm can be
employed.

Specifically, it is assumed that the  user  sends the pilot signal $x_{t}=1$ during
$T$ time slots. Defining $\boldsymbol{\Xi}=[\bm{\theta}_{1},...,\bm{\theta}_{T}]$
and stacking the $T$ observations in (\ref{jofieio-2}), we obtain
\begin{align}
	{\bf Y} & =[\mathbf{y}_{1},\ldots,\mathbf{y}_{T}]\nonumber \\
	& =\sqrt{P}{\bf G}\boldsymbol{\Xi}+\mathbf{N},\label{eq:io}
\end{align}
where $\mathbf{N}=[\mathbf{n}_{1},\ldots,\mathbf{n}_{T}]$. Based on (\ref{eq:io}), we elaborate the estimation process of both stages.
\begin{itemize}
	\item \textbf{First Stage:} We rewrite (\ref{eq:io})
	as
	\begin{align}
		{\bf Y} & =\sqrt{P}\mathbf{A}_{\mathrm{B}}(\boldsymbol{\omega}_{\mathrm{BH}})\mathbf{B}^{\mathrm{T}}\boldsymbol{\Xi}+\mathbf{N}\in\mathbb{C}^{N\times T},\label{eq:op}
	\end{align}
	where $\mathbf{B}=(\boldsymbol{\Lambda}_{\mathrm{BR}}\mathbf{A}_{\mathrm{R}}^{\mathrm{H}}(\boldsymbol{\omega}_{\mathrm{RH}})\mathrm{diag}(\mathbf{h}_{r}))^{\mathrm{T}}$.
	Any AoA estimation
	algorithms can be used to determine $\boldsymbol{\omega}_{\mathrm{BH}}$
	from (\ref{eq:op}). For example, the authors of \cite{Khaled} and \cite{TianLin}
	approximated (\ref{eq:op}) as a  sparse signal recovery problem similar to
	Problem (\ref{Problem-4}), and adopted the OMP technique to solve the formulated problem.
	Since $N\gg L_{\mathrm{BR}}$,  (\ref{eq:op}) readily satisfies 	the condition of the number of measurements required for successful recovery, i.e., $N\geq L_{\mathrm{BR}}\log(\frac{G_{\mathrm{B}}}{L_{\mathrm{BR}}})$.
	The main issue with  the on-grid CS method is the mismatch between
	the estimated and the actual angles. The estimation error can be improved
	by enlarging the dimension of the dictionary, but this also reduces
	the orthogonality of the dictionary, and leads to a higher computational complexity. Furthermore, the authors of \cite{xiaowei} and \cite{Gui}
	first transformed the signal in (\ref{eq:op}) from the spatial domain to
	the angle domain by utilizing a DFT matrix, and then estimated the AoAs
	by searching for the non-zero elements in the angle domain. However, this
	method is limited to massive MIMO scenarios, where the steering
	vectors  lie approximately in
	the DFT matrix space.  The estimation error caused by the slight power
	leakage can also be improved using the angle rotation technique \cite{Gui}.
	Moreover, to address the mismatch issue, the atomic norm minimization
	technique was adopted in \cite{jiguan} to estimate the off-grid angles
	 by employing convex optimization tools.
	\item \textbf{Second Stage: } Assuming that the estimate
	$\hat{\boldsymbol{\omega}}_{\mathrm{BH}}$ of $\boldsymbol{\omega}_{\mathrm{BH}}$ is perfect,
	we obtain
	\begin{align}
		{\bf Y}_{\mathrm{R}} & =\frac{1}{\sqrt{P}}(\mathbf{A}_{\mathrm{B}}^{\dagger}(\hat{\boldsymbol{\omega}}_{\mathrm{BH}}){\bf Y})^{\mathrm{T}}\nonumber \\
		& \approx\boldsymbol{\Xi}^{\mathrm{T}}\mathbf{B}+\mathbf{N}_{b}\in\mathbb{C}^{T\times L_{\mathrm{BR}}},\label{eq:YB}
	\end{align}
	where $\mathbf{N}_{b}=\frac{1}{\sqrt{P}}(\mathbf{A}_{\mathrm{B}}^{\dagger}(\hat{\boldsymbol{\omega}}_{\mathrm{BH}})\mathbf{N})^{\mathrm{T}}$ and $\mathbf{B}$ is the matrix to be estimated.
	
	\textbf{\textit{ Correlation-ignored Approach}} \textbf{\emph{\cite{Khaled,jiguan}}}.
	
	If one ignores the correlation among the cascaded gains and spatial
	frequencies, the estimation of ${\bf Y}_{\mathrm{R}}$ can be obtained
	by constructing $L_{\mathrm{BR}}$ $L_{\mathrm{RU}}$-dimensional AoA
	estimation problems \cite{Khaled,jiguan}. In particular,
	the $l$-th column of (\ref{eq:YB}) is given by
	\begin{align}
		[{\bf Y}_{\mathrm{R}}]_{:,l} & =\boldsymbol{\Xi}^{\mathrm{T}}\mathrm{diag}(\mathbf{h}_{r})\mathbf{a}_{\mathrm{R}}^{*}(\boldsymbol{\omega}_{\mathrm{RH},l})\alpha_{l}+[\mathbf{N}_{b}]_{:,l}\nonumber \\
		& =\boldsymbol{\Xi}^{\mathrm{T}}\mathbf{A}_{\mathrm{R}}(\triangle\boldsymbol{\omega}_{\mathrm{Rh_{r}},l})\alpha_{l}\boldsymbol{\beta}+[\mathbf{N}_{b}]_{:,l},\label{eq:t1}
	\end{align}
which depends on $\triangle\boldsymbol{\omega}_{\mathrm{Rh_{r}},l}=[\boldsymbol{\omega}_{\mathrm{Rh_{r}},1}^{\mathrm{T}}-\boldsymbol{\omega}_{\mathrm{RH},l}^{\mathrm{T}},...,\boldsymbol{\omega}_{\mathrm{Rh_{r}},L_{\mathrm{RU}}}^{\mathrm{T}}-\boldsymbol{\omega}_{\mathrm{RH},l}^{\mathrm{T}}]^{\mathrm{T}}$
	and   $\alpha_{l}\boldsymbol{\beta}$. Thus, (\ref{eq:YB})
	can be decomposed into $L_{\mathrm{BR}}$ $L_{\mathrm{RU}}$-dimensional
	AoA estimation problems along with determining $L_{\mathrm{BR}}L_{\mathrm{RU}}$
	cascaded path gains at the RIS. With the estimates of $\triangle\boldsymbol{\omega}_{\mathrm{Rh_{r}},l}$
	and $\alpha_{l}\boldsymbol{\beta}$ from (\ref{eq:t1}), the $l$-th
	column of $\mathbf{B}$ is estimated as
	\begin{equation}
		[{\bf B}]_{:,l}=\mathbf{A}_{\mathrm{R}}(\triangle\widehat{\boldsymbol{\omega}}_{\mathrm{Rh_{r}},l})\widehat{\alpha_{l}\boldsymbol{\beta}}.\label{eq:t1-1}
	\end{equation}
	
	\textbf{\textit{Correlation-based Approach \cite{Gui}}}.
	
	If one considers the fact that the cascaded gains and spatial frequencies
	are interrelated, the estimation complexity can be further reduced
	to solve one $L_{\mathrm{RU}}$-dimensional AoA estimation problem
	plus $L_{\mathrm{BR}}-1$ one-dimensional AoA estimation problems.
	In particular, $[{\bf Y}_{\mathrm{R}}]_{:,l}$ in (\ref{eq:t1}) is
	rewritten as
	\begin{align}
		[{\bf Y}_{\mathrm{R}}]_{:,l} & =\triangle\alpha_{l}\boldsymbol{\Xi}^{\mathrm{T}}\mathrm{diag}\left(\mathbf{a}_{\mathrm{R}}^{\mathrm{H}}(\triangle\boldsymbol{\omega}_{\mathrm{RH},l})\right)[{\bf B}]_{:,1}+[\mathbf{N}_{b}]_{:,l},\label{eq:yu}
	\end{align}
	where $\triangle\alpha_{l}=\frac{\alpha_{l}}{\alpha_{1}}$ and $\triangle\boldsymbol{\omega}_{\mathrm{RH},l}=\boldsymbol{\omega}_{\mathrm{RH},l}-\boldsymbol{\omega}_{\mathrm{RH},1}$.
	The first column of ${\bf B}$ is first estimated from (\ref{eq:t1})
	using an $L_{\mathrm{RU}}$-dimensional AoA estimation problem. Substituting
	the estimated $[{\bf B}]_{:,1}$ into (\ref{eq:yu}), then $[{\bf B}]_{:,l}$
	for $l=2,...,L_{\mathrm{BR}}$ can be determined by only estimating
	$\triangle\alpha_{l}$ and $\triangle\boldsymbol{\omega}_{\mathrm{RH},l}$
	through a one-dimensional AoA estimation problem. Therefore, only
	$L_{\mathrm{BR}}+L_{\mathrm{RU}}-1$ equivalent cascaded paths need
	to be estimated at the RIS. Compared with  the correlation-ignored approach, the correlation-based approach enjoys a much lower computational complexity since it needs to solve the $L_{\mathrm{RU}}$-dimensional AoA estimation problem only once.
	
	We characterize the pilot overhead at this stage based on the condition
	for successful recovery using the CS technique: $T\geq L_{\mathrm{RU}}\log(\frac{G_{\mathrm{R}}}{L_{\mathrm{RU}}})$.
\end{itemize}
\textbf{\emph{Method III: Beam Training for Channel Estimation}} \cite{Changshengwcl,Boyu,Zegrar}.

A different method for estimating  sparse channels is beam training, which
is suitable for line-of-sight (LoS)-dominant channels where the  LoS path
is much stronger than  the non-LoS (NLoS) components. Specifically, the composite
LoS-dominant user-RIS-BS channel can be formulated as
\begin{equation*}
	{\bf H}{\bf \Theta}_{t}\mathbf{h}_{r}=\alpha\beta\mathbf{a}_{\mathrm{B}}\left(\omega_{\mathrm{BH}}\right)\mathbf{a}_{\mathrm{R}}^{\mathrm{H}}\left(\boldsymbol{\omega}_{\mathrm{RH}}\right)\mathrm{diag}(\bm{\theta}_{t})\mathbf{a}_{\mathrm{R}}\left(\boldsymbol{\omega}_{\mathrm{RH}_{r}}\right).\label{eq:los}
\end{equation*}

It is readily seen that the maximum beam gain is obtained when the beams at the BS and the RIS align with the user-RIS-BS channel.
\begin{itemize}
	\item \textbf{BS Beam:} Let the vector $\mathbf{f}_{\mathrm{B}}\in\mathbb{C}^{N\times1}$
	be the combiner at the BS. The maximization of $|\mathbf{f}_{\mathrm{B}}^{\mathrm{H}}\mathbf{a}_{\mathrm{B}}\left(\omega_{\mathrm{BH}}\right)|$
	as a function of $\mathbf{f}_{\mathrm{B}}$ returns $\mathbf{f}_{\mathrm{B},\rm{opt}}=\mathbf{a}_{\mathrm{B}}^{\mathrm{H}}\left(\omega_{\mathrm{BH}}\right)$.
	\item \textbf{RIS Beam:} The optimal phase shifts at the RIS that maximize $|\mathbf{a}_{\mathrm{R}}^{\mathrm{H}}\left(\boldsymbol{\omega}_{\mathrm{RH}}\right)\mathrm{diag}(\bm{\theta}_{t})\mathbf{a}_{\mathrm{R}}\left(\boldsymbol{\omega}_{\mathrm{Rh}_{r}}\right)|$
	are given by $\bm{\theta}_{\rm{opt}}=\mathbf{a}_{\mathrm{R}}\left(\boldsymbol{\omega}_{\mathrm{RH}}-\boldsymbol{\omega}_{\mathrm{Rh}_{r}}\right)$.
\end{itemize}
To estimate the AoAs and the AoDs of the LoS-dominant user-RIS-BS channel, appropriate training
codebooks are first constructed for each beam by discretizing the
entire angular domain. Then,  an exhaustive beam training method can be used: the RIS applies the different beams of the codebook one-by-one in consecutive time slots. Then, the  BS  determines
the optimal BS and RIS beam directions that provide the maximum received
signal power/SNR. However, such an exhaustive beam training method is pilot-consuming
and requires at least $T=M_{x}$ time slots \cite{Changshengwcl}. Therefore,
various methods have been proposed to improve the efficiency of beam training,
such as the multi-beam training method \cite{Changshengwcl}, the ternary-tree
hierarchical search \cite{Boyu}, and the two-way tree stage search \cite{Zegrar}.

\begin{figure}
	\centering 
	\includegraphics[width=3.3in]{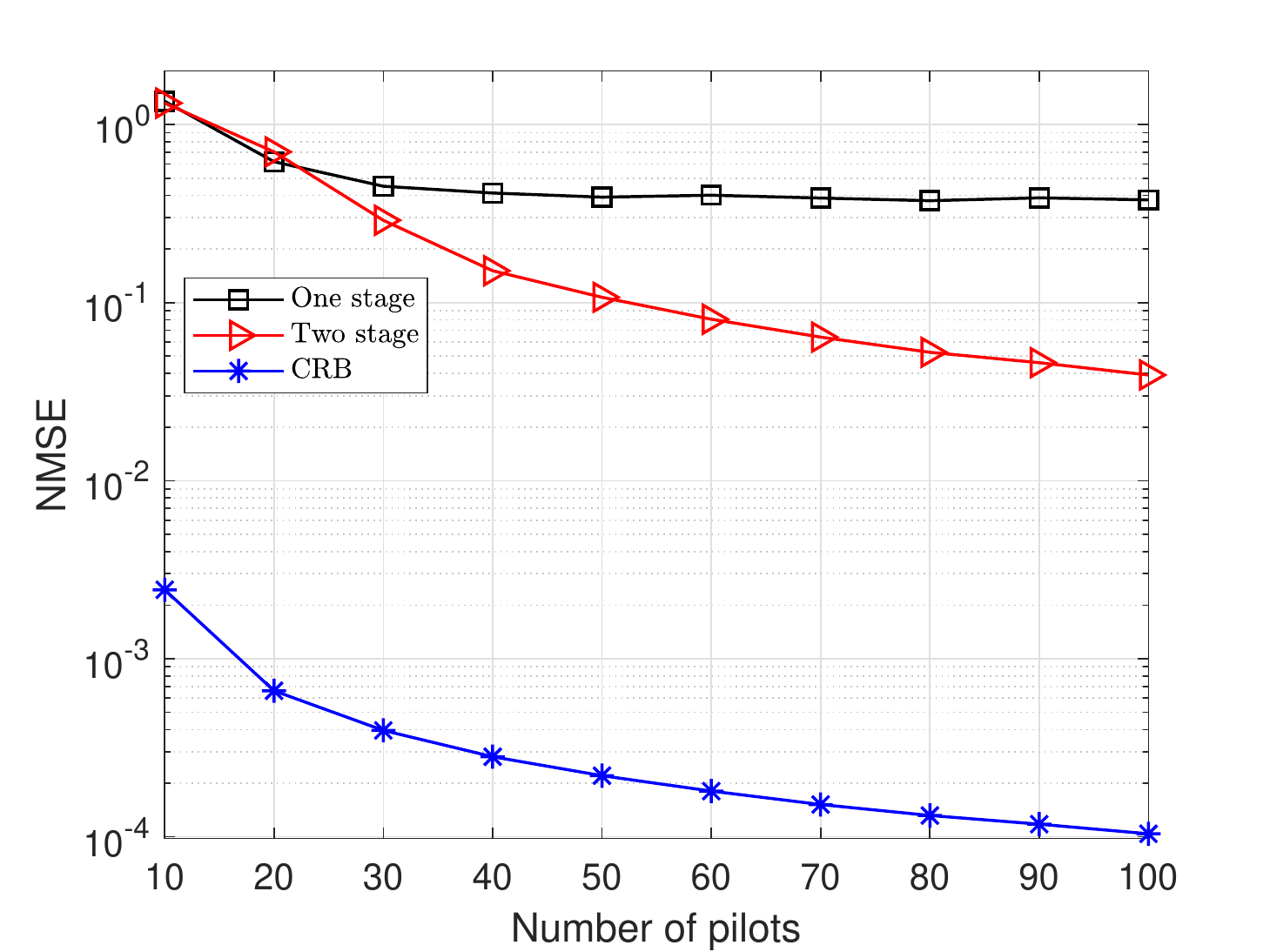}
	\caption{NMSE versus the pilot overhead, when $N=100$, $M=10\times10$, $L_{\mathrm{BR}}=L_{\mathrm{RU}}=4$
		and SNR$=0$ dB.}
	\vspace{-0.3cm}
	\label{nmse-geo}
\end{figure}

\textbf{Simulation Results:} As an example, Fig.~\ref{nmse-geo} shows the normalized mean square error (NMSE) of the cascaded channel matrix that is obtained by using the one-stage and two-stage estimation methods. The  NMSE is defined as
$\mathrm{NMSE}=\mathbb{E}\{\left\| {\widehat {\bf{G}} - {\bf{G}}} \right\|_F^2/\left\| {\bf{G}} \right\|_F^2\}.$ The OMP algorithm is adopted to solve the AoA estimation problems in both cases.
It can be seen that the two-stage method outperforms the one-stage method. This is because the two-stage method reduces the power leakage among the cascaded multipath components compared with (\ref{eq:fe-1}) by dividing the cascaded paths into several groups in (\ref{eq:t1}), each of which corresponds to one BS-RIS path and does not leak  power to other groups. However, there is a gap between the two methods and the CRB. This is due to  the inherent mismatch between the on-grid angle estimated by the OMP and the actual continuous angle.

\subsubsection{Multi-user case}

In the multi-user case, all users share the common channel ${\bf H}$.
This  fact can be utilized to reduce the pilot overhead
and the computational complexity as proposed  in \cite{Jiechen,xiaowei,Gui}.

\textbf{\emph{Method I: Double-sparse Based Channel Estimation}} \cite{Jiechen,xiaowei}.

The authors of \cite{Jiechen} investigated the row-column-block
sparsity of the multi-user cascaded channels and proposed a joint
multi-user channel estimation method:
\begin{itemize}
	\item \textbf{Common column-block sparsity due to the presence of common scatters near the
		BS:} Note that all users share the same AoA array responses at the BS. Once the common AoAs are
	estimated from a specific user using the methods introduced in Subsection
	\ref{structured-single}, the impact of the corresponding array responses can be
	eliminated from the received signals of all users.
	\item \textbf{Common row-block sparsity due to scaling property:} Based on
	\cite{Zhaorui}, the signals from all users propagate  through the same RIS,
	so the cascaded channels of different users are scaled by a diagonal
	array, as ${\bf G}_{k}={\bf G}_{1}{\rm diag}({\bf \tilde{h}}_{k})$
	defined in the text right after (\ref{grsrgt}). Then, all cascaded channels are approximated
	by the common row-block sparse representation as ${\bf G}_{k}=\mathbf{A}_{B}\mathbf{X}\mathbf{A}_{R}^{\mathrm{H}}{\rm diag}({\bf \tilde{h}}_{k})$,
	where the sparse matrix $\mathbf{X}$ contains the cascaded gains.
	The optimization of $\mathbf{X}$ and ${\bf \tilde{h}}_{k}$ is formulated
	into a multi-user joint sparse matrix recovery problem, and solved by
	using the iterative reweighted algorithm \cite{Jiechen,fangjun2016}.
\end{itemize}
The pilot overhead of this method is $T\geq K\left\lceil M/(L_{\mathrm{BR}}L_{\mathrm{RU}})\right\rceil $
\cite{Jiechen}.

\textbf{\emph{Method II: Exploiting the Common RIS-BS Channel}} \cite{Gui}.

The above approach still requires estimation of the unstructured vector
${\bf \tilde{h}}_{k}$, for $k=2,...,K$, the dimension of which is
proportional to $M$. In order to reduce the number of
parameters and pilot overhead, a geometric   version of ${\bf \tilde{h}}_{k}$
is constructed in \cite{Gui}:
\begin{itemize}
	\item First, the spatial frequencies and gains of the cascaded channel for
	user 1 are estimated using the method discussed in Subsection \ref{structured-single}
	with pilot overhead $T_{1}\geq L_{\mathrm{RU}}\log(\frac{G_{\mathrm{R}}}{L_{\mathrm{RU}}})$.
	\item Then, the cascaded channels for the remaining users are reformulated as ${\bf G}_{k}={\bf H}_{\mathrm{c}}{\rm diag}(\mathbf{h}_{\mathrm{c},k})$,
	for $k=2,...,K$, where ${\bf H}_{\mathrm{c}}={\bf H}{\rm diag}(\bar{\beta}\mathbf{a}_{\mathrm{R}}(\bar{\boldsymbol{\omega}}))^{-1}$
	can be constructed based on the parameters estimated from user 1.
	Please refer to \cite{Gui} for the detailed construction of ${\bf H}_{\mathrm{c}}$.
	The remaining unknow vector $\mathbf{h}_{\mathrm{c},k}={\rm diag}(\bar{\beta}\mathbf{a}_{\mathrm{R}}(\bar{\boldsymbol{\omega}}))\mathbf{h}_{r,k}$
	belongs to a standard AoA estimation model and can be estimated via
	various AoA estimation or CS methods. The required pilot overhead
	here is $T_{2}\geq(K-1)\frac{L_{\mathrm{RU}}}{L_{\mathrm{BR}}}\log(\frac{G_{\mathrm{R}}}{L_{\mathrm{RU}}})$.
\end{itemize}
 The overall pilot overhead for the scheme in \cite{Gui} is given by $T\geq L_{\mathrm{RU}}\log(\frac{G_{\mathrm{R}}}{L_{\mathrm{RU}}})+(K-1)\frac{L_{\mathrm{RU}}}{L_{\mathrm{BR}}}\log(\frac{G_{\mathrm{R}}}{L_{\mathrm{RU}}})$.

Table \ref{tableone} summarizes the pilot overhead of various algorithms
using  CS-based techniques. The dimension of the dictionary is generally
set to 2-4 times the number of antennas. Therefore, we can
substitute \textbf{$G_{\mathrm{B}}=4N$} and \textbf{ $G_{\mathrm{R}}=4M$}
into the third column to obtain a more intuitive relationship between
the pilot overhead and the number of antennas in the fourth column.

\begin{table*}
	\centering  \caption{Pilot Overhead of Various Channel Estimation Algorithms.}\label{tableone}
	\begin{tabular}{|c|c|c|c|}
		\hline
		\multicolumn{2}{|c|}{\textbf{Algorithm}} & \textbf{Pilot Overhead} & \textbf{Pilot Overhead, $G_{\mathrm{B}}=4N$ and $G_{\mathrm{R}}=4M$}\tabularnewline
		\hline
		\multirow{2}{*}{{\makecell[c]{Single \\ user}}} & {\makecell[c]{One stage\\ \cite{Peilan,hengliu} }}  & $  \frac{L_{\mathrm{BR}}L_{\mathrm{RU}}}{N}\log(\frac{G_{\mathrm{B}}G_{\mathrm{R}}^{2}}{L_{\mathrm{BR}}L_{\mathrm{RU}}}) $ & $ \frac{L_{\mathrm{BR}}L_{\mathrm{RU}}}{N}\log(\frac{64NM^{2}}{L_{\mathrm{BR}}L_{\mathrm{RU}}}) $\tabularnewline
		\cline{2-4} \cline{3-4} \cline{4-4}
		& {\makecell[c]{Two stage\\ \cite{Khaled,TianLin,jiguan} }}& $ L_{\mathrm{RU}}\log(\frac{G_{\mathrm{R}}}{L_{\mathrm{RU}}}) $ & $ L_{\mathrm{RU}}\log(\frac{4M}{L_{\mathrm{RU}}}) $\tabularnewline
		\hline
		\multirow{2}{*}{{\makecell[c]{Multiple \\ user}}}& \cite{Jiechen} & $ KM/(L_{\mathrm{BR}}L_{\mathrm{RU}}) $ & $ KM/(L_{\mathrm{BR}}L_{\mathrm{RU}}) $\tabularnewline
		\cline{2-4} \cline{3-4} \cline{4-4}
		& \cite{Gui} & $ L_{\mathrm{RU}}\log(\frac{G_{\mathrm{R}}}{L_{\mathrm{RU}}})+(K-1)\frac{L_{\mathrm{RU}}}{L_{\mathrm{BR}}}\log(\frac{G_{\mathrm{R}}}{L_{\mathrm{RU}}}) $ & $ L_{\mathrm{RU}}\log(\frac{4M}{L_{\mathrm{RU}}})+(K-1)\frac{L_{\mathrm{RU}}}{L_{\mathrm{BR}}}\log(\frac{4M}{L_{\mathrm{RU}}}) $\tabularnewline
		\hline
	\end{tabular}
\end{table*}

\section{Transmission Design}\label{transmissiondesign}
Based on the channels estimated by using the methods introduced in Section \ref{channelestimation}, the reflection coefficients of the RIS and the beamforming vectors at the BS can be jointly optimized to achieve the desired objectives, such as maximizing the sum spectral efficiency or energy efficiency, and minimizing the total energy consumption, symbol-error probability, or transmission delay, etc.

The same system model in Fig. \ref{systemodel} is considered, but we focus our attention on downlink transmission for illustrative purposes. The baseband  channels from the BS to the RIS, from the RIS to  user $k$, and from the BS to  user $k$ are denoted by ${\bf{H}}^{\rm{H}}\in {\mathbb{C}}^{M \times N}$, ${\bf{h}}_{r,k}^{\rm{H}}\in {\mathbb{C}^{1 \times M}}$ and ${\bf{h}}_{d,k}^{\rm{H}}\in {\mathbb{C}}^{1 \times N}$, respectively.
The transmitted data symbols for user $k$ is denoted by $s_k$ with ${\mathbb{E}}\{s_k\}=0$ and ${\mathbb{E}}\{s_k^2\}=1$.  The beamforming vector for user $k$ is ${\bf{w}}_k\in {\mathbb{C}}^{N \times 1}$. Then, the transmitted data at the BS is given by ${\bf{x}} = \sum\nolimits_{j = 1}^K {{{\bf{w}}_j}{s_j}}$, and the received signal at user $k$ is expressed as
\begin{equation}\label{frdgtrs}
  {y_k} = \left( {{\bf{h}}_{d,k}^{\rm{H}} + {\bf{h}}_{r,k}^{\rm{H}}{\bf{\Theta }}{{\bf{H}}^{\rm{H}}}} \right)\sum\nolimits_{j = 1}^K {{{\bf{w}}_j}{s_j}}  + {n_k},
\end{equation}
where $n_k \sim {\cal C}{\cal N}\left( {0,{\sigma_k^2}} \right)$ is the AWGN at user $k$ and ${\bf{\Theta }}$ is the reflection coefficient matrix defined in Section \ref{channelestimation}.  Using ${\bf{G}}_k^{\rm{H}} = {\rm{diag}}\left( {{\bf{h}}_{r,k}^{\rm{H}}} \right){{\bf{H}}^{\rm{H}}}$ and ${\bm{\theta }}= {\left[ {{\theta_{1}}, \cdots ,{\theta_{M}}} \right]^{\rm{T}}}$, the signal-to-interference-plus-noise-ratio (SINR) at user $k$ is given by
\begin{equation}\label{scafe}
{\rm{SIN}}{{\rm{R}}_k} = \frac{{{{\left| {\left( {{\bf{h}}_{d,k}^{\rm{H}} + {{\bm{\theta }} ^{\rm{T}}}{\bf{G}}_k^{\rm{H}}} \right){{\bf{w}}_k}} \right|}^2}}}{{\sum\nolimits_{i = 1,i \ne k}^K {{{\left| {\left( {{\bf{h}}_{d,k}^{\rm{H}} + {{\bm{\theta }} ^{\rm{T}}}{\bf{G}}_k^{\rm{H}}} \right){{\bf{w}}_i}} \right|}^2}}  + \sigma _k^2}}.
\end{equation}
Hence, it is observed from (\ref{scafe}) that the cascaded channel ${\bf{G}}_k^{\rm{H}}$ along with the direct channel ${\bf{h}}_{d,k}^{\rm{H}}$ are sufficient for designing the transmission of the BS and the RIS. The data rate of user $k$ is ${R_k} = {\log _2}\left( {1 + {\rm{SIN}}{{\rm{R}}_k}} \right)$.

For ease of writing, the collection of all beamforming vectors is denoted by ${\bf{W}} = \left[ {{{\bf{w}}_1}, \cdots ,{{\bf{w}}_K}} \right]$. The vast majority of the existing transmission design problems can be expressed in the following general form
\begin{equation}
\begin{aligned}
&\underset {\mathbf{W},{\bf{\Theta }}}{\min}\;\;f\left( {{\bf{W}},{\bf{\Theta }}} \right)\\
&\text{s.t.} \quad\text{C1}: {g_i}\left( {{\bf{W}},{\bf{\Theta }}} \right)\ge D_i,i = 1, \cdots ,I,\\
&\quad\quad\;\text{C2}:  {\theta _m} \in {{\cal S}_1}\ {\rm{or}}\ {{\cal S}_2},\forall m=1,2,\dots,M,\\
\end{aligned}
\label{opt-model}
\end{equation}
where $f\left( {{\bf{W}},{\bf{\Theta }}} \right)$  and ${g_i}\left( {{\bf{W}},{\bf{\Theta }}} \right)$ can be any functions that depend on $\mathbf{W}$ and ${\bf{\Theta }}$, and $I$ denotes the number of constraints in C1.  In constraint C2, ${\cal S}_1$  and ${\cal S}_2$ denote the set of continuous and discrete phase shifts respectively, which are given by
\begin{align}
 {{\cal S}_1} &= \left\{ {\left. \theta  \right|\theta  = {e^{j\varphi }},\varphi  \in \left[ {0,2\pi } \right)} \right\}, \label{s1}\\
   {{\cal S}_2} &= \left\{ {\left. \theta  \right|\theta  = {e^{j(l - 1)\Delta \varphi }},l = 1, \cdots ,L } \right\},\label{s2}
\end{align}
where $L = {2^b}$ with $b$ being the number of bits used to quantize the continuous phase shifts, and $\Delta \varphi  = {{2\pi } \mathord{\left/
 {\vphantom {{2\pi } L}} \right.
 \kern-\nulldelimiterspace} L}$. The formulation of constraints C1 and C2 implies that only the phase shifts of the reflecting elements are optimized. Some research works, e.g., \cite{yifei2020,zijian,tongbaitwc}, studied the case when both the phase shifts and the amplitudes of the reflecting elements can be optimized subject to the constraint $|\theta _m|\le 1, \forall m$. The possibility of optimizing both the phase shifts and the amplitudes simultaneously usually results in a more complex hardware design. Hence, we limit our overview to design methods in which only the phase shift can be adjusted. It is worth noting that some works considered practical RIS reflection models with phase-dependent amplitude \cite{Abeywickrama2020,Wenhao2020,Hongyu2021} or reflection models that account for the mutual coupling among the reflecting elements \cite{Gabriele,Xuewen,Andrea}. For simplicity, we only consider the case study with phase-independent and element-independent model  for ${{\cal S}_1}$ and ${{\cal S}_2}$.

In the following, we discuss the existing contributions on transmission design by focusing our attention on two aspects: 1) Optimization techniques; 2) CSI availability.

\subsection{Optimization Techniques}\label{algorithms}

There are two main difficulties in solving the optimization problem in (\ref{opt-model}): 1) The optimization problem is non-convex/NP-hard due to the unit modulus constraint or the discrete-valued phase shifts; 2) The phase shifts and the beamforming vectors are coupled as shown in the SINR expression in (\ref{scafe}). As a result, a globally optimal solution is difficult to obtain. Instead, most of the existing works have aimed at finding highly efficient and locally optimal solutions with low computational complexity. Fortunately, several works have confirmed that the suboptimal solutions can  achieve  improved system performance compared with the performance of systems in the absence of RISs.

If the phase shifts are given, the optimization problem reduces to a conventional beamforming design problem, which has been extensively studied in the literature. Inspired by this consideration, alternating optimization (AO) algorithms are usually applied to decouple the optimization variables. Due to the usually complicated expressions for the data rate/SINR, advanced algorithms such as the weighted minimum mean-square error (WMMSE) algorithm \cite{qingqing} or fractional programming (FP) \cite{kaimingshen} are often  used to convert the original intractable problem into a new tractable but approximated problem. Then, the beamforming vectors at the BS are typically obtained by  using existing beamforming design methods. In the following, we focus our attention on the optimization of the phase shifts of the RIS.

1) \textbf{Optimization Techniques for Continuous Phase Shifts.}

 Existing techniques for  optimizing the design of the continuous phase shifts can be classified into the following categories.

(1) \emph{Relaxation and projection \cite{jiechenaccess,guizhoutvt,yuzhangcl}}: The unit modulus constraint on the phase shift can be rewritten as ${{\cal S}_1} = \left\{ {\left. \theta  \right|\left| \theta  \right| = 1,\theta  \in \mathbb{C}} \right\}$. The idea of this technique is first to relax the non-convex constraint $\mathcal{S}_1$ to the convex constraint  ${{\tilde {\cal S}}_1} = \left\{ {\left. \theta  \right|\left| \theta  \right| \le 1,\theta  \in \mathbb{C}} \right\}$, and then to project the obtained solution onto the unit-modulus constraint ${{\cal S}_1}$. Accordingly, given the solution $\theta_m$ of the relaxed problem, the final solution is $\theta_m^\star={e^{j{\varphi _m}}}$, where $\varphi_m$ is the phase of $\theta_m$.

(2) \emph{Semidefinite relaxation (SDR) \cite{qingqingwutwc,miaocuiwcl,zhengchuwcl,peilanwangtvt,gangyangtwc,guizhouwclhardware,mingzeng2021}}: The SDR method is the most common method for optimizing the phase shifts under constraint $\mathcal{S}_1$, i.e., for continuous phase shifts. Define ${\bf{V}} = {\bm{\theta }}{{\bm{\theta }}^{\rm{H}}}$. Then, the unit modulus constraint can be  equivalently written as ${\bf{V}} \succ {\bf{0}}$ and ${\rm{rank}}\left( {\bf{V}} \right) = 1$. Because of the rank one constraint, the transformed problem is still non-convex. Based on the SDR method, the non-convex rank one constraint is removed. The obtained relaxed problem is a convex semidefinite program (SDP), which can be readily solved by using CVX \cite{grant2014cvx}. In general,  the obtained relaxed problem is not a rank-one solution, i.e., ${\rm{rank}}\left( {\bf{V}} \right) \ne 1$.  In this case, the Gaussian randomization method \cite{Sidiropoulos} is utilized to obtain a rank-one solution.

(3)\emph{ Majorization-Minimization (MM) algorithm \cite{chongwenhuangtwc,guimulticast,shongshencl,limengdong2020,cunhuajsac,zhangjietsp}}: The MM algorithm is another widely used technique for optimizing the phase shifts of the RIS. The MM algorithm is an iterative optimization method that approximates a difficult problem as a series of more tractable subproblems that are solved iteratively. Assume that the solution of the subproblem at the $t$-th iteration is ${\bm{\theta }}^t$ and the corresponding objective function is $f({\bm{\theta }}^t)$ \footnote{When the beamforming vectors in $\bf{W}$ are given, the objective function in (\ref{opt-model}) is denoted by $f({\bm{\theta }})$, and the functions in constraint C1 are denoted by $g_i({\bm{\theta }}), \forall i$.}. Based on the MM algorithm, a surrogate objective function $\tilde f\left( {\left. {\bm{\theta }} \right|{{\bm{\theta }}^t}} \right)$ is constructed, which fulfills the following three conditions: 1) $\tilde f\left( {\left. {\bm{\theta }}^t \right|{{\bm{\theta }}^t}} \right) = f\left( {{{\bm{\theta }}^t}} \right)$; 2) ${\left. {{\nabla _{\bm{\theta }}}\tilde f\left( {\left. {\bm{\theta }} \right|{{\bm{\theta }}^t}} \right)} \right|_{{\bm{\theta }} = {{\bm{\theta }}^t}}} = {\left. {{\nabla _{\bm{\theta }}}f\left( {{{\bm{\theta }}^t}} \right)} \right|_{{\bm{\theta }} = {{\bm{\theta }}^t}}}$; 3) $\tilde f\left( {\left. {\bm{\theta }} \right|{{\bm{\theta }}^t}} \right) \ge f\left( {\bm{\theta }} \right)$. If these conditions are fulfilled, the sequence of the solutions obtained by solving each subproblem will converge. By replacing the original objective function with the constructed function $\tilde f\left( {\left. {\bm{\theta }} \right|{{\bm{\theta }}^t}} \right)$ and removing the constant terms, the subproblem to be solved in each iteration is given by
 \begin{subequations}\label{asasedwsg}
\begin{align}
&{\mathop {\max }\limits_{{\bm{\theta }}}  \  {\mathop{\rm Re}\nolimits} \left\{ {{\bm{\theta }} ^{\rm{H}}{\bf{q}}^t } \right\}}
\\
&\textrm{s.t.}\quad \left| {{\theta _m}} \right| = 1, m = 1, \cdots ,M, \label{aadshxceur}
\end{align}
\end{subequations}
where ${\bf{q}}^t$ is a  constant complex vector at the $t$-th iteration. The  optimal solution to the optimization problem  in (\ref{asasedwsg}) is
\begin{equation}\label{dcsd}
{{\bm{\theta }}^{t+1}} = {e^{j\arg ({{\bf{q}}^t})}}.
\end{equation}
This procedure is iterated until convergence according to any criterion of convergence.  If the phase shifts of the RIS appear in the constraints of the optimization problem, the pricing-based method can be utilized \cite{cunhuajsac}.

(4) \emph{Manifold approach \cite{cunhuatwc,xianghaoyu,peilan2021,huiminspl,huayan2020}}: There exist different kinds of manifold methods. In this paper, we consider the complex circle manifold (CCM) method \cite{cunhuatwc} as an example. The constraint space in ${\cal S}_1$ can be regarded as the product of $M$ complex circles, which is a sub-manifold of $\mathbb{C}^{M}$ given by
\begin{equation}\label{johoii}
  {{\cal S}^M} \buildrel \Delta \over = \left\{ { {\bf{x}} \in {\mathbb{C}^M}:\left| {{x_l}} \right| = 1,l = 1,2, \cdots ,M} \right\},
\end{equation}
where $x_l$ is the $l$-th element of vector $\bf{x}$. The main idea of the CCM method is to derive the gradient descent algorithm based on the manifold space given in (\ref{johoii}). The optimization problem aims at optimizing the phase shifts to minimize  the objective function $\hat f\left( {\bm{\theta }} \right)$. The main steps can be summarized as follows.

 \emph{(a) Computation of the gradient in Euclidean space}: The search direction for the minimization problem is the opposite of the gradient of  $\hat f\left( {\bm{\theta }} \right)$, which is given by ${{\bm{\eta }}^t} =  - {\left. {{\nabla _{\bm{\theta }}}\hat f({\bm{\theta }})} \right|_{{\bm{\theta }} = {{\bm{\theta }}^t}}}$;

  \emph{(b) Computation of the Riemannian gradients}: The Riemannian gradient of  $\hat f\left( {\bm{\theta }} \right)$ at $\bm{\theta } = {{\bm{\theta }}^t}$ should lie in the tangent space ${\cal T}_{{\bm{\theta}^t}}{\cal S}^M$ \cite{cunhuatwc}. Then, the Riemannian gradient of $\hat f\left( {\bm{\theta }} \right)$ at ${\bm{\theta}^t}$ is obtained by projecting ${{\bm{\eta }}^t}$ onto ${\cal T}_{{\bm{\theta}^t}}{\cal S}^M$, which yields ${\bf{P}}_{{\cal T}_{{\bm{\theta}^t}}{\cal S}^M}({\bm{\eta} ^t})={\bm{\eta} ^t}-{\rm{Re}}\{\bm{\eta} ^{t*}\odot {\bm{\theta}^t}\}\odot{\bm{\theta}^t}$;

  \emph{(c) Update over the tangent space}: Update the point ${\bm{\theta}^t}$ on the tangent space ${{\cal T}_{{\bm{\theta}^t}}{\cal S}^M}$ as $\bar {\bm{\theta}^t}={\bm{\theta}^t}+\beta{\bf{P}}_{{\cal T}_{{\bm{\theta}^t}}{\cal S}^M}({\bm{\eta} ^t})$, where $\beta$ is a constant step size;

  \emph{(d) Retraction operator}: This step aims to map $\bar {\bm{\phi}^t}$  onto the manifold ${\cal S}^M$  using the
retraction operator ${\bm{\theta}^{t+1}}=\bar {\bm\theta}^t  \odot \frac{1}{{\left| \bar{ \bm\theta}^t  \right|}}$. Through iterating steps (a) to (d) until convergence, the final solution is obtained.

(5) \emph{Element-wise block coordinate descent (BCD) \cite{xianghao2019,shuowenjsac,qingqingjsac2020,Yasaman2021}}: The idea of the element-wise BCD algorithm is simple. At the $m$-th iteration, one reflection coefficient $\theta_m$ is optimized by keeping fixed the other reflecting coefficients ${\theta _{m'}}, m' \ne m,m = 1, \cdots ,M$. The algorithm ends after $M$ iterations when all the reflection coefficients are optimized one-by-one while keeping the other fixed. The element-wise BCD algorithm is simple since it is simpler to optimize a single variable rather than optimizing $M$ variables simultaneously. However, the complexity may be high when the number of reflecting elements is large.

(6) \emph{Rank-one equivalents \cite{fumin,xiaoyan}}: Similar to the SDR method,  by defining ${\bf{V}} = {\bm{\theta }}{{\bm{\theta }}^{\rm{H}}}$, the unit modulus constraint can be  written as ${\bf{V}} \succ {\bf{0}}$ and ${\rm{rank}}\left( {\bf{V}} \right) = 1$. The rank-one constraint can be equivalently transformed to
\begin{equation}\label{refes}
 {\rm{tr}}\left( {\bf{V}} \right) - {\left\| {\bf{V}} \right\|_2} = 0.
\end{equation}
Also, ${\rm{tr}}\left( {\bf{V}} \right) = \sum\nolimits_{m = 1}^M {{\lambda _m}} $ and ${\left\| {\bf{V}} \right\|_2} = {\lambda _1}$, where $\lambda _m$ denotes the $m$-th largest singular value of ${\bf{V}}$. Since ${\bf{V}} \succ {\bf{0}}$ and ${\bf{V}}$ is a non-zero matrix, the equality ${\rm{tr}}\left( {\bf{V}} \right) - {\left\| {\bf{V}} \right\|_2} = 0$ holds only when $\lambda_1>0$ and $\lambda_m=0, m=2,\cdots, M$. Then, at the $(t+1)$-th iteration of the iterative algorithm, a lower-bound for $\left\| {\bf{V}} \right\|_2$ at the point ${\bf{V}}^t$ can be derived as
\begin{equation}\label{wdefergt}
 {\left\| {\bf{V}} \right\|_2} \ge {\left\| {{{\bf{V}}^t}} \right\|_2} + \left\langle {\left( {{\bf{V}} - {{\bf{V}}^t}} \right),{{\left. {{\partial _{\bf{V}}}{{\left\| {\bf{V}} \right\|}_2}} \right|}_{{\bf{V}} = {{\bf{V}}^t}}}} \right\rangle  \buildrel \Delta \over = f\left( {{\bf{V}};{{\bf{V}}^t}} \right),
\end{equation}
where ${{{\left. {{\partial _{\bf{V}}}{{\left\| {\bf{V}} \right\|}_2}} \right|}_{{\bf{V}} = {{\bf{V}}^t}}}}$ is a subgradient of ${\left\| {\bf{V}} \right\|_2}$ with respect to ${\bf{V}}$ at ${{\bf{V}} = {{\bf{V}}^t}}$, which is equal to ${{\bf{u}}_1}{\bf{u}}_1^{\rm{H}}$ with ${\bf{u}}_1$ denoting the  eigenvector that corresponds to the largest singular value of ${\bf{V}}^t$.

Based on (\ref{wdefergt}), the constraint in (\ref{refes}) can be approximated with the following convex constraint
\begin{equation}\label{swdew}
{\rm{tr}}\left( {\bf{V}} \right) - f\left( {{\bf{V}};{{\bf{V}}^t}} \right) \le \varepsilon,
\end{equation}
 where $\varepsilon$ is a very small positive constant. Then,  using (\ref{wdefergt}) and (\ref{swdew}), one has ${\rm{0}} \le {\rm{tr}}\left( {\bf{V}} \right) - {\left\| {\bf{V}} \right\|_2} \le {\rm{tr}}\left( {\bf{V}} \right) - f\left( {{\bf{V}};{{\bf{V}}^t}} \right) \le \varepsilon $. Hence, when $\varepsilon$ tends to zero, ${\rm{tr}}\left( {\bf{V}} \right)$ will approach $\left\| {\bf{V}} \right\|_2$, which ensures that the rank-one constraint is fulfilled.

(7) \emph{Alternating direction method of multipliers (ADMM) based algorithm \cite{yiqing2021,huiyuan2021,qingqingwutwc}}: An auxiliary variable $\bm{\omega}$ is introduced such that ${\bm{\omega}}  = {\bm{\theta }}$, which can be regarded as a copy of $\bm{\theta }$. The feasible region of constraint C1 is denoted by $\cal B$, which, by using the indicator function, can be formulated as follows
\begin{equation}\label{dbgtfrg}
 {{\mathbb{I}}_{\cal B}}\left( {{\bf{W}},{\bm{\theta }}} \right) = \left\{ \begin{array}{l}
f\left( {{\bf{W}},{\bm{\theta }}} \right),\ {\rm{if}}\  \left\{ {{\bf{W}},{\bm{\theta }}} \right\} \in {\cal B}\\
\infty,\qquad \quad\  {\rm{otherwise.}}
\end{array} \right.
\end{equation}
Similarly, the feasible region that corresponds to constraint C2, i.e.,  ${\cal S}_1$ can be written as follows
\begin{equation}\label{dbgqdwqdwfrg}
{{\mathbb{I}}_{{\cal S}_1}}\left( {\bm{\omega}}  \right) = \left\{ \begin{array}{l}
0,\ {\rm{if}}\  {\bm{\omega}} \in {{\cal S}_1}\\
\infty,  {\rm{otherwise.}}
\end{array} \right.
\end{equation}
Then, the equivalent ADMM reformulation for the optimization problem in  (\ref{opt-model}) is
\begin{equation}
\begin{aligned}
&\underset {\mathbf{W},{\bm{\theta }},{\bm{\omega}}}{\min}\;\;{{\mathbb{I}}_{\cal B}}\left( {{\bf{W}},{\bm{\theta }}} \right)+ {{\mathbb{I}}_{{\cal S}_1}}\left( {\bm{\omega}}  \right)\\
&\text{s.t.} \quad {\bm{\omega}}  = {\bm{\theta }}.
\end{aligned}
\label{ADMMopt-model}
\end{equation}
The augmented Lagrangian of the optimization problem in (\ref{ADMMopt-model}) is
\begin{equation}\label{dweewf}
  {\cal L}_\xi= {{\mathbb{I}}_{\cal B}}\left( {{\bf{W}},{\bm{\theta }}} \right)+ {{\mathbb{I}}_{{\cal S}_1}}\left( {\bm{\omega}}  \right)+\frac{\xi }{2}\left\| {{\bm{\theta }} - {\bm{\omega }} + {\bm{\lambda }}} \right\|_2^2,
\end{equation}
where $\xi>0$ is a constant penalty parameter, and ${\bm{\lambda }} = {\left[ {{\lambda _1}, \cdots ,{\lambda _M}} \right]^T}$ is the dual variable vector of the constraint ${\bm{\omega}}  = {\bm{\theta }}$. Based on the ADMM algorithm, the variables $\mathbf{W},{\bm{\theta }}$ and ${\bm{\omega}}$ are alternately optimized.

The ADMM algorithm is an iterative approach. In the $t$-th iteration, given $\mathbf{W}^{t},{\bm{\theta }^t}$ and ${\bm{\omega}}^t$, the variables are updated as follows.

\emph{(a) Updating $\bm{\theta }$:}  The subproblem for updating $\bm{\theta }$ is
\begin{equation}
\begin{aligned}
&\underset {\bm{\theta}}{\min}\;\;f\left( {{\bm{\theta }}} \right)+\frac{\xi }{2}\left\| {{\bm{\theta }} - {\bm{\omega }^t} + {\bm{\lambda }}^t} \right\|_2^2\\
&\text{s.t.} \quad {g_i}\left( {{\bm{\theta }}} \right)\ge D_i,i = 1, \cdots ,I.
\end{aligned}
\label{odwedewel}
\end{equation}
Note that the unit-modulus constraint for $\bm{\theta }$ is not included in this subproblem, which significantly reduces the complexity of computing $\bm{\theta }$.

\emph{(b) Updating $\bf{W}$:} The subproblem for updating $\bf{W}$ is
\begin{equation}
\begin{aligned}
&\underset {\bf{W}}{\min}\;\;f\left( \bf{W} \right) \\
&\text{s.t.} \quad {g_i}\left( {\bf{W}} \right)\ge D_i,i = 1, \cdots ,I.
\end{aligned}
\label{odwswqdwewel}
\end{equation}

\emph{(c) Updating $\bm{\omega }$:} The subproblem for updating $\bm{\omega }$ is
\begin{equation}\label{zarfgae}
 \bm{\omega }^{t+1}=\arg \mathop {\min }\limits_{{\bm{\omega }} \in {{\cal S}_1}} \left\| {{{\bm{\theta }}^{t + 1}} + {{\bm{\lambda }}^t} - {\bm{\omega }}} \right\|_2^2.
\end{equation}
The objective of the optimization problem in (\ref{zarfgae}) is to project ${{\bm{\theta }}^{t + 1}} + {{\bm{\lambda }}^t}$ onto the feasible set ${\cal S}_1$, whose solution is $\bm{\omega }^{t+1} = {e^{j\arg ({{\bm{\theta }}^{t + 1}} + {{\bm{\lambda }}^t})}}.$

\emph{(d) Updating $\bm{\lambda }$:}  The update of $\bm{\lambda }$ is $\bm{\lambda }^{t+1}=\bm{\lambda }^{t}+{\bm{\theta }}^{t + 1}- \bm{\omega }^{t+1}$.

(8) \emph{Penalty convex-concave procedure (CCP) \cite{guizhoutsprobust,leizhangtccn2021,yuanbinchen}}: The unit modulus constraint can be equivalently rewritten as $1 \le {\left| {{\theta _m}} \right|^2} \le 1, \forall m$. Using the successive convex approximation (SCA) method, the non-convex constraint $1 \le {\left| {{\theta _m}} \right|^2} $ can be converted into a series of convex constraints, i.e., $1 \le 2{\rm{Re}}(\theta _m^*\theta _m^t)-{\left| {{\theta _m^t}} \right|^2} $, where $\theta _m^t$ is the solution in the $t$-th iteration. By introducing $2M$ slack variables ${\bf{b}} = \left[ {{b_1}, \cdots ,{b_{2M}}} \right]$, the phase shift optimization problem can be rewritten as
\begin{equation}
\begin{aligned}
&\underset {\bm{\theta},\bf{b}\ge \bf{0}}{\min}\;\;f\left( {{\bm{\theta }}} \right)- {\lambda ^t}\sum\nolimits_{m = 1}^{2M} {{b_m}} \\
&\text{s.t.} \quad {g_i}\left( {{\bm{\theta }}} \right)\ge D_i,i = 1, \cdots ,I, \\
&\quad\quad\; {\left| {{\theta _m^t}} \right|^2}- 2{\rm{Re}}(\theta _m^*\theta _m^t)  \le b_m-1,\forall m,\\
&\quad\quad\; {\left| {{\theta _m}} \right|^2} \le 1+b_{m+M}, \forall m,
\end{aligned}
\label{odwdswdwcvfel}
\end{equation}
where $\lambda ^t$ is the regularization factor to control the feasibility of the constraints in the $t$-th iteration. After some transformations, the optimization problem in (\ref{odwdswdwcvfel}) can be solved by CVX, and the detailed procedure to solve this problem, which can be found in references \cite{guizhoutsprobust,leizhangtccn2021,yuanbinchen}, is omitted here for brevity.

(9) \emph{Barrier function penalty \cite{jiaye2020,xiaolinghutcom}}:  The unit modulus constraint can be equivalently written as ${\rm{tr}}\left( {{\bm{\theta }}{{\bm{\theta }}^{\rm{H}}}} \right) = M$ and ${\left\| {\bm{\theta }} \right\|_\infty } \le 1$. Since ${\left\| {\bm{\theta }} \right\|_\infty }$ is non-differentiable, the $l_p$ norm with large $p$ can be used to approximate it, i.e., ${\left\| {\bm{\theta }} \right\|_\infty } = \mathop {\lim }\limits_{p \to \infty } {\left\| {\bm{\theta }} \right\|_p}$. To deal with the constraint ${\left\| {\bm{\theta }} \right\|_p} \le 1$, the logarithmic barrier function $F(x)$ can be used to approximate the penalty of violating the $l_p$ constraint, as
\begin{equation}\label{dfeaf}
 F(x) = \left\{ \begin{array}{l}
 - \frac{1}{\kappa }\ln (x),\  x > 0,\\
\infty,\qquad\quad\   x \le 0,
\end{array} \right.
\end{equation}
where $\kappa>0$ is the barrier function penalty factor.
For simplicity, constraint C1 is ignored. Accordingly, the phase shift optimization problem can be reformulated as
\begin{equation}
\begin{aligned}
&\underset {\bm{\theta}}{\min}\;\;G\left( {{\bm{\theta }}} \right)=f\left( {{\bm{\theta }}} \right)+F\left( {1 - {{\left\| {\bm{\theta }} \right\|}_p}} \right)\\
&\text{s.t.} \quad {\rm{tr}}\left( {{\bm{\theta }}{{\bm{\theta }}^{\rm{H}}}} \right) = M.
\end{aligned}
\label{odwdFEZfweel}
\end{equation}
Due to the non-convex constraint, the optimization problem in (\ref{odwdFEZfweel}) is still non-convex. To circumvent this issue, a possible solution is to utilize the gradient and projection method, which provides a low complexity but suboptimal solution. Specifically, the gradient of the objective function $G\left( {{\bm{\theta }}} \right)$ can be formulated  as
\begin{equation}\label{dwdefefa}
{\nabla _{\bm{\theta }}}G({\bm{\theta }}) = \frac{{\left\| {\bm{\theta }} \right\|_p^{1 - p}}}{{2\kappa \left( {1 - {{\left\| {\bm{\theta }} \right\|}_p}} \right)}}{\bm{\xi}}  + {\nabla _{\bm{\theta }}}f({\bm{\theta }}),
\end{equation}
where ${\bm{\xi}}  = {\left[ {{\theta _1}{{\left| {{\theta _1}} \right|}^{p - 2}}, \cdots ,{\theta _M}{{\left| {{\theta _M}} \right|}^{p - 2}}} \right]^T}$.

Since the problem formulation in (\ref{odwdFEZfweel}) is a minimization problem, the search direction is opposite to the direction of the gradient in (\ref{dwdefefa}). Let ${\bm\theta}^{(i)}$ denote $\bm\theta$ at the $i$-th iteration, the search direction in the $i$-th iteration is ${{\bf{d}}_{{\rm{gd}}}^{(i)}} =  - {\nabla _{\bm{\theta }}}G({\bm{\theta }})|_{{\bm{\theta }}={\bm{\theta }}^{(i)}}$
Then, this search direction ${{\bf{d}}_{{\rm{gd}}}^{(i)}}$ is projected onto the tangent plane of ${\rm{tr}}\left( {{\bm{\theta }}{{\bm{\theta }}^{\rm{H}}}} \right) = M$, as
\begin{equation}\label{wqsdfgbh}
  {{\bf{d}}_p^{(i)}} = {{\bf{d}}_{{\rm{gd}}}^{(i)}} - \frac{{\left({\bf{d}}_{\rm gd}^{(i)}\right)^T\left({{\bm{\theta }}^{(i)}}\right)^*{\bm{\theta }}^{(i)}}}{{{{\left\| {\bm{\theta }}^{(i)} \right\|}^2}}}.
\end{equation}
Then, the update of $\bm\theta $ in the $(i+1)$-th iteration is
\begin{equation}\label{defefr}
{{\bm{\theta }}^{(i + 1)}} = (1 - {\alpha ^ \star }){{\bm{\theta }}^{(i)}} + {\alpha ^ \star }\sqrt M \frac{{{{\bm{d}}_p^{(i)}}}}{{{{\left\| {{{\bm{d}}_p^{(i)}}} \right\|}^2}}},
\end{equation}
where the parameter ${\alpha ^\star}$ is obtained by
\begin{equation}\label{drdfa}
 {\alpha ^ \star } = \arg \mathop {\max }\limits_\alpha  f\left( {(1 - \alpha ){{\bm{\theta }}^{(i)}} + \alpha \sqrt M \frac{{{{\bm{d}}_p^{(i)}}}}{{{{\left\| {{{\bm{d}}_p^{(i)}}} \right\|}^2}}}} \right).
\end{equation}

(10) \emph{Accelerated projected gradient (APG) \cite{Mingjiewcl,Sileispl,9148781,9422787,perovic2021maximum}}: For simplicity, constraint C1 is ignored and  only  the optimization of the phase shifts is considered.  A projection operator ${{\mathbb{P}}_{{{\cal{S}}_1}}}$ is defined as
\begin{equation}\label{dEFEA}
{{\bm{\hat \theta }}}={{\mathbb{P}}_{{{\cal{S}}_1}}}(\bm\theta)\Leftrightarrow {{\hat \theta }_m} = \left\{ \begin{array}{l}
{{{\theta _m}} \mathord{\left/
		{\vphantom {{{\theta _m}} {\left| {{\theta _m}} \right|}}} \right.
		\kern-\nulldelimiterspace} {\left| {{\theta _m}} \right|}},{\rm{if }}\ {\theta _m} \ne 0\\
1,\qquad\quad {\rm{otherwise}}.
\end{array} \right.
\end{equation}
Then, the update of the phase shifts in the $(i+1)$-th iteration is given by
\begin{equation}\label{xvfaew}
  {{\bm{\theta }}_{i + 1}} = {{\mathbb{P}}_{{{\cal{S}}_1}}}\left( {{{\bf{z}}_i} - \frac{1}{{{\gamma _i}}}{{\left. {{\nabla _{\bm{\theta }}}f\left( {\bm{\theta }} \right)} \right|}_{{\bm{\theta }} = {{\bf{z}}_i}}}} \right),
\end{equation}
where ${\bf{z}}_i={{\bm{\theta }}_{i }}+\alpha_i\left( {{\bm{\theta }}_{i }}-{{\bm{\theta }}_{i-1}} \right) $, and $\alpha_i$ is updated as
\begin{equation}\label{ftyhff}
 {\alpha _i} = \frac{{{\xi _{i - 1}} - 1}}{{{\xi _i}}},{\xi _i} = \frac{{1 + \sqrt {1 + 4\xi _{i - 1}^2} }}{2}.
\end{equation}
In (\ref{xvfaew}), $\gamma _i$ is obtained  by using  the backtracking line search method \cite{beck2009fast}.

(11) \emph{Gradient descent approach \cite{zhi2021ergodic,Papazafeiropoulostwc,chongwenhuangtwc,Perovic}}: When the objective function $f(\bm{\theta})$ is differentiable, the optimization problem can be solved by using the gradient descent method. Specifically, let $\bm{\theta}^t$ be the phase shift vector at the $t$-th iteration. Then, the optimization variable $\bm{\theta}$ at the $(t+1)$-th iteration is updated as
\begin{equation}\label{sDfe}
{{\bm{\theta }}^{t + 1}} = \exp \left( {j\arg \left( {{{\bm{\theta }}^t} - \mu {{\left. {{\nabla _{\bm{\theta }}}f\left( {\bm{\theta }} \right)} \right|}_{{\bm{\theta }} = {{\bm{\theta }}^t}}}} \right)} \right),
\end{equation}
where $\mu$ is the step size and the $\arg$ operator is used for satisfying the unit-modulus constraint.

(12) \emph{Heuristic methods \cite{zhi2020power,kangdawclstatis,zhi2021twoarxiv,jianxindaicl2021}}: When the objective function is analytically involving, the above-mentioned algorithms may not be applicable or the computation of the gradient may be time-consuming. Possible solutions to circumvent this issue include the use of heuristic methods such the genetic algorithms (GA) or the particle swarm optimization (PSO) methods. More details can be found in \cite{zhi2020power,kangdawclstatis,zhi2021twoarxiv,jianxindaicl2021}.

(13) \emph{Deep reinforcement learning \cite{Chongwenjsac,Kemingwcl,Helinyang}}: Machine learning methods can also be applied to optimize the phase shifts of the RIS. A suitable approach is the use of deep reinforcement learning. In fact, unlike supervised learning methods that require a large number of training labels, deep reinforcement learning based methods do not need training labels and can learn and operate in an online manner. Examples of application of deep reinforcement learning to the optimization of RIS-aided communications can be found in  \cite{Chongwenjsac,Kemingwcl,Helinyang}.

\begin{figure}
	\centering 
	\includegraphics[width=3.3in]{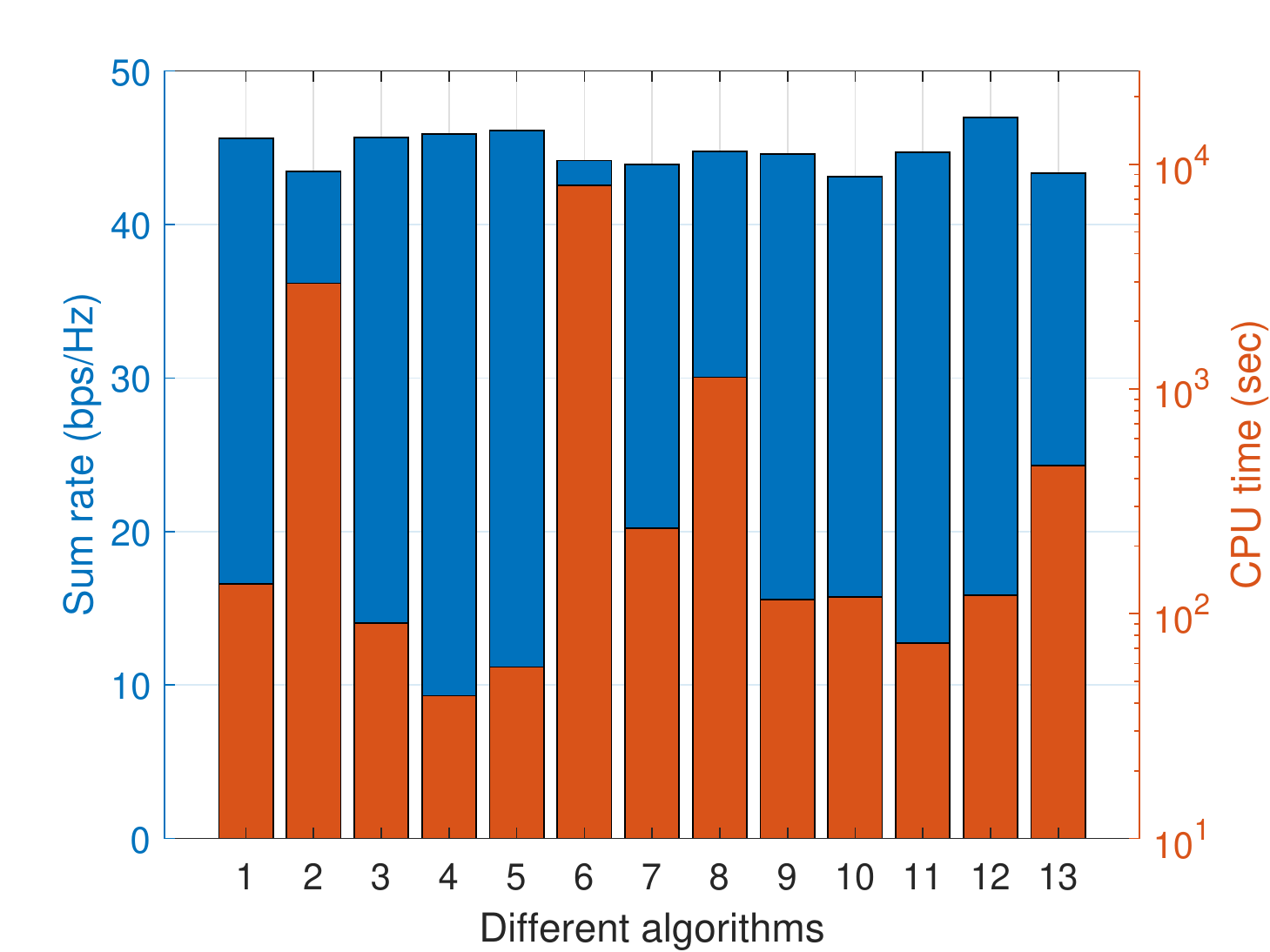}
	\caption{Sum rate and CPU time consumption of different algorithms, when $N=10$, $M=100$, $K=4$  and SNR=$5$ dB.}
	\vspace{-0.3cm}
	\label{transceiver}
\end{figure}

\textbf{Simulation results:} Fig. \ref{transceiver} illustrates the performance of the different algorithms discussed in this article in terms of sum rate and CPU run time. All algorithms are represented by the numbers they are introduced above. It can be seen that most of the algorithms for which a closed-form solution for the phase shifts can be found at each iteration (algorithms 3-5, 7, 9-11) provide a high sum rate with  a low CPU time (around 100 seconds). However, the  time-consuming algorithms (algorithms 1,2,6,8), which are implemented by using CVX, are more flexible to address optimization problems with complex constraints, such as quality of service (QoS) constraints.

2) \textbf{Optimization Techniques for Discrete Phase Shifts}

 Due to hardware limitations, it is challenging to implement continuous-valued phase shifts in practice \cite{qingqingwutcomdiscrete,Boyadi2020}. Hence, it is important to study the optimization of RISs subject to discrete-valued phase shifts, which leads to an NP-hard optimization problem. The existing works in this area can be classified into the following categories.

(1) \emph{Rounding method \cite{jiechenaccess,qingqingwutcomdiscrete,mingminzhaotwctwotime,jieyuan2021,yiqing2021}}: The main idea of this method is first to obtain a continuous solution that fulfills the unit modulus constraint. We denote this solution as ${ {\hat \theta }_m}, \forall m$. Then, the obtained solution is rounded to the nearest discrete value in ${\cal S}_2$ as
\begin{equation}\label{dewfee}
 {\theta _m^\star} = \arg \mathop {\min }\limits_{\phi  \in {{\cal S}_2}} \left| {{{\hat \theta }_m} - \phi } \right|, \forall m.
\end{equation}

(2) \emph{Binary mode selection  method \cite{qingqingwutcomdiscrete,Shaokanghu2021}}: Note that ${\theta_m=e^{j\varphi_m }}, \forall m$,  the objective function of the optimization problem can be transformed into a function of $\varphi_m$, ${\rm{cos}}(\varphi_m)$ and ${\rm{sin}}(\varphi_m)$ \footnote{The objective function may also include the phase differences. For illustration purposes, this is ignored in this paper.}. For each reflecting element $m$, we introduce the binary vector ${{\bf{x}}_m} = {\left[ {{x_{1,m}}, \cdots ,{x_{L,m}}} \right]^T}$ such that $\sum\nolimits_{l = 1}^L {{x_{l,m}}}  = 1,{x_{l,m}} \in \left\{ {0,1} \right\}$. Hence, ${{\bf{x}}_m}$ can be regarded as a mode selection vector. Define
${\bf{a}} = {\left[ {0,\Delta \varphi , \cdots ,(L - 1)\Delta \varphi } \right]^T}$, ${\bf{b}} = {\left[ {1,\cos \left( {\Delta \varphi } \right), \cdots ,\cos \left( {\left( {L - 1} \right)\Delta \varphi } \right)} \right]^T}$ and ${\bf{c}} = {\left[ {0,\sin \left( {\Delta \varphi } \right), \cdots ,\sin \left( {\left( {L - 1} \right)\Delta \varphi } \right)} \right]^T}$. Then, we have
 \begin{equation}\label{wefvref}
   {\varphi _m} = {{\bf{a}}^T}{{\bf{x}}_m},\cos \left( {{\varphi _m}} \right) = {{\bf{b}}^T}{{\bf{x}}_m},\sin \left( {{\varphi _m}} \right) = {{\bf{c}}^T}{{\bf{x}}_m}.
 \end{equation}
Based on (\ref{wefvref}), the original optimization problem is converted into a binary variable optimization problem, and the branch and bound (BnB) method can be utilized to obtain the globally optimal solution \cite{qingqingwutcomdiscrete}. However, the BnB method has an exponential computational complexity. To reduce the complexity, the authors of \cite{Shaokanghu2021} proposed to apply the SCA method. Specifically, the binary constraint on $x_{l,m}$ can be equivalently transformed into the following two continuous constraints:
 \begin{equation}\label{dewdwed}
   {x_{l,m}} - x_{l,m}^2 \le 0,0 \le {x_{l,m}} \le 1.
 \end{equation}
The first constraint in (\ref{dewdwed})  is the difference between two convex functions, and the SCA method can be used.

 (3) \emph{Negative square penalty (NSP) \cite{mingjieshao2019,liyoutsp}}: The NSP method  \cite{mingjieshao2019} was adopted to solve the discrete phase shift in \cite{liyoutsp}.  Specifically, the discrete phase shift optimization problem can be expressed as
 \begin{equation}\label{dfre}
 \mathop {\min \;}\limits_{{\bm{\theta }} \in {\tilde{\cal S}_2}} \;f\left( {\bm{\theta }} \right)
 \end{equation}
where ${\tilde{\cal S}_2} = \left\{ {\left. \theta  \right|\theta  = {e^{j\left( {\frac{{2\pi }}{L}l + \frac{\pi }{L}} \right)}},l = 0, \cdots ,L - 1} \right\}$. Based on the  NSP methods, we introduce the following problem
\begin{equation}\label{cscfedra}
 \mathop {\min \;}\limits_{{\bm{\theta }} \in {{\cal \bar S}_2}} \;F\left( {\bm{\theta }} \right) \buildrel \Delta \over = f\left( {\bm{\theta }} \right) - \lambda {\left\| {\bm{\theta }} \right\|^2},
\end{equation}
where ${{{{\cal\bar S}}_2}}={\rm{conv}\tilde{\cal  S}_2}$ is the convex hull of $\tilde{\cal  S}_2$, and $\lambda$ is a penalty parameter.
The penalty term $\lambda {\left\| {\bm{\theta }} \right\|^2}$ pushes each $\theta_m$ to an extreme point of  ${{{{\cal\bar S}}_2}}$. It is important to note that the constraint of the optimization in (\ref{cscfedra}) is convex, which is easier to handle than the original problem formulation in (\ref{dfre}). The authors of \cite{mingjieshao2019} provided the conditions on $\lambda$ for the problem in (\ref{cscfedra}) to be equivalent to the original problem in (\ref{dfre}). The MM method is utilized to solve the reformulated problem in (\ref{cscfedra}). Let $\bm{\theta}^i$ denote the value of $\bm\theta$ at the $i$-th iteration. Then, for any $\bm{\theta}$, we have
\begin{equation}\label{ffdwad}
  F\left( {\bm{\theta }} \right) \le f\left( {\bm{\theta }} \right) + \lambda {\left\| {{{\bm{\theta }}^i}} \right\|^2} - 2{\rm  Re}\left\{ {{{\bm{\theta }}^{\rm{H}}}{{\bm{\theta }}^i}} \right\} \buildrel \Delta \over = G\left( {\left. {\bm{\theta }} \right|{{\bm{\theta }}^i}} \right).
\end{equation}
Then, at the $i$-th iteration, the optimization problem is reformulated as follows
\begin{equation}\label{cscqsqwdra}
 \mathop {\min \;}\limits_{{\bm{\theta }} \in {{\cal \bar S}_2}} \;G\left( {\left. {\bm{\theta }} \right|{{\bm{\theta }}^i}} \right).
\end{equation}
The optimization problem in (\ref{cscqsqwdra}) can be tackled by using the APG method, which can be written as
\begin{equation}\label{gfewadwa}
 {\bm{\theta }}^{i+1}={{\mathbb{P}}_{{{\cal{\bar S}}_2}}}\left( {{{\bf{z}}^i} - \frac{1}{{{\beta _i}}}{{\left. {{\nabla _{\bm{\theta }}}G\left( {\left. {\bm{\theta }} \right|{{\bm{\theta }}^i}} \right)} \right|}_{{\bm{\theta }} = {{\bm{\theta }}^i}}}} \right),
\end{equation}
where the same notation as for the APG method introduced for the case of continuous phase shifts is used. Specifically, the projection onto the convex set  ${\cal{\bar S}}_2$  is denoted by ${{\mathbb{P}}_{{{\cal{\bar S}}_2}}}(u)$ whose closed-form expression is \cite{mingjieshao2019}
\begin{equation*}
  {{\mathbb{P}}_{{{\cal{\bar S}}_2}}}(u)={e^{j\frac{{2\pi l}}{L}}}\left( {\left[ {{\rm Re}\left( {\tilde u} \right)} \right]_0^{\cos \left( {\pi /L} \right)} + j\left[ {{\rm Im}\left( {\tilde u} \right)} \right]_{ - \sin \left( {\pi /L} \right)}^{\sin \left( {\pi /L} \right)}} \right),
\end{equation*}
 where
 \begin{equation*}
   n = \left\lfloor {\frac{{\angle u + \pi /L}}{{2\pi /L}}} \right\rfloor ,\tilde u = u{e^{ - j\frac{{2\pi n}}{L}}}.
 \end{equation*}

 (4) \emph{Heuristic methods \cite{jianxindaicl2021}}: Heuristic methods  such as the PSO are effective methods to address the discrete phase shift optimization problem. More details can be found in  \cite{jianxindaicl2021}.

\subsection{Various Levels of CSI Availability}

\begin{figure}
	\centering 
	\includegraphics[width=3.3in]{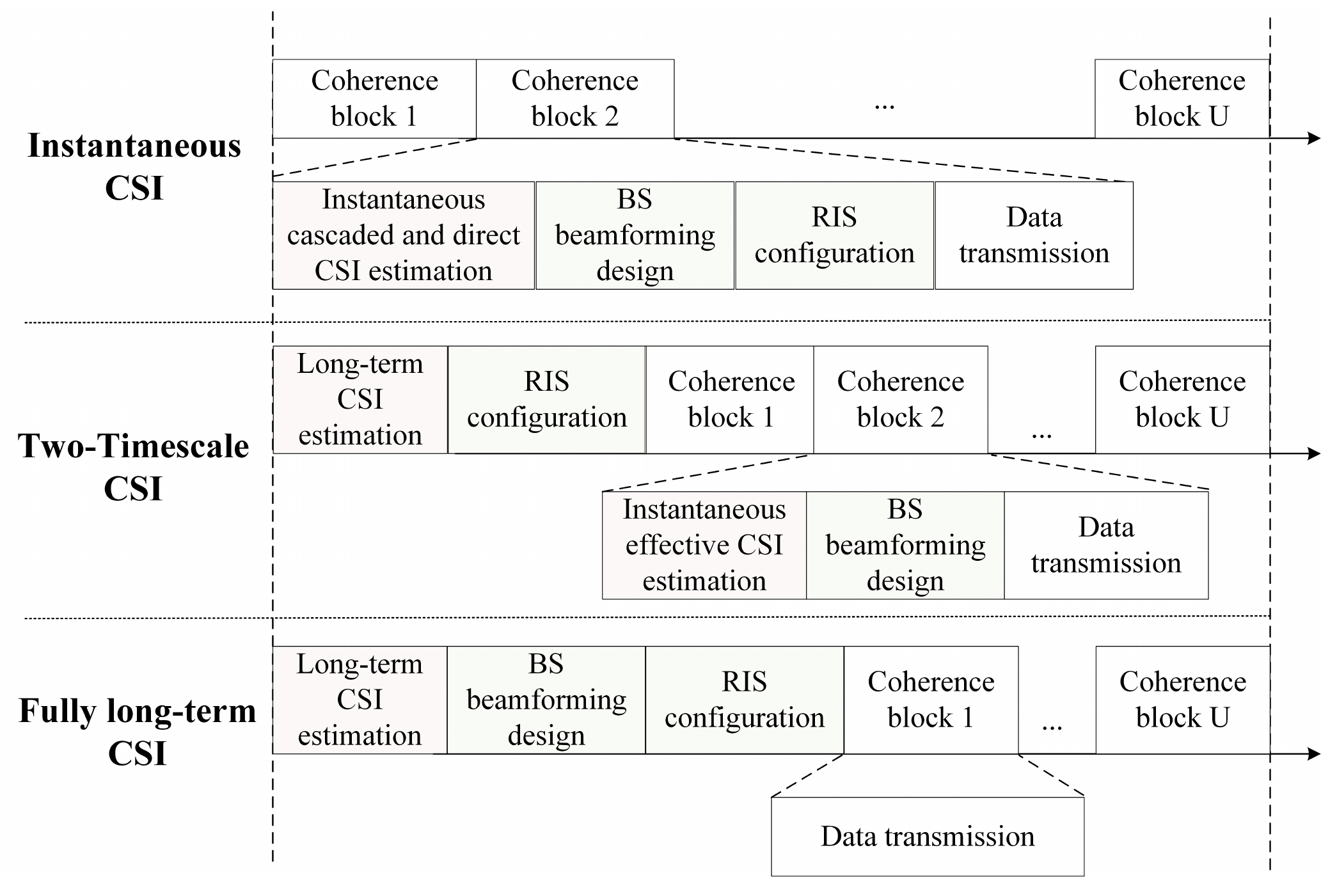} \caption{Illustration of transmission designs based on the various levels of CSI availability. A total of $U$ channel coherence blocks are considered, during which the long-term CSI is fixed.}
	\vspace{-0.3cm}
	\label{figure_protocol}
\end{figure}

Depending on the levels of CSI availability, existing contributions on transmission design of RIS-aided communication systems can be classified into three categories: 1) Systems designs based on instantaneous CSI; 2) Systems designs based on the so-called two-timescale CSI; 3) Systems designs that rely on fully long-term CSI. The operation protocol for each category is shown in Fig. \ref{figure_protocol} and the details for each category are given next.

\emph{\textbf{1) Instantaneous CSI}}

For the case of instantaneous CSI, the overall CSI of the system is assumed to be available at the BS.  The RIS-related channels can be the cascaded channels ${{\bf{G}}_k}, \forall k$, or the individual channels ${\bf{H}}^{\rm{H}}$ and ${\bf{h}}_{r,k}^{\rm{H}}, \forall k$. The existing works on the transmission design  can be classified into two categories: (1) Perfect instantaneous CSI; (2) Imperfect instantaneous CSI.

\emph{(1) Perfect instantaneous CSI}:
Most of the existing works have considered transmission design based on the assumption that the instantaneous CSI is perfectly available. Based on this assumption, the performance gains provided by introducing an RIS in various wireless applications have been investigated, such as mmWave/terahertz  systems \cite{peilanwangtvt,guizhoutvt,Pradhan2020,pan2020sum,ziweiwan,yijinpan2021}, multicell systems \cite{cunhuatwc,hailiangxie,Buzzi2021}, physical layer security systems \cite{miaocuiwcl,shongshencl,jiechenaccess,hongsheng2020,zhengchuwcl,limengdong2020,xinrongguan2020}, simultaneous wireless information and power transfer (SWIPT) \cite{cunhuajsac,qingqingjsac2020,qingqingwuwclweiswipt,huiyuan2021,pan2021self,Lyubin,yuanzheng,feng2021waveform}, mobile edge computing networks \cite{tongbai2020,bai2021empowering,Fasheng2021,xiaoyan,tongbaitwc,zhengchu,zhiyangli,yashuicao}, multicast networks \cite{guimulticast,linsong}, cognitive radio networks \cite{jieyuan2021,leizhangtvt,xingrongcognitive}, non-orthogonal multiple access \cite{fumin,beixiongzheng2020,yiqing2021,hongwang2020,jianyue,xidongtcom,gangyangtwc,jiakuo,mingzeng2021,xidongmutwc}, two-way communications \cite{zhangjietsp,yuzhangcl}, and full-duplex (FD) communication\cite{Caiyl2021FD}.
 In these works, the AO method was adopted to alternately optimize the beamforming vectors at the BS and the phase shifts at the RIS, and the phase shift optimization problem was addressed  using the algorithms summarized in Subsection \ref{algorithms}.

\emph{(2) Imperfect instantaneous CSI}:
As discussed in Section \ref{channelestimation}, channel estimation errors are inevitable. If the estimated CSI is naively regarded as perfect in the transmission design, the obtained solution will likely violate the QoS requirements. This issue is further aggravated in RIS-aided communication systems due to the additional RIS-related channels to be estimated. Hence, it is imperative to consider robust transmission designs by taking into account the channel estimation errors. Most of the early contributions in this area studied the case when the channels from the RIS to the users are imperfect \cite{guizhoucunhua,xianghaoyu2020,dongfangxu2020tcom,qunwang,jiezhizhang,zhangzhenglulv} while regarding the channels from the BS to the RIS as perfect. This approach requires estimation of the BS-RIS and RIS-user channels separately, which is challenging to implement in practice. Instead, the authors of \cite{guizhoutsprobust} proposed a framework of robust transmission design for  RIS-aided multiuser systems based on imperfect CSI of the cascaded channels. Two models were used to characterize the cascaded channel estimation error, namely, the bounded CSI error model and the statistical CSI error model. Specifically, the cascaded and direct channels can be written as
\begin{equation}\label{vgfsgfe}
 {{\bf{h}}_{d,k}} = {{{\bf{\hat h}}}_{d,k}} + {{{\bf{\tilde h}}}_{d,k}},{{\bf{G}}_k} = {{{\bf{\hat G}}}_k} + {{{\bf{\tilde G}}}_k},
\end{equation}
where ${{{\bf{\hat h}}}_{d,k}}$ and ${{{\bf{\hat G}}}_k}$ are the estimated channels, and ${{{\bf{\tilde h}}}_{d,k}}$ and $ {{{\bf{\tilde G}}}_k}$ are the corresponding channel estimation errors.
Next, we discuss the two CSI error models.

\emph{A. Bounded CSI Error Model}:
In this case,  the channel error is modeled as follows
\begin{equation}\label{cvdfdw}
 {\left\| {{{{\bf{\tilde G}}}_k}} \right\|_2} \le {\varepsilon _{c,k}},{\left\| {{{{\bf{\tilde h}}}_{d,k}}} \right\|_2} \le {\varepsilon _{d,k}},
\end{equation}
where ${\varepsilon _{c,k}}$ and ${\varepsilon _{d,k}}$ quantify the level of channel uncertainty. Under this model, the authors of \cite{guizhoutsprobust} jointly optimized the active beamforming at the BS and the phase shifts at the RIS so as to minimize the total power consumption under the unit-modulus constraint for the phase shifts and by ensuring that the data rate of each user is above a threshold for all possible channel error realizations. By defining
\begin{align}
  {\Omega _{{{\bf{h}}_{d,k}}}} &= \left\{ {{{{\bf{\tilde h}}}_{d,k}} \in {{\mathbb{C}}^{N \times 1}}:{{\left\| {{{{\bf{\tilde h}}}_{d,k}}} \right\|}_2} \le {\varepsilon _{d,k}}} \right\}, \\
  {\Omega _{{{\bf{G}}_k}}} &= \left\{ {{{{\bf{\tilde G}}}_k} \in {{\mathbb{C}}^{N \times M}}:{{\left\| {{{{\bf{\tilde G}}}_k}} \right\|}_2} \le {\varepsilon _{c,k}}} \right\},
\end{align}
the robust design problem can be formulated as
\begin{equation}\label{odwdweddsxsedfel}
\begin{aligned}
&\underset {\bf{W},\bm{\theta}}{\min}\;\; \left\| {\bf{W}} \right\|_2^2\\
&\text{s.t.} \quad  {\theta _m} \in {{\cal S}_1}\ {\rm{or}}\ {{\cal S}_2},\forall m=1,2,\dots,M, \\
&\quad\quad R_k({\bf{W}},{\bm{\theta}})\!\ge\! D_k, \forall {{{\bf{\tilde h}}}_{d,k}}\in{\Omega _{{{\bf{h}}_{d,k}}}}, \forall {{{\bf{\tilde G}}}_{k}}\in{\Omega _{{{\bf{G}}_k}}}, \forall k,
\end{aligned}
\end{equation}
where $D_k$ is the minimum data requirement of user $k$, ${\cal S}_1$ and ${{\cal S}_2}$ are defined in (\ref{s1}) and (\ref{s2}), respectively.

The key difficulty in solving the optimization problem in (\ref{odwdweddsxsedfel}) lies in how to deal with an infinite number of possible channel errors. The most popular technique is to use the $\cal S$-procedure, which transforms the worst-case constraints into a more tractable form with linear matrix inequalities. The AO and penalty CCP techiques were adopted to solve Problem (\ref{odwdweddsxsedfel}) in \cite{guizhoutsprobust}. Later, similar techniques were used in RIS-aided cognitive radio networks \cite{leizhangtccn2021,jieyuan2021}, physical layer security systems \cite{shenghong2021twc}, and secure cognitive radio communications \cite{limengtcomsecure}. In \cite{xianghaotcom2021}, the penalty-based alternating minimization method was proposed, which is guaranteed to converge to a stationary point and is a Karush-Kuhn-Tucker (KKT) solution of the considered problem.

\emph{B. Statistical CSI Error Model}: For the statistical CSI error model, the channel error is modeled as a random variable. When the linear minimum mean square error (LMMSE) is used for channel estimation, the channel estimation error generally follows a complex Gaussian distribution. Hence, the channel error can be modeled as
\begin{align}
{\rm{vec}}\left( {{{{\bf{\tilde G}}}_k}} \right) &\sim  {\cal{CN}}\left( {0,{{\bm{\Sigma}} _{c,k}}} \right), {{\bm{\Sigma}}_{c,k}} \succ 0, \forall k, \\
  {{{\bf{\tilde h}}}_{d,k}}  &\sim {\cal{CN}}\left( {0,{{\bm{\Sigma}} _{d,k}}} \right),  {{\bm{\Sigma}} _{d,k}} \succ 0, \forall k,
\end{align}
where ${{\bm\Sigma} _{c,k}} \in {{\mathbb{C} }^{MN \times MN}}$ and ${{\bm\Sigma} _{h,k}} \in {{\mathbb{C} }^{N \times N}}$ are positive definite error covariance matrices. Due to the randomness of the channel error, the design objective is mainly to optimize the beamforming vectors at the BS and the phase shifts at the RIS while ensuring a minimum non-outage probability. Specifically, the robust design problem can be formulated as
\begin{equation}
\begin{aligned}
&\underset {\bf{W},\bm{\theta}}{\min}\;\; \left\| {\bf{W}} \right\|_2^2\\
&\text{s.t.} \quad {\theta _m} \in {{\cal S}_1}\ {\rm{or}}\ {{\cal S}_2},\forall m=1,2,\dots,M, \\
&\quad\quad \Pr \left\{ {{R_k}({\bf{W}},{\bm\theta} ) \ge {D_k}} \right\} \ge 1 - {\rho _k},\forall k,
\end{aligned}
\label{oddwqqqsedfel}
\end{equation}
where ${\rho _k}$ is the maximum outage probability.

The main challenge in solving the optimization problem in (\ref{oddwqqqsedfel}) lies in the fact that the rate outage probability constraints do not admit a closed-form expression. One method to circumvent this issue is the Bernstein-Type inequality \cite{kunwangyu}, which transforms an intractable outage probability constraint into a tractable linear matrix inequality \cite{guizhoutsprobust}.  This method has been applied to RIS-aided physical layer security systems \cite{shenghong2021twc,hong2020outage} and RIS-aided vehicular communications \cite{yuanbinchen}. Another method that can be applied is the constrained stochastic SCA algorithm proposed in \cite{minimingtsp}, which is shown to guarantee the desired outage probability performance of the users. In addition, the robust design of RIS-aided cognitive radio networks was considered in \cite{leizhangtccn2021}, where  it was ensured that the probability that the interference power perceived at the primary users (PUs) is below an interference temperature (IT) limit was larger than a threshold. Both the  triangle inequality and the inverse Chi-square distribution techniques were invoked to derive a  tractable approximate constraint.

\emph{\textbf{2) Two-Timescale CSI}}

As illustrated in Fig. \ref{figure_protocol}, transmission designs based on the knowledge of the instantaneous CSI require that the BS needs to estimate the cascaded channel ${\bf{G}}_k$ and direct channel ${\bf{h}}_{d,k}$ in each channel coherence block, in which the  number of time slots required for channel training is often proportional to the number of reflecting elements, e.g., for the unstructured channel models discussed in Section \ref{channelestimation}. Since the RIS is not endowed with power amplification and signal processing capabilities, it is expected to be equipped with hundreds or even thousands of reflecting elements for ensuring the desired coverage. As a result, the channel training overhead may be excessive, and there may be only a few or even no time slots left for data transmission. Most recently, in instantaneous CSI cases, it was revealed in \cite{Kundu2021,zhi2021twoarxiv,Zappone2021} that the net data rate that accounts for the penalty due to channel estimation overhead first increases and then decreases with the number of reflecting elements. It is well known that time division duplexing (TDD) is the preferred option in massive MIMO systems since the channel training overhead depends on the number of users, and  is not related to the number of BS antennas. In RIS-aided wireless systems, no matter whether  TDD  or frequency division duplexing (FDD) is used, the required channel training overhead is always proportional to the number of reflecting elements under the assumption of unstructured channel models. Furthermore, in each coherence block with a duration of at most several hundred milliseconds, the BS needs to optimize the  phase shifts and active beamforming vectors, which requires the BS to have sufficient computational capabilities. In addition, the phase shifts computed at the BS need to be sent to the RIS controller for updating the phase shifts in each coherence block. This would incur high feedback overhead.

To address these issues, a novel two-timescale beamforming design was first proposed in \cite{yuhan2019} and its transmission protocol is shown in Fig. \ref{figure_protocol}. The main idea is that the active beamforming vectors at the BS are designed based on the instantaneous effective/aggragated BS-user channels that are the superposition of the direct and RIS-reflected channels, while the phase shifts at the RIS are designed based on long-term CSI, such as their distribution parameters including channel mean and channel covariance matrices. In each coherence block, only the instantaneous effective channel of each user needs to be estimated and the channel training overhead is equal to the number of users, which is the same as for legacy massive MIMO systems without RISs. Furthermore, since the long-term CSI  remains usually invariant for a large number of channel coherence blocks, the phase shifts of the RIS can be updated at a much lower rate than the fast fading fluctuations, which significantly reduces the computational burden and feedback overhead.

To better understand, let us consider the system model in Fig. \ref{systemodel}. All the channels are assumed to be subject to correlated Rician fading
\begin{align}
  {\bf{H}} &= \sqrt \beta  \left( {\sqrt {\frac{\delta }{{1 + \delta }}} {\bf{\bar H}} + \sqrt {\frac{1}{{1 + \delta }}} {\bf{\tilde H}}} \right),\nonumber \\
  {{\bf{h}}_{r,k}} &= \sqrt {{\alpha _k}} \left( {\sqrt {\frac{{{\varepsilon _k}}}{{1 + {\varepsilon _k}}}} {{{\bf{\bar h}}}_{r,k}} + \sqrt {\frac{1}{{1 + {\varepsilon _k}}}} {{{\bf{\tilde h}}}_{r,k}}} \right),\forall k,\nonumber\\
  {{\bf{h}}_{d,k}} &= \sqrt {{\gamma _k}} \left( {\sqrt {\frac{{{\varpi _k}}}{{1 + {\varpi_k}}}} {{{\bf{\bar h}}}_{d,k}} + \sqrt {\frac{1}{{1 + {\varpi _k}}}} {{{\bf{\tilde h}}}_{d,k}}} \right),\forall k,\nonumber
\end{align}
where $\beta$, $\alpha _k$ and $\gamma _k$ are the large-scale path loss coefficients, $\delta$, $\varepsilon _k$ and $\varpi _k$ are the Rician factors, ${{\bf{\bar H}}}$, ${{{{\bf{\bar h}}}_{r,k}}}$, and ${{{{\bf{\bar h}}}_{d,k}}}$ are LoS components, ${\bf{\tilde H}} \sim {\cal CN}\left( {{\bf{0}},{{\bf{R}}_{{{\mathrm{HR}}}}} \otimes {{\bf{R}}_{{\mathrm{HB}}}}} \right)$, ${{{\bf{\tilde h}}}_{r,k}} \sim {\cal CN}\left( {0,{{\bf{R}}_{{\mathrm{h}_\mathrm{r}\mathrm{R},k}}}} \right)$ and ${{{\bf{\tilde h}}}_{d,k}} \sim {\cal CN}\left( {0,{{\bf{R}}_{{\mathrm{h}_\mathrm{d}\mathrm{B},k}}}} \right)$ are  NLoS components with ${{\bf{R}}_{{{\mathrm{HR}}}}}$, ${{\bf{R}}_{{\mathrm{HB}}}}$, ${{\bf{R}}_{{\mathrm{h}_\mathrm{r}\mathrm{R},k}}}$ and ${{\bf{R}}_{{\mathrm{h}_\mathrm{d}\mathrm{B},k}}}$  being  the corresponding spatial covariance matrices. When the Rician factors are equal to zero, the channels become   correlated Rayleigh channels. When the spatial covariance matrices are identity matrices, the channels are independent and identically distributed Rician channels. When  both conditions hold, the channels reduce to independent and identically distributed Rayleigh channels. The spatial covariance matrices can be estimated  by using the method in \cite{Changliang} or the model proposed in \cite{Bjornson2021}. Also, the LoS components mainly depend on location/angle information, which can be estimated  using the method in \cite{Gui}.  In general, the NLoS components vary rapidly and need to be estimated at each coherence block, while the LoS components change slowly and usually remain constant over a number of channel coherence blocks.

Let ${{\bf{q}}_k} \buildrel \Delta \over =  {{\bf{h}}_{d,k}} + {\bf{H}}{\bm\Theta} {{\bf{h}}_{r,k}}$ be the effective channel of user $k$  and let ${\bf{Q}} = \left[ {{{\bf{q}}_1}, \cdots ,{{\bf{q}}_K}} \right]$ be the collection of all effective channels. In the downlink, the  beamforming vector ${{\bf{w}}_k}$ is  a function of the effective channels ${{\bf{Q}}}$, i.e., ${{\bf{w}}_k} = {z_k}\left( {\bf{Q}} \right)$.
By substituting ${{\bf{w}}_k}$ into the optimization problem in (\ref{opt-model}), the original problem can be converted into an optimization problem that depends only on the optimization variable $\bm{\theta}$
\begin{equation}
\begin{aligned}
&\underset {\bm{\theta }}{\min}\;\;{\mathbb{E}}\left\{ {f\left( {\bm{\theta }} \right)} \right\}\\
&\text{s.t.} \quad  {\theta _m} \in {{\cal S}_1}\ {\rm{or}}\ {{\cal S}_2},\forall m=1,2,\dots,M, \\
&\quad\quad{\mathbb{E}}\left\{ {g_i}\left( { {\bm{\theta }}} \right)\right\}\ge D_i,i = 1, \cdots ,I,
\end{aligned}
\label{opdewl}
\end{equation}
where the expectation is taken over the NLoS components of the channels. The optimization problem in the uplink  can be formulated \textit{mutatis mutandis}.

In the following, we discuss the existing works from two aspects: (1) Downlink transmission; (2) Uplink transmission.

(1) \emph{Downlink transmission}: In the two-timescale scheme, to reduce the complexity of solving the optimization problem in  (\ref{opdewl}), most existing works have  adopted maximum-ratio-transmission (MRT) at the BS:
\begin{equation}\label{freadw}
  {{\bf{w}}_k} = \sqrt {{P_k}} \frac{{{{\bf{q}}_k}}}{{\left\| {{{\bf{q}}_k}} \right\|}},
\end{equation}
where $P_k$ is the transmission power of user $k$.

For the single-user case, the MRT precoder is optimal. This setup has been studied in \cite{yuhan2019,yuhangjia,Papazafeiropouloswcl2021,huayanguowcl}.  Specifically, by assuming Rician fading for the RIS-related channels and Rayleigh fading for the direct channel, an upper bound for the ergodic data rate was derived in \cite{yuhan2019}  using Jensen's inequality, based on which the optimal phase shift was obtained in closed form as a function of the LoS components of the channels. The work in \cite{yuhan2019} was further extended in \cite{yuhangjia} to the case where there is a co-channel BS transmitting interference signals. The authors of \cite{Papazafeiropouloswcl2021} first derived an approximated closed-form expression of the coverage probability of a multi-RIS-aided system in the presence of correlated Rayleigh fading, by using the deterministic equivalent analysis, and then optimized the phase shifts relying on the projected gradient method. Instead of using Jensen's inequality, the stochastic gradient descent method was used in \cite{huayanguowcl} for solving the phase shift optimization problem, which is applicable to any channel distribution.

In the multiuser case, MRT at the BS was adopted in \cite{jianxindaicl2021,al2021reconfigurable}. In particular, by taking into account the use of low-resolution digital-analog converters (DACs) at the BS, the authors of \cite{jianxindaicl2021} derived an approximate expression for the ergodic data rate of RIS-aided massive MIMO systems. Because of the complicated expression of the obtained data rate, the heuristic PSO method  was invoked to solve the rate maximization problem  under the assumptions of either continuous or discrete  phase shifts. Simulation results showed that three quantization bits for the phase shifts are sufficient to obtain similar performance as for the continuous phase shifts. By considering the channel estimation error of the effective channel, the authors of \cite{al2021reconfigurable} derived deterministic equivalents for the sum rate, where a LoS deterministic RIS-BS channel and Rician fading RIS-user/BS-user channels were assumed. The projected gradient ascent-based algorithm was used for solving the phase shift optimization problem.

In addition to the MRT, another beamforming vector that can be utilized at the BS is the optimal linear precoder    studied in \cite{Nadeem2020}, which is the optimal solution to the problem of maximizing the minimum data rate. Specifically, the authors of \cite{Nadeem2020} developed deterministic approximations of the minimum data rate using random matrix theory (RMT) tools by assuming a deterministic BS-RIS channel and correlated Rayleigh RIS-user channels.

In \cite{mingminzhaotwctwotime} and \cite{mingmin}, more sophisticated algorithms, such as deep unfolding and the stochastic successive convex approximation, were utilized to jointly optimize the beamforming vectors and the phase shifts. Although the algorithms in \cite{mingminzhaotwctwotime} and \cite{mingmin} are applicable to any channel distributions, the computational complexity would be excessive when the number of antennas at the BS is large, and thus not suitable for massive MIMO systems.

(2) \emph{Uplink transmission}: 
Similar to the downlink, maximum-ratio-combining (MRC) at the BS is widely used in the uplink. In this case, the decoding vector is
\begin{equation}\label{fadewd}
 {{\bf{v}}_k} = {{\bf{q}}_k}.
\end{equation}
The MRC scheme has been studied in \cite{zhi2020power,kangdawclstatis,jianxindai2021,zhi2021twoarxiv,Papazafeiropoulostwc,wang2021massive}. The achievable data rate was derived in \cite{zhi2020power} for RIS-aided massive MIMO systems assuming Rician fading for the BS-RIS and RIS-user channels. Based on the obtained result,  the scaling laws with respect to the number of RIS elements and the number of BS antennas were analyzed. The work was extended in \cite{kangdawclstatis} to the case study where the direct link between the BS and the users are present, and the authors provided the conditions under which RIS-aided massive MIMO systems outperform conventional massive MIMO systems. In \cite{jianxindai2021}, the achievable data rate was analyzed in the presence of transceiver hardware impairments (HWIs) and RIS phase noise over Rician fading RIS-related channels, and it was demonstrated that the hardware imperfections do not affect the power scaling law.
Most recently, the authors of \cite{zhi2021twoarxiv} adopted the LMMSE estimator to estimate the effective channel, and derived a closed-form expression for the approximate achievable data rate by taking into account the channel estimation errors. A framework for the power scaling law analysis was introduced in \cite{zhi2021twoarxiv} under various channel fading distributions. Due to the complicated data rate expression, the phase shift optimization problem was solved by using the heuristic GA method in \cite{zhi2020power,kangdawclstatis,jianxindai2021,zhi2021twoarxiv}. In addition to the effective channel estimation error, the impact of both the transceiver HWIs and the RIS phase noise on the  achievable data rate was analyzed in \cite{Papazafeiropoulostwc} for transmission over a deterministic BS-RIS channel and other correlated Rayleigh fading channels. In \cite{wang2021massive}, the achievable data rate for imperfect effective channels was derived where all channels were assumed to undergo correlated Rayleigh fading. The authors of \cite{wang2021massive} demonstrated that the channel hardening and favorable propagation conditions still hold for the effective channels.

Besides the MRC, another low-complexity decoding vector is the zero-forcing (ZF) detector. Let  ${\bf{V}} = \left[ {{{\bf{v}}_1}, \cdots ,{{\bf{v}}_K}} \right]$ denote the collection of all users' decoding vectors. Then, the ZF detector at the BS is
\begin{equation}\label{feewdd}
  {\bf{V}} = {\bf{Q}}{\left( {{{\bf{Q}}^{\rm{H}}}{\bf{Q}}} \right)^{ - 1}}.
\end{equation}
The two-timescale design for RIS-aided massive MIMO with ZF was studied in \cite{zhi2021ergodic,kangdazfimperfect}. By approximating the non-central Wishart distribution with a central Wishart distribution, a closed-form expression of the ergodic data rate was derived in \cite{zhi2021ergodic} under  the assumption of a BS-RIS channel that undergoes Rician fading, and LoS RIS-user channels and BS-user channels that undergo Rayleigh fading. Based on the derived expression of the ergodic data rate, the gradient ascent algorithm was proposed to optimize the phase shifts of the RIS. The simulation results in \cite{zhi2021ergodic} demonstrated that the ZF detectors significantly outperform the MRC detectors in RIS-aided massive MIMO systems. The reason is that the users share the same RIS-BS channel and then their cascaded channels are highly correlated over  Rician fading channels. Therefore, by effectively eliminating the interference,  the ZF detector provides a higher ergodic rate as compared with the MRC. The analysis in \cite{zhi2021ergodic} was extended to the case with imperfect effective CSI \cite{kangdazfimperfect}, where closed-form expressions for the uplink data rate were derived. The analytical results showed that the rate of all the users scales at least on the order of $O\left( {{{\log }_2}\left( {MN} \right)} \right)$. The low-complexity MM algorithm was proposed to optimize the RIS phase shifts.

 In the uplink, the optimal decoding vector is the LMMSE receiver given by
\begin{equation}\label{fdrafwED}
{{\bf{v}}_k} = {\left( {\sum\nolimits_{i = 1}^K {{P_i}{{\bf{q}}_i}{\bf{q}}_i^{\rm{H}}}  + {\sigma ^2}{{\bf{I}}_N}} \right)^{ - 1}}{{\bf{q}}_k}.
\end{equation}
Using the LMMSE receiver,  the authors of \cite{Papazafeiropoulos2021} analyzed the approximate SINR of each user  by leveraging the deterministic equivalent method, where the BS-RIS channel is assumed to be deterministic and the other channels follow a correlated Rayleigh distribution. In \cite{Papazafeiropoulos2021},  both RIS phase noise and transceiver hardware impairments were taken into account,  and the phase shifts of the RIS were optimized  using the gradient ascent algorithm  to maximize the minimum SINR.

\emph{\textbf{3) Fully long-term CSI}}

 In the two-timescale design, the instantaneous effective CSI needs to be estimated in each coherence block. To further reduce the channel estimation overhead, an appealing approach is based on transmission designs that require only long-term CSI. The operation procedure is shown in Fig. \ref{figure_protocol}. Specifically, at the beginning of the transmission, the BS estimates or measures the long-term CSI, based on which the BS computes the beamforming vector and the phase shifts  that are used in all the subsequent coherence blocks until the long-term CSI changes. By adopting the same channel fading distribution as in the two-timescale CSI subsection, the long-term CSI problem  is formulated as
\begin{equation}
\begin{aligned}
&\underset {\bm{\theta }, \bf{W}}{\min}\;\;{\mathbb{E}}\left\{ {f\left({\bf{W}}, {\bm{\theta }} \right)} \right\}\\
&\text{s.t.} \quad  {\theta _m} \in {{\cal S}_1}\ {\rm{or}}\ {{\cal S}_2},\forall m=1,2,\dots,M, \\
&\qquad{\mathbb{E}}\left\{ {g_i}\left( { {\bf{W}}, {\bm{\theta }}} \right)\right\}\ge D_i,i = 1, \cdots ,I,
\end{aligned}
\label{opdqsqwqwl}
\end{equation}
where the expectation is taken over the NLoS components of the channels.  In contrast to the two-timescale  optimization problem in (\ref{opdewl}), the beamforming vectors at the BS are designed based on long-term CSI. The main difficulty in solving Problem (\ref{opdqsqwqwl}) lies in the lack of explicit expressions for the  objective function and constraints, i.e., the expectation in (\ref{opdqsqwqwl}) cannot in general be formulated in a closed-form expression. In general, three optimization techniques can be utilized to circumvent this issue: (1) Applying Jensen's inequality; (2) Utilizing large system analysis; (3) Leveraging deep reinforcement learning methods.

(1) \emph{Jensen's inequality}: Jensen's inequality is an effective and simple method to tackle the randomness in the channels, and has been widely used in the existing literature \cite{hu2020statistical,Jinghewangtccn,xugan2021,zhangjietvt2021,caihongluo2021,menghua2021}. The main idea of this method is to derive an upper bound for the function $f(\cdot)$ in (\ref{opdqsqwqwl}) that is a function of the optimization variables and the long-term CSI. In these works, usually, the function $f(\cdot)$ in (\ref{opdqsqwqwl}) is the data rate. Once the upper-bound is obtained, the many optimization methods overviewed in Subsection \ref{algorithms} can be invoked for solving the approximated phase shift optimization problem. Based on this approach, various research works can be found in the literature. The single-user case was studied in \cite{hu2020statistical} and \cite{Jinghewangtccn} for Rician fading channels and correlated Rician fading channels, respectively. Given the beamforming vectors, the optimal phase shifts were derived in closed form in \cite{hu2020statistical}, and the SDR method was used in \cite{Jinghewangtccn} to solve the phase shift optimization problem. The authors of \cite{xugan2021} derived an upper bound for the data rate in both downlink and uplink multiuser scenarios, based on which the ADMM method was used for solving the phase shift optimization problem. The more complex multicell networks scenario in the presence of  interference channels was studied in \cite{zhangjietvt2021} and \cite{caihongluo2021} assuming Rician fading channels, and the phase shifts were optimized  using the GA and CCM methods, respectively. Jensen's inequality was used  in \cite{menghua2021} to derive an upper bound of the ergodic data rate in RIS-aided UAV communication systems, based on which the trajectory of the UAV and the phase shifts of the RIS were jointly optimized.

(2) \emph{Large system analysis}: Another widely used method to tackle the computation of the expectation in (\ref{opdqsqwqwl}) relies on large system analysis \cite{junzhang2021,zhang2021large,liyoutwc2021,kaizhexu}. The main idea is to apply the replica method
in large dimension random matrix theory to derive a deterministic approximation of the ergodic data rate, based on which various optimization techniques can be used to optimize the phase shifts. Specifically, the authors of \cite{junzhang2021} and \cite{zhang2021large} considered the RIS-aided single-user case under correlated Rician fading channels and the double-scattering channel model, respectively. The gradient decent method and element-wise BCD method were used for determining the RIS phase shifts in \cite{junzhang2021} and \cite{zhang2021large}, respectively. Recently, the authors of \cite{liyoutwc2021} and \cite{kaizhexu} studied RIS-aided MIMO multiple-access channels. Specifically, in \cite{liyoutwc2021}, the BS-RIS channel is assumed to be deterministic and the RIS-user channels are distributed according to a spatially correlated Rayleigh fading distribution. The MM method is used to solve the phase shift optimization problem. In \cite{kaizhexu}, all the channels are assumed to be subject to correlated Rician fading and the gradient decent method is used to solve the phase shift optimization problem.

(3) \emph{Deep reinforcement learning}: For example, a novel deep deterministic policy gradient (DDPG) based algorithm was proposed in \cite{hongrenddpg} for solving the optimization problem in (\ref{opdqsqwqwl}). The idea is that the BS first estimates the long-term CSI, based on which  the BS randomly generates a set of instantaneous CSI samples based on the channel distribution in an offline manner. Then, the generated data set is employed for DDPG training. The final trained  solutions can be used in subsequent coherence blocks. Compared with the above two methods, the appealing feature of the DDPG approach is the low training overhead, while providing  almost the same performance as the existing methods.

In \cite{xiaolinghutcom} and \cite{HuXiaoling2020}, the authors considered the case study in the presence of imperfect long-term CSI. Specifically, in \cite{xiaolinghutcom}, the effective angles from the BS to the user are first derived, and are then used for optimizing the BS beamforming and the RIS phase shifts by accounting for the errors in angle estimation. In  \cite{HuXiaoling2020}, the authors exploited user location information to obtain the angle information and derived the corresponding angle error distribution based on the distribution of estimated location information. The achievable data rate was then derived based on the statistical information of the angle error distribution.

\begin{figure}
	\centering 
	\includegraphics[width=3.3in]{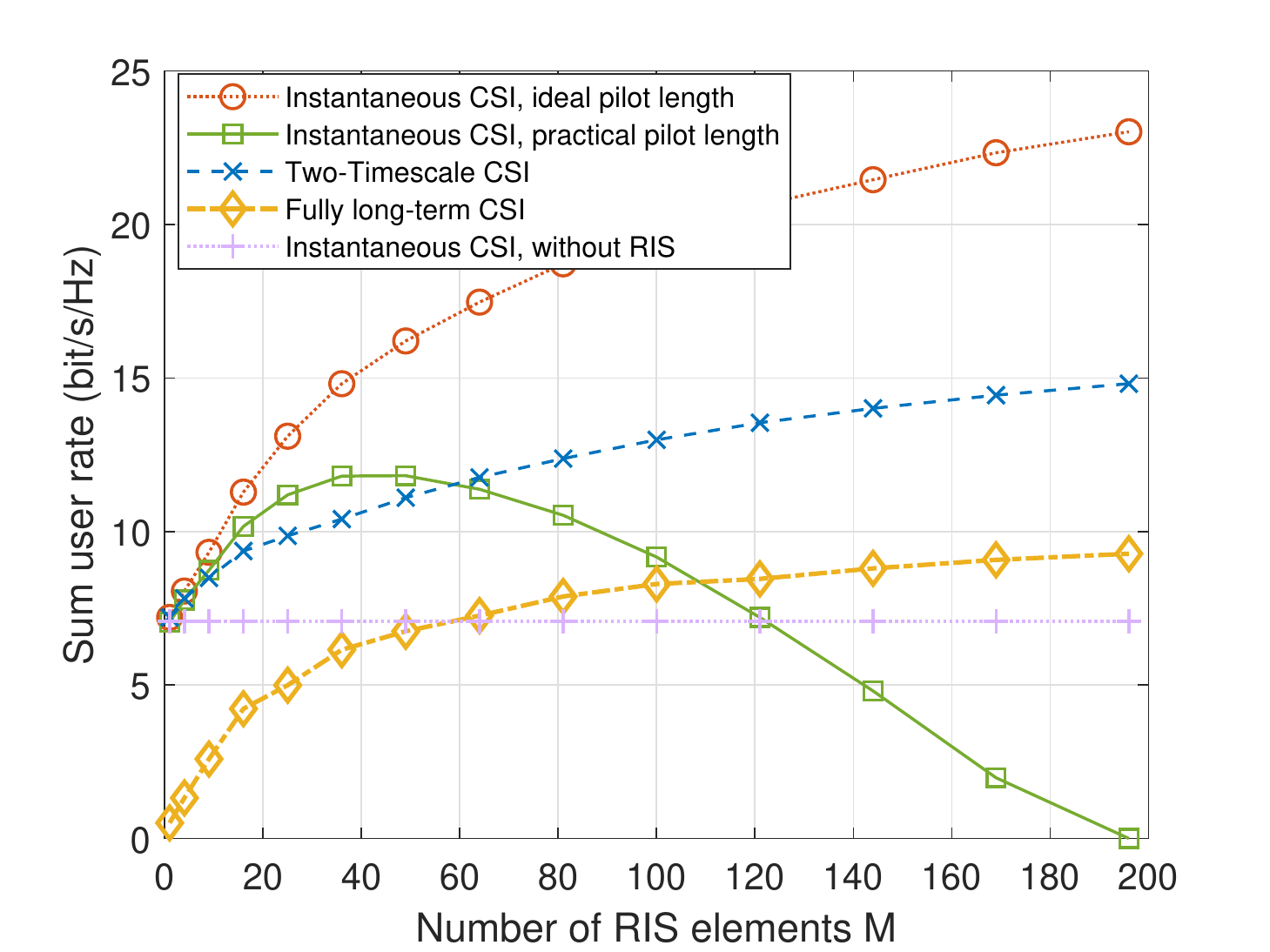}
	\caption{Comparison of achievable rate under three different levels of CSI availability in an RIS-aided massive MIMO system, where $K=4$, $U=50$, $\delta=1$, $\varepsilon_k=20,\forall k$ and $N=100$. The number of time slots in each coherence block is $T_c=196$\cite{ngo2013energy,zhang2014power}. For brevity, we set $\varpi_k=0, \forall k$ and consider the rank-$1$ LoS BS-RIS channel $\overline{\mathbf{H}}=\mathbf{a}_N\mathbf{a}_M^H$.}
	\vspace{-0.3cm}
	\label{protocol_compare}
\end{figure}

\textbf{Simulation results:} Fig. \ref{protocol_compare} illustrates the performance of transmission schemes based on instantaneous CSI, two-timescale CSI, and fully long-term CSI. We consider the  uplink transmission in massive MIMO systems where low-complexity MRC  is employed at the BS. The channel estimation overhead for long-term CSI estimation is ignored. The  pilot lengths required for acquiring the necessary instantaneous CSI and two-timescale CSI  are equal to $T_{\rm{ins}}=K + M + \max \left( {K - 1,(K - 1)\left\lceil {M/N} \right\rceil } \right)$ \cite{Zhaorui} and $T_{\rm{two}}=K$\cite{zhi2021twoarxiv}, respectively. The fully long-term CSI scheme does not require estimation of the instantaneous (effective) CSI, which leads to a pilot length equal to $T_{\rm{ful}}=0$.

As far as the instantaneous CSI case is concerned, in the $u$-th coherence block, the cascaded and direct channels of user $k$ and the phase shifts of the RIS are denoted by $ \mathbf{G}_k^u$, $\mathbf{h}_{d,k}^u$ and $\bm{\theta}^u$, respectively. Then, in the $u$-th coherence block, the instantaneous CSI-based MRC decoding vector  is set as $(\mathbf{w}_k^u{(\bm\theta^u)})^H=(    \mathbf{G}_k^u\bm{\theta}^u  + \mathbf{h}_{d,k}^u  )^H$ and the corresponding SINR of user $k$ is
\begin{align}\label{SINR_instantaneous}
&\mathrm{SINR}_{{\rm{ins}},k}^u\left(   \bm{\theta}^u   \right)=\nonumber\\
&\frac{P_k\left| \left(\mathbf{w}_k^u{\left(\bm\theta^u\right)}\right)^H \left({\mathbf{G}}_k^{u} \bm{\theta}^{u}+{\mathbf{h}}_{d,k}^{u}\right)\right|^{2}}
{\sum\limits_{i=1,i\neq k}^{K} P_i\left|\left(\mathbf{w}_k^u{\left(\bm\theta^u\right)}\right)^H \left({\mathbf{G}}_i^{u} \bm{\theta}^{u}+{\mathbf{h}}_{d,i}^{u}\right)\right|^{2}+\sigma^{2} \left\|    \mathbf{w}_k^u{\left(\bm\theta^u\right)}      \right\|^{2}},
\end{align}
and the corresponding average sum user rate is given by
\begin{align}
&R_{\rm{ins}}\!\!=\!\!\left(1\!-\!\frac{  T_{\rm{ins}}   }{T_{c}}\right)\!\! \frac{1}{U}\!\! \sum_{u=1}^{U}\sum_{k=1}^{K} \log _{2}\left(1\!+\!
\mathrm{SINR}_{{\rm{ins}},k}^u(\bm{\theta}^u)
\right),
\end{align}
where the factor $1-\frac{    T_{\rm{ins}}}{T_{c}}$ accounts for the rate loss due to the pilot overhead.

As far as the two-timescale CSI scheme is concerned, the effective channel of user $k$ in the $u$-th coherence block is denoted by $\mathbf{q}_k^u =   {\bf{H}}^u{\bm\Theta} {{\bf{h}}^u_{r,k}} +  {{\bf{h}}^u_{d,k}} $, where the phase shift matrix ${\bm\Theta}$ remains constant for all coherence blocks $1\leq u \leq U$. In the $u$-th coherence block, the instantaneous CSI-based MRC decoding vector is set equal to $ (\mathbf{w}_k^u{(\bm\Theta)})^H  =(    {\bf{H}}^u{\bm\Theta} {{\bf{h}}^u_{r,k}} +  {{\bf{h}}^u_{d,k}}     )^H$ and then the SINR of user $k$ is given by
\begin{align}\label{SINR_twotimescale}
&\mathrm{SINR}_{{\rm{two}},k}^u (\bm{\Theta}) =\nonumber\\
&\frac{P_k\left| \left(\mathbf{w}_k^u{\left(\bm\Theta\right)}\right)^H \left({\bf{H}}^u{\bm\Theta} {{\bf{h}}^u_{r,k}} +  {{\bf{h}}^u_{d,k}}  \right)\right|^{2}}
{\sum\limits_{i=1,i\neq k}^{K} P_i\left|\left(\mathbf{w}_k^u{\left(\bm\Theta\right)}\right)^H \left({\bf{H}}^u{\bm\Theta} {{\bf{h}}^u_{r,i}} +  {{\bf{h}}^u_{d,i}}  \right)\right|^{2}+\sigma^{2} \left\|    \mathbf{w}_k^u{\left(\bm\Theta\right)}      \right\|^{2}}.
\end{align}
It is worth noting that the phase shift  ${\bm\Theta}$ needs to be optimized only once  in the considered series of coherence blocks. However, in the instantaneous CSI case, the phase shift   needs to be optimized $U$ times.
The average sum user rate of the two-timescale scheme is given by
\begin{align}
&R_{\rm{two}}=\left(1-\frac{  K   }{T_{c}}\right) \frac{1}{U} \sum_{u=1}^{U}\sum_{k=1}^{K} \log _{2}\left(1+
\mathrm{SINR}_{{\rm{two}},k}^u(\bm{\Theta})
\right).
\end{align}

As for the fully long-term CSI case is concerned, the long-term CSI-based MRC beamformer is set equal to $(\mathbf{w}_k{(\bm\Theta)})^H  =  \left( \mathbb{E}\left\{{\bf{H}}^u{\bm\Theta} {{\bf{h}}^u_{r,k}} +  {{\bf{h}}^u_{d,k}} \right\}   \right) ^H$ which only contains large-scale LoS channel components and remains constant for all coherence blocks. Therefore, in the $u$-th coherence block, the SINR of user $k$ is given by
\begin{align}
&\mathrm{SINR}_{{\rm{ful}},k}^u (\bm{\Theta}) =\nonumber\\
& \frac{P_k\left| \left(\mathbf{w}_k{\left(\bm\Theta\right)}\right)^H \left({\bf{H}}^u{\bm\Theta} {{\bf{h}}^u_{r,k}} +  {{\bf{h}}^u_{d,k}}  \right)\right|^{2}}
{\sum\limits_{i=1,i\neq k}^{K} P_i\left|\left(\mathbf{w}_k{\left(\bm\Theta\right)}\right)^H \left({\bf{H}}^u{\bm\Theta} {{\bf{h}}^u_{r,i}} +  {{\bf{h}}^u_{d,i}}  \right)\right|^{2}+\sigma^{2} \left\|    \mathbf{w}_k{\left(\bm\Theta\right)}      \right\|^{2}},
\end{align}
and the corresponding average sum rate is
\begin{align}
&R_{\rm{ful}}= \frac{1}{U} \sum_{u=1}^{U}\sum_{k=1}^{K} \log _{2}\left(1+ \mathrm{SINR}_{{\rm{ful}},k}^u (\bm{\Theta})
\right).\label{rate_fully_longterm}
\end{align}

The phase shifts of the above three schemes are obtained by using the GA method with the objective of optimizing the sum data rate.
However, in the instantaneous CSI scheme, the phase shift of the RIS, ${\bm \theta}^u$, $1\leq u \leq U$, needs to be designed in each coherence block. In the two-timescale and the fully long-term CSI schemes, we only need to design the variable $\mathbf{\Theta}$ once.

Fig. \ref{protocol_compare} clearly unveils the performance tradeoffs of the considered schemes as a function of the level of CSI available.
The instantaneous CSI scheme with an ideal pilot length equal to $T_{\rm{ins}}^{(\rm{ideal})}=K=T_{\rm{two}}$ offers, as expected, the best performance. If, the actual pilot training overhead is taken into account, however, we note that the average rate first increases and then decreases as a function of the number $M$ of RIS elements. On the other hand, the fully long-term CSI scheme has the lowest pilot overhead, but it offers a relatively low achievable rate since the decoding vector at the BS cannot be adjusted according to the instantaneous effective CSI. By contrast, the two-timescale scheme offers good performance while maintaining a low pilot overhead.

\section{RIS-aided Radio Localization}\label{radiolocalization}

In wireless communication networks, radio localization
\cite{liu2007survey} offers a viable alternative for obtaining
user location information in GPS-denied environments \cite{abu2018error}.
Radio localization techniques are based on the general idea that the radio signals provide information on the position of network nodes. More precisely,  the location of agent
nodes (e.g., mobile devices or vehicles) are estimated with the aid of
known-position anchor nodes (e.g., BSs or APs) and the exchange of radio signal between the anchor and agent nodes. In general, the position of the agent node is estimated by using a two-step approach\cite{caffery2000wireless}.
First, the
distance/angle-related measurements are extracted from the received signal, then
the location is estimated from the measurements, such as the time of arrival (ToA), the time difference of arrival (TDoA),
the received signal strength (RSS), the AoA, and the AoD.

Third generation (3G) communication systems can provide a positioning accuracy of the order of tens of meters by
using TDoA measurements\cite{wymeersch20175g}. This accuracy
is improved to some extent in fourth generation (4G) systems. Existing studies
have shown that the position errors in 5G mmWave communication
systems are on the order of centimeters\cite{shahmansoori2017position,guerra2018single,abu2018error}.
Due to thriving  new applications such as smart factories, automated/assisted
driving, and augmented reality, the requirements in terms of positioning accuracy  for
5G/6G communication networks are becoming increasingly stringent. In addition, the reliability of the localization provided by 5G/6G communications is of particular importance. Since 5G/6G systems can be deployed in  high-frequency mmWave and THz bands, the links are vulnerable to obstacles. Since LoS propagation is usually required
for accurately estimating the location, existing localization methods
result in prohibitively large estimation errors if the LoS link is blocked.

The use of RISs  can yield reliable and high-precision
position estimates at a low cost and high energy efficiency. The RISs can be
integrated into existing radio localization systems
to co-work with other anchor nodes, and thus provide better positioning estimates of the agent nodes. The benefits of RISs for positioning are as follows. First, an RIS can establish
a virtual LoS  link when the LoS link is blocked. Thus,
the RIS can restore the positioning capability of the network when the GPS or the BS signals are weak. Second, RISs can be regarded, unlike active anchor nodes, as quasi-passive anchor nodes that need no power amplifiers and radio frequency chains. Therefore, they can provide high-precision estimate of the locations with low hardware cost
and low energy consumption. Third, RISs can be constructed with
a large physical aperture. Hence, they can offer a higher
angular resolution, which is appealing for radio localization. Fourth, unlike a non-reconfigurable scatterer in the environment, RISs can provide a high beamforming gain by tuning the phase shifts of the reflecting elements. As a result, the use of RISs provides several promising opportunities
for assisting localization systems in next generation wireless networks.

Current state-of-the-art research on RIS-aided radio localization
has mainly focused on three aspects: performance analysis, development of algorithms, and the interplay between communication and localization. As far as the performance analysis is concerned, the authors of \cite{HeJiguang2020VTC}
studied the theoretical performance of a 2D
RIS-aided mmWave positioning system by computing the CRB in the far field region. The authors of \cite{liu2021reconfigurable}
derived the Fisher information matrix (FIM) and the CRB for evaluating the performance of a three-dimensional (3D) RIS-assisted positioning system, where a near-field channel model was utilized.
The authors of \cite{elzanaty2021reconfigurable} derived the CRB for evaluating the localization and orientation
performance of synchronous and asynchronous signalling schemes in an RIS-assisted localization system in the  near-field region.
As far as the localization algorithms, the authors of \cite{wang2021jointxxx}
considered an RIS-assisted 3D localization system, where the AoAs and
AoDs were estimated by using the MLE algorithm, and  user localization was obtained
via a  Taylor series algorithm. The localization accuracy of an indoor localization
system was improved in \cite{zhang2020towards} and \cite{zhang2021metalocalization} by choosing proper RIS reflection coefficients that maximize the differences of RSS values among adjacent locations.
As far as the interplay between communication
and localization is concerned, the localization accuracy and the data rate were balanced by
optimizing the time allocation of an RIS-aided wireless communication
system in \cite{wang2021joint}. In particular, the CRB and the effective
achievable data rate were used as the performance metrics.   The authors
of \cite{he2020adaptive} proposed an adaptive RIS phase shifter design
based on hierarchical codebooks and feedback from the mobile user
for enabling accurate localization and high data rate transmission.

Under the spherical wave-front channel model, there exist only a few contributions that studied the RIS-aided radio localization problem.
A near-field codebook was developed in \cite{cui2021channel} for extremely large-scale RIS (XL-RIS) beam training by dividing the two-dimensional (2D) plane into several sampled points in the $x$-$y$ coordinate system.
Also, the authors of \cite{Friedlander.2019b} showed that the characteristics of the transmitted signal (transmit antenna type, size, orientation etc.) may profoundly affect the received signals in the near field, which needs to be taken into consideration for high-precision localization.
In \cite{AbuShaban.62021}, the FIM was analysed for an uplink localization system using an RIS-based lens, and the PEB and OEB were also evaluated by exploiting the wavefront curvature.
The theoretical localization performance of multipath-aided localization in both LoS and NLoS conditions was characterized
  in \cite{Rahal.2021924}, where it was shown  that using the RIS's reflected signal wavefront curvature in the near field is enough to deduce the user's position even in the absence of a direct path.

In the following, we discuss RIS-aided localization with focus on two channel conditions: 1) Far-field channel model; 2) Near-field channel model.

\subsection{Far-Field Localization Techniques}

\begin{figure}
\centering 
\includegraphics[width=3.2in]{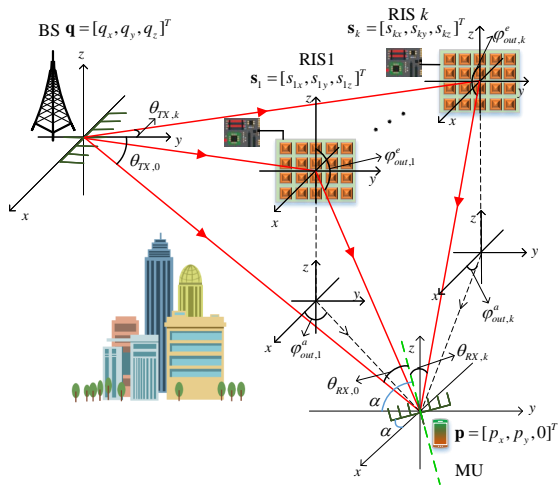} \caption{Illustration of an RIS-aided localization system.}
\vspace{-0.3cm}
 \label{Positionsystemodel}
\end{figure}
\subsubsection{System Model}
In Fig. \ref{Positionsystemodel}, as an illustrative example we consider a mmWave
MIMO system for user localization in a 3D space. The considered system model consists of a  BS  equipped with a  ULA with $N_{t}$ antennas, a mobile user (MU) with a ULA with $N_{r}$
antennas, and  $K-1$ RISs.  Each RIS is equipped with a square UPA with $L^{2}$ reflecting elements. The
carrier frequency is $f_{c}$ with a corresponding wavelength equal to
$\lambda$ and bandwidth equal to $B$. A 3D Cartesian coordinate system is used to describe the locations of the anchor nodes. The ULA of the BS is placed parallel to the $x$ axis without
rotation, and its center is located at $\mathbf{q}=[q_{x},q_{y},q_{z}]^{\rm T}\in\mathbb{R}^{3\times1}$.
The UPA of the $k$-th RIS is placed parallel to the $yz$-plane,
and its center is located at $\mathbf{s}_{k}=[s_{kx},s_{ky},s_{kz}]^{\rm T}\in\mathbb{R}^{3\times1}$.
The ULA of the MU is placed on the $xy$-plane, and is rotated by $\alpha\in[0,\pi)$
radians relative to the $x$-axis. The center of the ULA of the MU is
located at $\mathbf{p}=[p_{x},p_{y},0]^{\rm T}\in\mathbb{R}^{3\times1}$.
The symbols $q_{z}$ and $s_{kz}$ denote the heights of the BS and the
$k$-th RIS with respect to the MU on the ground. The locations of the
BS and the RIS are assumed to be known. Then, the objective of the positioning system is to estimate
the location $(p_{x},p_{y})$ of the MU and its rotation
angle $\alpha$. To simplify the notation and to facilitate the analysis, we collect the parameters
to be estimated in a vector   $\tilde{\bm{{\eta}}}=[p_{x},p_{y},\alpha]^{\rm T}$.

A narrow-band channel model is considered. The overall  channel from the MU to the BS is expressed as the combination of the direct BS-MU channel and the channels reflected by all the RISs. In mathematical terms, we have
\begin{equation}
	\mathbf{H}=\mathbf{H}_{\rm{BM}}+\sum\nolimits_{k=1}^{K-1}\mathbf{H}_{{\rm{RIS}},k},
\end{equation}
where $\mathbf{H}_{\rm{BM}}\in\mathbb{C}^{N_{r}\times N_{t}}$ is the
BS-MU channel  and $\mathbf{H}_{{\rm{RIS}},k}\in\mathbb{C}^{N_{r}\times N_{t}}$
is the $k$-th RIS's reflected channel. For the sake of analysis,
we only consider the LoS paths in the channels of $\mathbf{H}_{{\rm {BM}}}$
and $\mathbf{H}_{{\rm {RIS}},k}$, which can be further expressed
as \begin{subequations} \label{HLOSandHRISk}
	\begin{align}
		\mathbf{H}_{{\rm {BM}}} & =h_{0}\mathbf{a}_{{\rm {RX}}}(\theta_{{\rm {RX},0}})\mathbf{a}_{{\rm {TX}}}^{{\rm {H}}}(\theta_{{\rm {TX},0}}),\label{ChannelHLOS}\\
		\mathbf{H}_{{\rm {RIS}},k} & =h_{k}\mathbf{a}_{{\rm {RX}}}(\theta_{{\rm {RX}},k})\mathbf{a}_{{\rm {RIS}},{\rm {OUT}}}^{\rm H}(\varphi_{{\rm {out}},k}^{a},\varphi_{{\rm {out}},k}^{e})\bm{{\Theta}}_{k}\nonumber \\
		& \quad\cdot\mathbf{a}_{{\rm {RIS}},{\rm {IN}}}(\varphi_{{\rm {in}},k}^{a},\varphi_{{\rm {in}},k}^{e})\mathbf{a}_{{\rm {TX}}}^{\rm H}(\theta_{{\rm {TX}},k}),\label{ChannelHRISk}
	\end{align}
\end{subequations}where $h_{0}$ and $\{h_{1},\cdots,h_{K-1}\}$ are
the complex channel gains of the BS-MU direct link and the BS-RIS-MU
reflected links, respectively. $\bm{{\Theta}}_{k}$ is the reflection
coefficient matrix of the $k$-th RIS. $\mathbf{a}_{{\rm {TX}}}(\theta_{{\rm {TX},0}})\in\mathbb{C}^{N_{t}\times1}$
and $\mathbf{a}_{{\rm {RX}}}(\theta_{{\rm {RX},0}})\in\mathbb{C}^{N_{r}\times1}$
are the antenna response vectors of the transmitter and the receiver,
in which $\theta_{{\rm {TX},0}}$ denotes the AoD and $\theta_{{\rm {RX},0}}$
denotes the AoA of the BS-MU direct link. Since the transmitter and
the receiver use ULAs, $\mathbf{a}_{{\rm {TX}}}(\theta)$ and $\mathbf{a}_{{\rm {RX}}}(\theta)$
can be expressed as \begin{subequations}
	\begin{equation}
		\mathbf{a}_{{\rm {TX}}}(\theta)=[1,e^{j\frac{2\pi}{\lambda}d\sin(\theta)},\cdots,e^{j(N_{t}-1)\frac{2\pi}{\lambda}d\sin(\theta)}]^{{\rm {T}}},
	\end{equation}
	\begin{equation}
		\mathbf{a}_{{\rm {RX}}}(\theta)=[1,e^{j\frac{2\pi}{\lambda}d\sin(\theta)},\cdots,e^{j(N_{r}-1)\frac{2\pi}{\lambda}d\sin(\theta)}]^{{\rm {T}}},
	\end{equation}
\end{subequations}where $d$ denotes the distance between adjacent
antennas and $\theta$ is the AoA or AoD. The $\mathbf{a}_{X}(\varphi_{{\rm {x}},k}^{a},\varphi_{{\rm {x}},k}^{e})$
in (\ref{ChannelHRISk}) denotes the array response vector of the
RIS array, which is equal to $\mathbf{a}_{{\rm {RIS}},{\rm {IN}}}(\varphi_{{\rm {in}},k}^{a},\varphi_{{\rm {in}},k}^{e})$
and $\mathbf{a}_{{\rm {RIS}},{\rm {OUT}}}(\varphi_{{\rm {out}},k}^{a},\varphi_{{\rm {out}},k}^{e})$
when $x$ is equal to IN and OUT, respectively. Since the RISs use
UPAs, $\mathbf{a}_{X}(\varphi_{{\rm {x}},k}^{a},\varphi_{{\rm {x}},k}^{e})$
can be expressed as
\begin{align*}
	& \mathbf{a}_{X}(\varphi_{{\rm {x}},k}^{a},\varphi_{{\rm {x}},k}^{e})=[1,e^{j\frac{2\pi}{\lambda}d\cos(\varphi_{{\rm {x}},k}^{e})},\cdots,e^{j\frac{2\pi}{\lambda}\left(L-1\right)d\cos(\varphi_{{\rm {x}},k}^{e})}]^{{\rm {T}}}\\
	& \otimes[1,e^{j\frac{2\pi}{\lambda}d\sin(\varphi_{{\rm {x}},k}^{e})\sin(\varphi_{{\rm {x}},k}^{a})}\cdots,e^{j\frac{2\pi}{\lambda}\left(L-1\right)d\sin(\varphi_{{\rm {x}},k}^{e})\sin(\varphi_{{\rm {x}},k}^{a})}]^{{\rm {T}}},
\end{align*}
where $\varphi_{{\rm {in}},k}^{a}$ and $\varphi_{{\rm {in}},k}^{e}$
are the azimuth AoA and elevation AoA at the $k$-th RIS from the
BS, while $\varphi_{{\rm {out}},k}^{a}$ and $\varphi_{{\rm {out}},k}^{e}$
are the azimuth AoD and elevation AoD at the $k$-th RIS to the MU.

The complex channel gains $h_{k},k=0,1,\cdots,K-1$ in (\ref{HLOSandHRISk})
follow the distribution of $\mathcal{CN}(0,10^{\frac{-{\rm {PL}}_{k}}{10}})$,
where ${\rm {PL}}_{0},{\rm {PL}}_{1},\cdots,{\rm {PL}}_{K-1}$ are
the corresponding path losses. Based on \cite{wankai} and \cite{5G}, the path loss of the BS-MU channel in \eqref{ChannelHLOS}
can be expressed in dB as
\begin{align}
	{\rm {PL}}_{0}=&10\log_{10}(64\pi^{3})+10\alpha_{0}\log_{10}d_{BM}\nonumber \\
	&+20\log_{10}f_{c}+\xi_{0},\label{pathlossLOS}
\end{align}
where $d_{BM}$ is the  distance (in meters) between the BS and the MU, $\alpha_{0}$
is the path loss exponent. $\xi_{0}\sim\mathcal{N}(0,\sigma_{SF_{0}}^{2})$
is the log-normal term accounting for the shadow fading, where $\sigma_{SF_{0}}^{2}$
denotes the log-normal shadowing variance.

The path loss of the RIS-reflected channel via the $k$-th RIS can be
expressed in dB as
\begin{align}
	{\rm {PL}}_{k}=&10\log_{10}(64\pi^{3})+10\alpha_{k}\log_{10}(d_{\rm{BR}}\cdot d_{\rm{RM}})\nonumber  \\
	&+40\log_{10}f_{c}+\xi_{k},\label{pathlossReflect}
\end{align}
where $d_{\rm{BR}}$ and $d_{\rm{RM}}$ are the   distances (in
meters) of the BS-RIS $k$ link and RIS $k$-MU link, respectively.
$\alpha_{k}$ is the path loss exponent of the reflected channel,
and $\xi_{k}$ is the log-normal term accounting for the shadow fading following  distribution
$\mathcal{N}(0,\sigma_{SF_{k}}^{2})$ with variance $\sigma_{SF_{k}}^{2}$.

For estimating the locations of users, we assume a pilot-based transmission scheme. The transmitted pilot signal is a continuous time-domain
waveform $x(t)$ with   bandwidth $B$ and duration of
$T_{o}$ seconds. Assuming $M_{t}$ pilot signals, i.e., $\mathbf{x}(t)=[x_{1}(t),x_{2}(t),\cdots,x_{M_{t}}(t)]^{\rm T}\in\mathbb{C}^{M_{t} \times 1}$, is transmitted at time $t$ with  corresponding beamforming matrix denoted by $\mathbf{F}=[\mathbf{f}_{1},\mathbf{f}_{2},\cdots,\mathbf{f}_{M_{t}}]\in\mathbb{C}^{N_{t}\times M_{t}}$ where  $M_{t}\ll N_{t}$, then the signal $\mathbf{y}(t)\in\mathbb{C}^{N_{r}\times1}$ received at the MU is given by
\begin{equation*}
	\mathbf{y}(t)   =\mathbf{H}_{\rm{BM}}\mathbf{F}\mathbf{x}(t-\tau_{0})+\sum_{k=1}^{K-1}\mathbf{H}_{\rm{RIS},k}\mathbf{F}\mathbf{x}(t-\tau_{k})+\mathbf{n}(t),
\end{equation*}
where  $\tau_{0}$ is the propagation delay of the
direct BS-MU path, and $\tau_{k},k=1,2,\cdots,K-1$ is the propagation
delay from the BS to the MU via the $k$-th RIS. The delays $\tau_{k},k=0,1,2,\cdots,K-1$
is also called TOAs. The received noise $\mathbf{n}(t)$
is Gaussian with zero mean and two-sided power spectral density equal to
$N_{0}/2$.

\subsubsection{Two-Step Localization Scheme}

The position and orientation of the MU can be estimated based on the observed signal $\mathbf{y}(t)$
 by using a two-step localization scheme. In the first step, the angles,
channel gains, and time delays are estimated using some existing methods.
 For example, the channel gains $\{\hat{h}_{k},\hat{h}_{0}\}$ can be estimated with the  LS  approach,
 while the angles $\{\hat{\theta}_{TX,0}, \hat{\varphi}_{out,k}^{a}, \hat{\varphi}_{out,k}^{e}, \hat{\theta}_{RX,0}, \hat{\theta}_{RX,k}\}$
 can be estimated by using array signal processing algorithms in the spatial domain, such as MUSIC and ESPRIT.
The estimated time delays $\{\hat{\tau}_{k}, \hat{\tau}_{0}\}$ can be extracted from the pilot signal.
In the second step, the position of the MU can be obtained by using multi-angulation (or triangulation) from the
angle-related measurements, or by using multi-lateration (or trilateration) methods from distance-related measurements,
 or by a combination of both.

As shown in Fig. \ref{Positionsystemodel}, the localization measurements are closely related with the coordinates
 and rotation angle of MU, which means that there exists a mapping from the MU's location to the measurements obtained in the first step, which can be described as follows.

\subsubsection*{ToA}

The propagation delay of the BS-MU path, which is related with the MU coordinate $\mathbf{p}$, is given by
\begin{align}
\hat{\tau}_{0} & =\left\Vert \mathbf{q}-\mathbf{p}\right\Vert /c+n_{t_{0}},\label{l1}
\end{align}
where $c$ is the speed of light and $n_{t_{0}}$ denotes the ToA
error. The ToA $\hat{\tau}_{k0}$ from the MU to the $k$-th RIS can be obtained by subtracting the propagation delay
of the $k$-th RIS-BS path from the propagation delay of the path
from BS to MU via the $k$-th RIS, which are expressed as
\begin{align}
\hat{\tau}_{k0}=\left\Vert \mathbf{p}-\mathbf{s}_{k}\right\Vert /c+n_{t_{k}}=\hat{\tau}_{k}-\left\Vert \mathbf{q}-\mathbf{s}_{k}\right\Vert /c+n_{t_{k}},\label{l2}
\end{align}
where $n_{t_{k}}$ denotes the ToA error.

The ToA measurements are related with the path lengths, which are further related with the MU position. Specifically, the $\hat{\tau}_{0}$ and $\hat{\tau}_{k0}$ can describe the estimated distance
of the direct path as well as the estimated distance from the MU to
the $k$-th RIS, which are expressed as follows
\begin{subequations}
\begin{align}
\hat{d}_{0}&=c\hat{\tau}_{0}=\left\Vert \mathbf{q}-\mathbf{p}\right\Vert +n_{d_{0}},\label{1l3} \\
\hat{d}_{k}&=c\hat{\tau}_{k0}=\left\Vert \mathbf{p}-\mathbf{s}_{k}\right\Vert +n_{d_{k}},\label{2l3}
\end{align}
\end{subequations}
where $n_{d_{0}}$ and $n_{d_{k}}$ are the estimation errors.
\subsubsection*{TDoA}
The location of the MU can be derived from the TDoA, which describes the  distance differences between different pairs of
the direct MU-BS path and the $k$ MU-RIS paths.

 Specifically, by taking the direct path from the BS to MU as the reference
 path, the TDoA can be expressed as
\begin{align}
\hat{\tau}_{td,k}=\hat{\tau}_{k0}-\hat{\tau}_{0}.\label{ll4}
\end{align}
Then, the distance difference corresponding to the TDoA is represented as
\begin{align}
c\hat{\tau}_{td,k}&=c\hat{\tau}_{k0}-c\hat{\tau}_{0}=\hat{d}_{d,k} =(\hat{d}_{k}-\hat{d}_{0})\nonumber \\
 & =\left\Vert \mathbf{p}-\mathbf{s}_{k}\right\Vert -\left\Vert \mathbf{q}-\mathbf{p}\right\Vert +n_{td_{k}},\label{l4}
\end{align}
where $n_{td_{k}}$ is the estimation error of $d_{d,k}$.

\subsubsection*{AoA and AoD}

The angle-related measurements (AoA and AoD) are estimated at the arrays of BS, MU, and RIS  using angle estimation algorithms. They are also closely related with the MU position, which can be described as
\begin{align}
\hat{\theta}_{TX,0} & =\arcsin\left(\frac{p_{x}-q_{x}}{\left\Vert \mathbf{p}-\mathbf{q}\right\Vert _{2}}\right)+n_{\theta_{TX,0}},\label{l5}\\
\hat{\varphi}_{out,k}^{a} & =\arcsin\left(\frac{p_{y}-s_{ky}}{\sqrt{\left(p_{x}-s_{kx}\right)^{2}+\left(p_{y}-s_{ky}\right)^{2}}}\right)+n_{a_{k}},\\
\hat{\varphi}_{out,k}^{e} & =\arccos\left(\frac{-s_{kz}}{\left\Vert \mathbf{p}-\mathbf{s}_{k}\right\Vert _{2}}\right)+n_{e_{k}},\\
\hat{\theta}_{RX,0} & =\arcsin\left(\frac{\left(p_{x}-q_{x}\right)\cos\alpha-\left(p_{y}-q_{y}\right)\sin\alpha}{\left\Vert \mathbf{p}-\mathbf{q}\right\Vert _{2}}\right)\nonumber \\
 & \quad+n_{\theta_{RX,0}},\\
\hat{\theta}_{RX,k} & =\arcsin\left(\frac{\left(p_{x}-s_{kx}\right)\cos\alpha-\left(p_{y}-s_{ky}\right)\sin\alpha}{\left\Vert \mathbf{p}-\mathbf{s}_{k}\right\Vert _{2}}\right)\nonumber \\
 & \quad+n_{\theta_{RX,k}}.
\end{align}
where $n_{\theta_{TX,0}}$,$n_{a_{k}}$,$n_{e_{k}}$, $n_{\theta_{RX,0}}$,
and $n_{\theta_{RX,k}}$ denote error associated to the AoA and AoD estimations. Besides,
 $\theta_{TX,k}$ are known values, which can be calculated in advance from the known coordinates of the BS and RISs.

\subsubsection*{Channel Gains}
Some recent works, e.g., \cite{HeJiguang2020VTC,elzanaty2021reconfigurable}, extract
the distance information from the channel gains $h_{k}$ according to the relations $h_{0}=\frac{\lambda}{4\pi}\frac{1}{d_{BM}}$
and $h_{k}=\frac{\lambda}{4\pi}\frac{1}{d_{BR}d_{RM}}$. However, the channel gains depend on the distance-dependent large-scale path loss as well as the distance-independent shadowing fading.
Thus, it may not be accurate to extract the distance information
from the channel gains. Therefore, this measurement is not  considered in the following numerical results.

In the second step, by collecting the equations mapping the MU position into the measurements illustrated above, a set of interlinked equations can be formulated. The position and rotation angle of the MU, i.e., the vector
$\tilde{\bm{{\eta}}}=[p_{x},p_{y},\alpha]^{\rm T}$ can be retrieved by jointly solving the set of  equations. The larger the number of measurements (equations) is available, the better
the estimate of the position is. The obtained equations are, however, nonlinear, and the measurements are affected by errors. Consequently, they are difficult to solve.
To circumvent this difficulty, wireless localization algorithms, such as the Chan algorithm \cite{chan1994simple} and
the Taylor series algorithm \cite{foy1976position} can be utilized.

\subsubsection{Position Error Bound (PEB) and Rotation Error Bound (REB) Analysis}

To evaluate the performance of the position and orientation estimation,
we analyze the  FIM  and  CRB. To provide a benchmark for the accuracy of the position estimation,
the CRB and the PEB/REB are investigated in \cite{HeJiguang2020VTC,he2020adaptive},
by assuming a  2D  localization problem formulation.
In the following, we derive the CRB and PEB/REB in a general 3D space. Further details can be found in \cite{Liuyu2021multiple}.

First, we construct the FIM based  on the channel parameters, including the
ToA, AoA, AoD, and the complex channel gains. The unknown parameters
are collected in the following vector
\begin{equation}
\bm{\eta}=[\bm{\tau},\theta_{{\rm{TX}},0},\bm{\theta}_{{\rm{RX}}},\bm{\varphi}_{{\rm{out}}}^{a},\bm{\varphi}_{{\rm{out}}}^{e},\mathbf{h}_{R},\mathbf{h}_{I}]^{\rm T},
\end{equation}
where \begin{subequations}
\begin{align}
\bm{\tau} & =[\tau_{0},\tau_{1},\cdots,\tau_{K-1}]^{\rm T},\\
\bm{\theta}_{{\rm{RX}}} & =[\theta_{{\rm{RX}},0},\theta_{{\rm{RX}},1},\cdots,\theta_{{\rm{RX}},K-1}]^{\rm T},\\
\bm{\varphi}_{{\rm{out}}}^{a} & =[\varphi_{{\rm{out}},1}^{a},\varphi_{{\rm{out}},2}^{a},\cdots,\varphi_{{\rm{out}},K-1}^{a}]^{\rm T},\\
\bm{\varphi}_{{\rm{out}}}^{e} & =[\varphi_{{\rm{out}},1}^{e},\varphi_{{\rm{out}},2}^{e},\cdots,\varphi_{{\rm{out}},K-1}^{e}]^{\rm T},\\
\mathbf{h}_{R} & =[h_{R,0},h_{R,1},\cdots,h_{R,K-1}]^{\rm T},\\
\mathbf{h}_{I} & =[h_{I,0},h_{I,1},\cdots,h_{I,K-1}]^{\rm T},
\end{align}
\end{subequations}
 and $\theta_{{\rm{TX}},0}$ denotes the AOD from the BS to the MU. The real parts of the channel gains $h_{R,k}={\rm Re}\{h_{k}\}$
are collected in the vector   $\mathbf{h}_{R}$  and the imaginary parts
$h_{I,k}={\rm Im}(h_{k})$ are collected in the vector $\mathbf{h}_{I}$. We denote the unbiased estimator of the unknown
channel parameters as $\hat{\bm{\eta}}$. Then, the mean-square error
bound is
\begin{equation}
\mathbb{E}[(\hat{\bm{\eta}}-\bm{\eta})(\hat{\bm{\eta}}-\bm{\eta})^{\rm H}]\ge\mathbf{J}_{{\rm \mathbf{\bm{\eta}}}}^{-1},
\end{equation}
where $\mathbf{J}_{\bm{{\eta}}}$ is the $(6K-1)\times(6K-1)$ dimensional
FIM. Assuming that the noise is a wide-sense stationary (WSS) Gaussian random vector and is written as $\bm{n}(t)$, the element in the $m$-th
row and $n$-th column of matrix $\mathbf{J}_{{\rm \mathbf{\bm{\eta}}}}$ can be written as \cite{1993Fundamentals}
\begin{align}
[\mathbf{J}_{{\rm \mathbf{\bm{\eta}}}}]_{mn} & =\mathbb{E}_{\mathbf{y}(t)\left|\mathbf{\bm{\eta}}\right.}\left[-\frac{\partial^{2}\ln p(\mathbf{y}(t)\left|\bm{{\eta}}\right.)}{\partial\eta_{m}\partial\eta_{n}}\right]\nonumber \\
 & \approx\frac{2}{N_{0}}\int_{0}^{T_{0}}\textrm{Re}\left\{ \frac{\partial\bm{\mu}^{\rm H}(t)}{\partial\eta_{m}}\frac{\partial\bm{\mu}(t)}{\partial\eta_{n}}\right\}.
\end{align}
where $\eta_{m}$ denotes the $m$-th element of $\bm{\eta}$, $T_{0}$ denotes the observation time, $\bm{\mu}(t)=\mathbf{y}(t)-\bm{n}(t)$, and $p(\mathbf{y}(t)\left|\bm{{\eta}}\right.)$ denotes the probability density function (PDF) (i.e, the likelihood function) of the random vector $\mathbf{y}(t)$ conditioned on the parameter vector $\bm{{\eta}}$.

The FIM of the position and orientation of the MU, i.e., $\bf \tilde \eta$, can be obtained by applying a
transformation of variables from the channel parameters
$\bm{\eta}$ to the location parameters $\tilde{\bm{{\eta}}}$. Specifically, the
FIM of $\tilde{\bm{{\eta}}}$ is obtained by applying the transformation
matrix $\mathbf{T}$ defined as
\begin{equation}
[\mathbf{T}]_{1:3,m}=[\frac{\partial\eta_{m}}{\partial p_{x}},\frac{\partial\eta_{m}}{\partial p_{y}},\frac{\partial\eta_{m}}{\partial\alpha}]^{\rm T}.\label{TransMatrix}
\end{equation}
The equation for the FIM of the position and orientation parameters in $\tilde{\bm{{\eta}}}$ is given by \cite{1993Fundamentals}
\begin{equation}
\mathbf{J}=\mathbf{T}\mathbf{J}_{\bm{\eta}}\mathbf{T}^{\rm H}.
\end{equation}
Once the FIM is obtained, the CRB is
obtained by inverting the matrix. Thus, the PEB
is equal to the square root of the trace of the first $2\times2$ sub-matrix
and can be formulated as
\begin{equation}
{\rm PEB}=\sqrt{\mathrm{tr}\left(\mathbf{J}_{1:2,1:2}^{-1}\right)},
\end{equation}
and the REB is equal to the square root of
the third entry of diagonal
\begin{equation}
{\rm REB}=\sqrt{\mathrm{tr}\left(\mathbf{J}_{3,3}^{-1}\right)}.
\end{equation}

\textbf{Simulation results:} To illustrate the position accuracy that can be obtained in the presence of RISs, we report some numerical results based on the measurements of the TDoA in (\ref{l4}). Specifically, we compare the root mean square error (RMSE) of the position estimation obtained by using Taylor and Chan's algorithms against the the CRB. The locations in metre of the BS and MU are ${\bf q}=[0,0,40]^T$ and ${\bf p}=[90,30,0]^T$, respectively,  and the locations of three RISs are ${\bf s1}=[60,45,15]^T$, ${\bf s2}=[50,50,5]^T$ and ${\bf s3}=[40,20,10]^T$. The setup with RISs is considered as a case study. The phase shift matrix of the RIS is set to a unit identity matrix for the sake of analysis, if the matrix is optimized, the estimation accuracy will be improved. The other simulation parameters are given in the caption of the figure, where $\sigma_{td}$ denotes the variance of the Gaussian estimation error $n_{td_k}$. It can be seen that Taylor's and Chan's algorithms achieve almost the same performance and have negligible performance loss compared with the CRB.
\begin{figure}
	\centering 
	\includegraphics[width=3.3in]{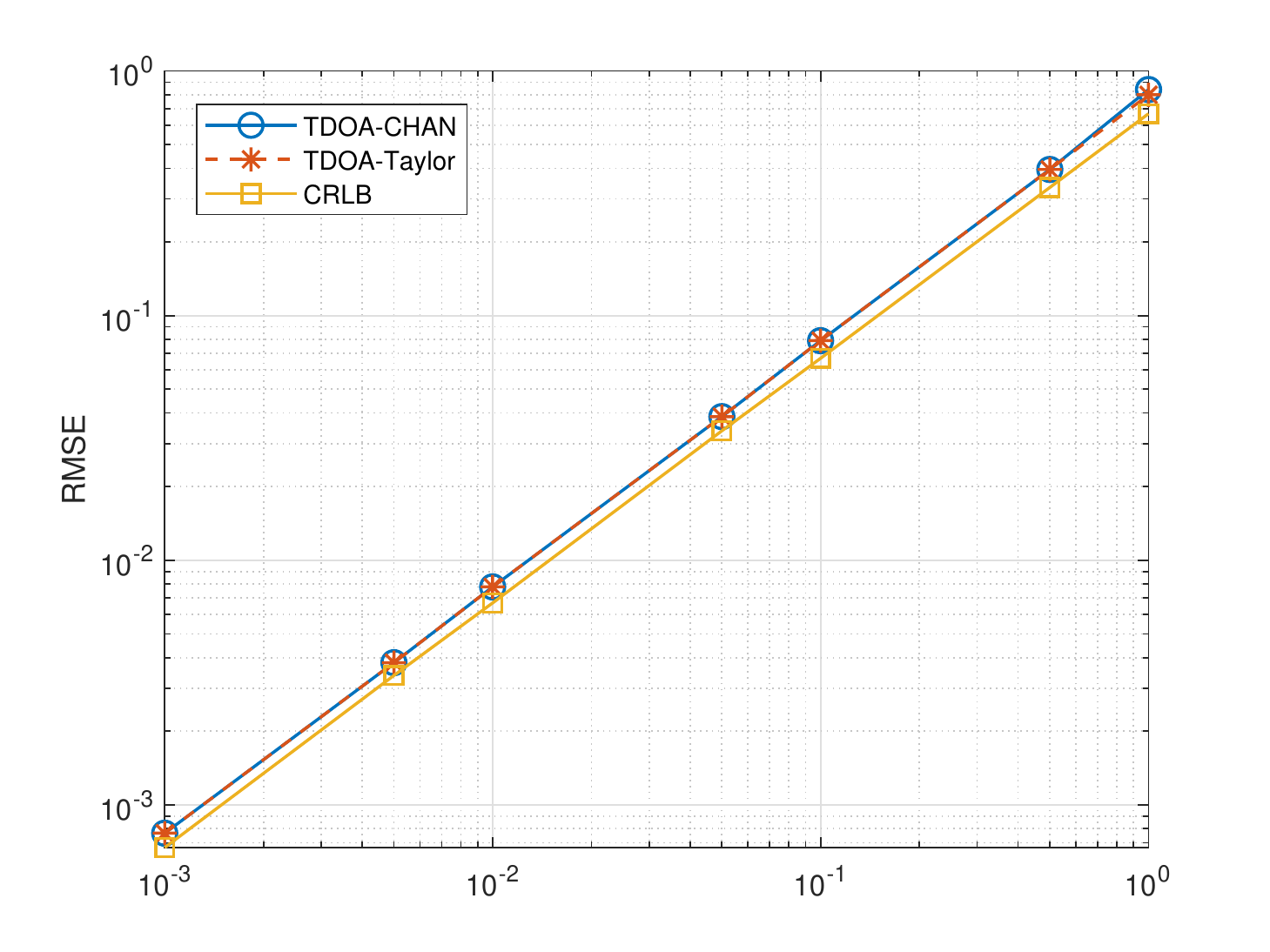}
	\caption{RMSE versus the standard variance $\sigma_{td}$, when $N_t=32$, $N_r=8$, $L^2=16\times16$, $f_c=9$ GHz, and $K=4$.}
	\vspace{-0.3cm}
	\label{tdoa-algorithm}
\end{figure}
\subsection{Near-Field Localization Techniques}

Thanks to the absence of power amplifiers, digital processing units, and RF chains for each scattering element, a large number of quasi passive reflecting elements can be integrated on the RIS panel at an affordable cost and power consumption.
Consequently, an RIS may evolve into a panel with a very large size,
 which is sometimes called  XL-RIS.

\begin{figure}
	\centering
	\includegraphics[width=0.25\textwidth]{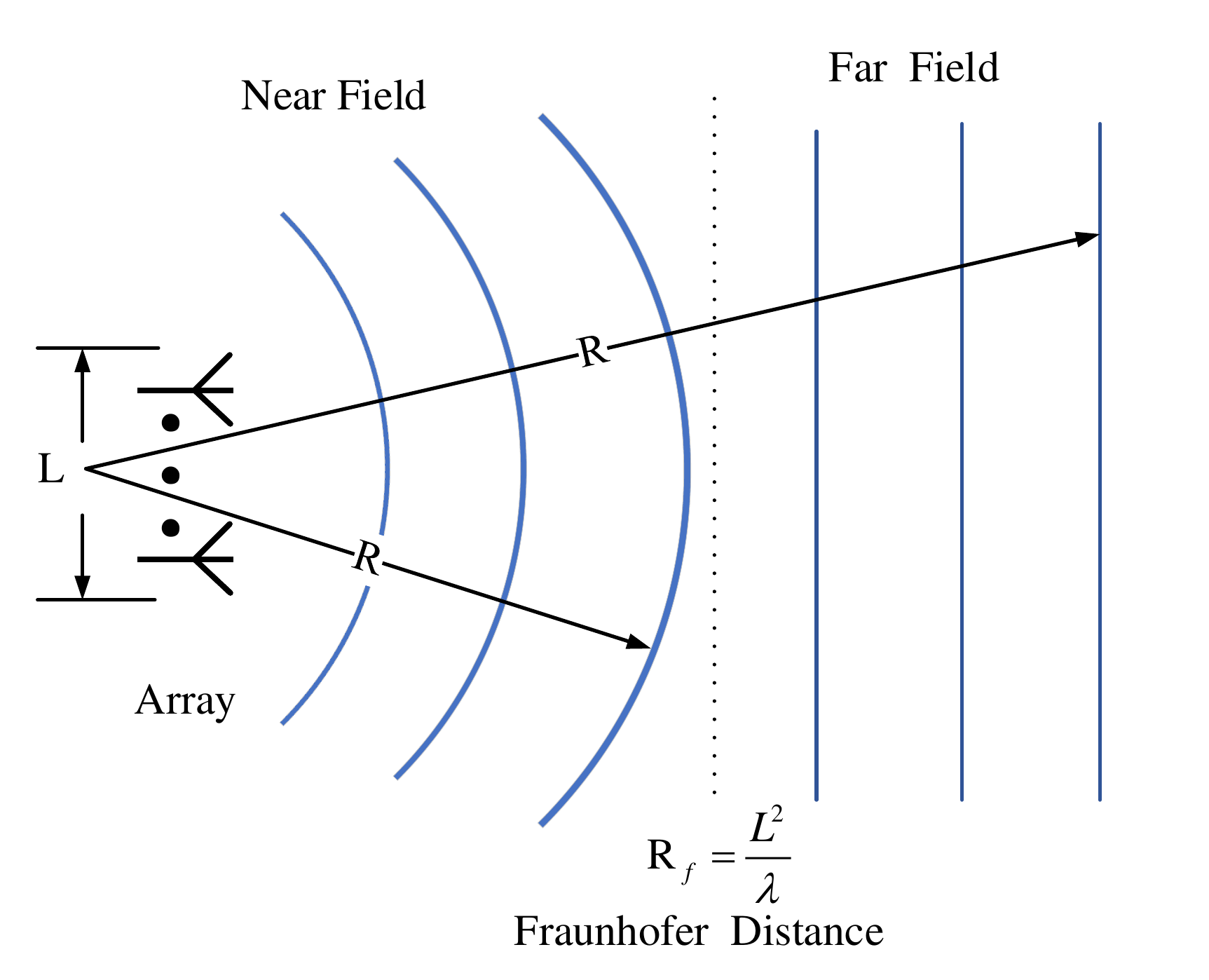} \vspace{-1em}
	\caption{Illustration of the Fraunhofer distance and the far-field planar wave approximation.}
	\label{FigFarfield}
\end{figure}

In this case, as illustrated in Fig. \ref{FigFarfield}, the RIS may be large enough and the observation point may be sufficiently close to it that the far-field planar wave assumption may not hold anymore. Specifically, the spherical curvature of the wavefront may not be ignored. In these cases, specifically, a signal formulation that accounts for the near-field is needed.
According to \cite{selvan2017fraunhofer}, the far-field of the RIS can be defined as the set of observation distances $R$ that are greater than the Fraunhofer distance $R_{f}$, i.e.,
\begin{equation}
	R\geq R_{f}=\frac{2L^{2}}{\lambda},\label{FDistance}
\end{equation}
where $L$ is the maximum aperture of the RIS UPA  and $\lambda$ is the carrier wavelength. This conventional definition of far-field region corresponds to a maximum phase error equal to $\pi/8$ over the entire length $L$.

According to (\ref{FDistance}), the Fraunhofer distance can be quite large if an XL-RIS is utilized and its size is sufficiently large with respect to the wavelength.
If the RIS is deployed at  high-frequency bands, such as mmwave/sub-THz,  the Fraunhofer distance also increases due to the smaller
 wavelength \cite{Dovelos.2021329,danufane2020pathloss,RenzoRIS2020}. For instance, let us consider a $100\times 100$-UPA RIS
 panel that operates at the carrier frequency of $200$ GHz.
The RIS elements are assumed to be spaced by one half-wavelength, i.e. $\frac{\lambda}{2} = 0.75$ mm,
and, thus,  the maximum aperture (the diagonal) of the RIS panel is $\sqrt{2}\times 100 \times \frac{\lambda}{2} \approx 10.6$ cm.
Then, the Fraunhofer distance for this scenario is $ R_{f}= 7.5$ meters.
Therefore, it is likely that some users are in the near-field when the RIS is deployed for indoor scenarios.
In this case, the spherical wavefront of the electromagnetic waves cannot be ignored.
Recent system-level simulations have shown that the near-field region may not be ignored in  outdoor scenarios either \cite{sihlbom2021reconfigurable}.

\subsubsection{Near-Field Channel Model}

In the near-field region, spherical wave-front provide an underlying generic parametric model for estimating the positions of MUs and scatterers \cite{Yin.2017}.
Under the assumption of a spherical wave-front, the steering responses of the RIS and the BS arrays need to be parameterized by
taking into account  the 3D location of the signal source rather than by simply  utilizing the AoDs and the AoAs.

As an example, let us consider the uplink transmission where  MU is equipped with one antenna, the RIS is an
 $M_x\times M_z$ UPA that lies on the XOZ plane, and the BS is equipped with an $N$-element ULA.
In the near-field, the channel from the MU to the RIS is
\begin{align}
	\bm{h}_{r}  =\sum\nolimits_{l=0}^{L_{\mathrm{RU}}}\tilde{\beta}_{l}\bm{a}_R(\bm{r}_{l}) =\bm{A}_{\mathrm{R}}(\bm{r}_{l})\boldsymbol{\tilde{\beta}},
\end{align}
where $L_{\mathrm{RU}}$ is the number of scatterers or spatial paths, the vector $\boldsymbol{\tilde{\beta}}=[\tilde{\beta}_{0},\ldots,\tilde{\beta}_{L_{\mathrm{RU}}}]^{\mathrm{T}}$
contains the complex path gains, and the matrix
 $\bm{A}_{\mathrm{R}}(\bm{r}_{l})=[\bm{a}_{\mathrm{R}}\left(\bm{r}_{0}\right),\ldots,\bm{a}_R(\bm{r}_{L_{\mathrm{RU}}})]$
 collects the steering vectors associated with the scatterers that produce the multiple propagation paths.
The LoS path from the MU to the RIS is denoted as  path $0$, and the location of the MU is denoted as $\bm{r}_{0} = [x_0,y_0,z_0]^T$.
The steering vector associated with the $l$-th scatterer located at ${\bm p}_{l} =[x_l,y_l,z_l]^T$ is denoted as $\bm{a}({\bm p}_{l})$, and is given by
\begin{align}
	\!\!\!\!\!\!\bm{a}({\bm p}_{l})&=[e^{-j\frac{2\pi}{\lambda}d_{1,1}({\bm p}_{l})},\cdots,e^{-j\frac{2\pi}{\lambda}d_{1,M_z}({\bm p}_{l})},\cdots,\nonumber \\
	& \qquad \quad e^{-j\frac{2\pi}{\lambda}d_{M_x,1}({\bm p}_{l})},\cdots,e^{-j\frac{2\pi}{\lambda}d_{M_x,M_z}({\bm p}_{l})}]^{\rm T},
\end{align}
where $d_{m,n}({\bm p}_{l})$ represents the distance from the scatterer located at ${\bm p}_{l}$ to the $(m,n)$-th RIS element.

We concentrate on the LoS channel between the RIS and the AP and assume that the communication channel is slowly varying and
 frequency-flat.
By taking the spherical nature of the wave propagation, the path length from each antenna elements of the BS to each elements of the RIS determines the phase shifts of the received signals.
Similar to \cite{Bohagen.2007} and \cite{Bohagen.2009}, we assume that the path loss of all the received paths is the same
 and is denoted by $\alpha$. Under this assumption, the elements of the BS-RIS LoS channel are collected in the matrix
 $\bm{\bm{H}}$ whose elements are
\begin{equation}
	\bm{\bm{H}}(m,n) = \alpha \exp\left(j \frac{2\pi}{\lambda}r_{m,n}\right),
\end{equation}
where $r_{m,n}$ denotes the path length between the $m$-th antenna element of the BS and the $n$-th RIS element.

The signal transmitted by the MU is denoted by $s$ and the received data at the BS is
\begin{equation}\label{NearFiedy}
	\bm{y} =  \bm{H}{\bf{\Theta }}\bm{h}_{r} s  + \bm{n},
\end{equation}
where $\bm{n} \sim {\cal C}{\cal N}\left( {{\bf{0}},{\sigma^2}\bf{I}} \right)$ and ${\bf{\Theta }}$ denote the AWGN and
the reflection coefficient matrix defined in Section II, respectively.

The set of unknown parameters are
\begin{align}
	\bm{\eta} &= [{\bm p}_{0},{\bm p}_{1} \cdots,{\bm p}_{L_{\mathrm{RU}}}, \bm\beta^T]^T,
\end{align}
where $\bm\beta =[\alpha\tilde{\beta}_{0},\ldots,\alpha\tilde{\beta}_{L_{\mathrm{RU}}}]^{\mathrm{T}}$
 denotes the vector of cascaded channel fading coefficients.

A commonly adopted localization method is the maximum-likelihood estimation.
The log-likelihood function of the received signal ${\bf y}$ is given by $f\left(\bm{y} |\bf{\Theta},\bm{\eta}\right) \propto   -|| \bm{y} - {\bf{H}}{\bf{\Theta }}{\bf{h}}_{r}^{\rm{H}}(\bm{\eta}) s ||^2$, where irrelevant constant terms are ignored.
Then, the maximum likehood estimation problem for the locations of the MU and the scatterers is given by
\begin{equation}
	\begin{aligned}
		&\underset {\bm{\eta},\bf{\Theta}}{\max}\;\;f\left(\bm{y} |\bf{\Theta},\bm{\eta}\right) =
		\underset {\bm{\eta},\bf{\Theta}}{\min}\;\;|| \bm{y} - {\bm{H}}{\bf{\Theta }}{\bm{h}}_{r}(\bm{\eta}) s ||^2\\
		&\text{s.t.} \quad\text{C1}: \bm{\eta} \in {{\cal S}_{\eta}},\\
		&\quad\quad\;\text{C2}:  {\theta _m} \in {{\cal S}_1}\ {\rm{or}}\ {{\cal S}_2},\forall m=1,2,\dots,M,\\
	\end{aligned}
	\label{opt-Prom}
\end{equation}
where ${\cal S}_{\eta}$ denotes the set of feasible locations.

\subsubsection{Near-Field Localization Scheme}



The location of the MU can be estimated via a two-stage near-field localization scheme.
As the phase shift of the RIS can be obtained as $\hat{\bf{\Theta}}$ by leveraging the methods discussed in Section III, here we only focus on the localization problem.
In the first stage, the coarse estimation of the positions of MU and scatterers is derived \cite{cui2021channel,Chen.2002}.
 In the second stage, the finer estimation of the positions of MU and scatterers can be estimated with the iterative methods, e.g., two-dimensional search \cite{DaiChannel}.
 The details of the two-stage near-field localization scheme are outlined below.

\emph{Stage 1}:
Suppose that the transmit signal is $s=1$.
As the received power from the scatterer-reflected paths is much weaker than the LoS path, we first try to find the location of MU based on the LoS path with the largest received power.
In the first stage, the location of the MU is estimated as follows:
\begin{equation}\label{MLEstObj}
	\hat{{\bm p}}_0 =  \arg \underset {{\bm p}_{0}}{\min}\;\;|| \bm{y} - {\bm{H}}\hat{\bf{\Theta}}\bm{a}_R({\bm p}_{0})||^2.
\end{equation}
Without any prior information for the source location, the objective function in (\ref{MLEstObj}) needs to  be evaluated over a set of grid locations.
For the sake of reducing complexity, a non-uniform grid may be used.
Specifically, the angle and distance are sampled in the polar domain using a uniform grid in \cite{cui2021channel}.
In general, the grids should be dense near the array and sparse away from the array\cite{Chen.2002}.

With the obtained location,  the corresponding channel gain can be obtained by using the projection method as
\begin{equation}
	\hat{\beta}_0 =  \frac{({\bm{H}}\hat{\bf{\Theta}}\bm{a}_R(\hat{{\bm p}}_{0}))^H \bm{y}}{||\bm{H}\hat{\bf{\Theta}}\bm{a}_R(\hat{{\bm p}}_{0})||^2}.
\end{equation}

Then, the locations of the scatterers are estimated as
\begin{equation}\label{MLEstObjL}
	\hat{{\bm p}}_l =  \arg \underset {{\bm p}_{l}}{\min}\;\;\left \| \ \bm{y}_{-l} - {\bm{H}}\hat{\bf{\Theta}}\bm{a}_R({\bm p}_{l})\right \|^2,
\end{equation}
where $\bm{y}_{-l} =  \bm{y} - \sum_{s=1}^{l-1}{\bm{H}}\hat{\bf{\Theta}}\hat{\beta}_{s}\bm{a}_R( \hat{{\bm p}}_{s}) $,
 and the corresponding channel gain is obtained as
\begin{equation}
	\hat{\beta}_l =  \frac{({\bm{H}}\hat{\bf{\Theta}}\bm{a}_R(\hat{{\bm p}}_{l}))^H\bm{y}_{-l} }{||\bm{H}\hat{\bf{\Theta}}\bm{a}_R(\hat{{\bm p}}_{l})||^2}.
\end{equation}

\emph{Stage 2}:
With the coarse stimation, a refined estimation can be further conducted in the second stage.
The position of the MU and the scatterers can be refined using a finer grid while keeping the estimation of other locations fixed. Specifically,
\begin{equation}
	\begin{aligned}
		&\underset {{{\bm p}}_l}{\min}\;\;|| \bm{y} - {\bm{H}}{\bf{\Theta }}{\bm{h}}_{r}({{\bm p}}_l,\bm{\eta}_{-l})||^2,
	\end{aligned}
	\label{opt-Prom2}
\end{equation}
where $\bm{\eta}_{-l} = [\bm{\hat{r}}_{0},\cdots,\bm{\hat{r}}_{l-1},\bm{\hat{r}}_{l+1},\cdots,\bm{\hat{r}}_{L_{\mathrm{RU}}}, \hat{\bm\beta}^T]^T$.
Several iterations are needed for high-resolution location estimates.

\textbf{Simulation results:}
For the near field localization case, we evaluate the localization performance in terms of the MSE, which is obtained by the proposed two-stage near-field localization scheme.
The locations of BS and RIS are ${\bf q}=[3,0,3]^T$ and ${\bf p}=[0,0.5,2]^T$, while the locations of the MU and scatterers are randomly generated in the $5$ m $\times$ $5$ m cell range.
The number of RIS elements is $21 \times 21$, and the phase shift matrix is obtained by maximizing the received signal power.
It can be seen that the MSE performance degrades with the number of scatterers due to the more scattered energy and the feedback estimation error propagation, while increasing the SNR can help improve the localization accuracy.
\begin{figure}
	\centering 
	\includegraphics[width= 3.3in]{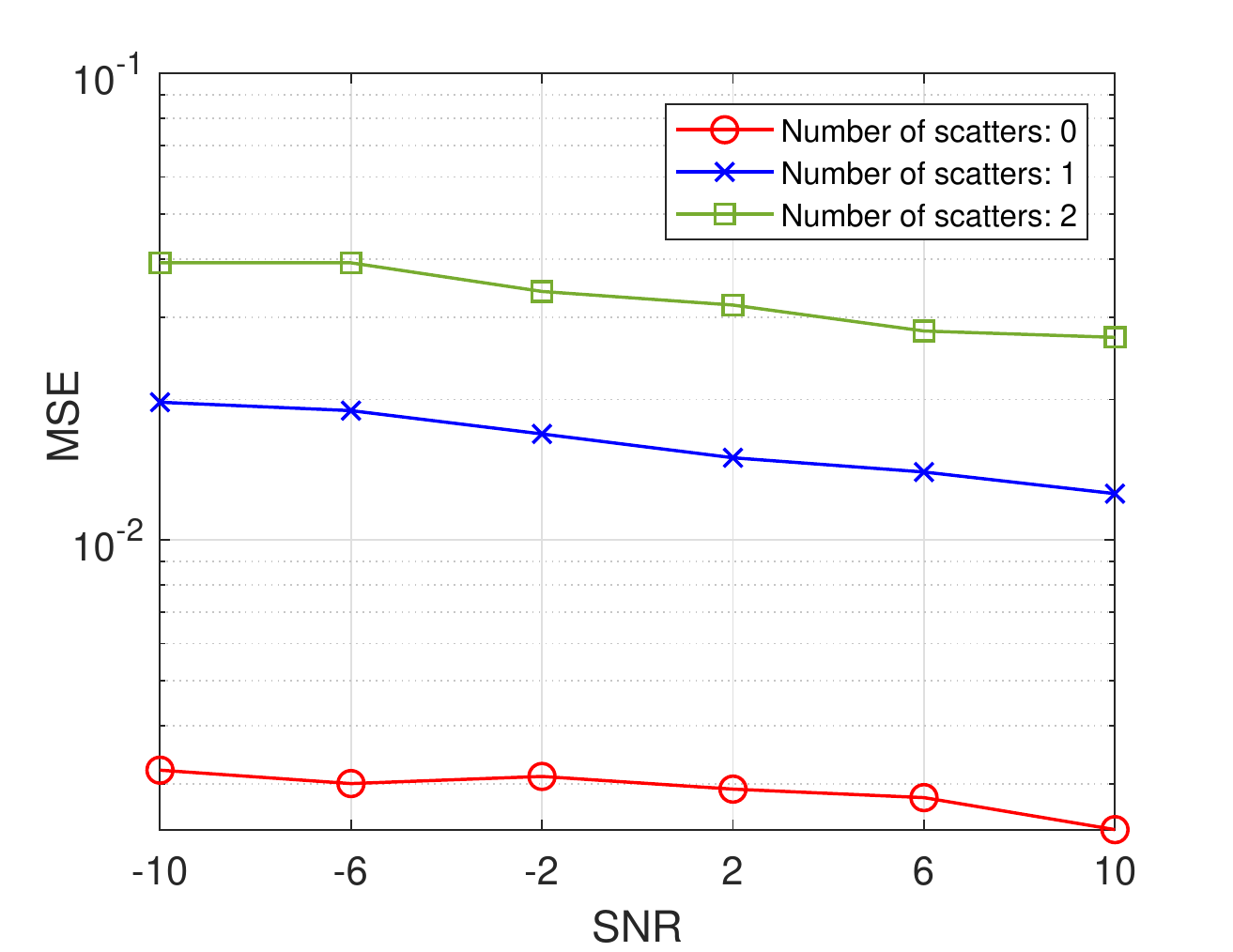}
	\caption{MSE versus the SNR, when $N_t=32$ and $f_c=240$ GHz.}
	\vspace{-0.3cm}
	\label{tdoa-algorithm}
\end{figure}

\section{Future Directions}\label{futuredirection}
In this section, we summarize some open research issues that are focused on the modeling, the analysis, and the optimization of RIS-aided wireless systems.

\subsection{Mobility}
Most existing works focus on quasi-static scenarios where the users and the RISs are assumed to be nearly stationary. However, it is also imperative to investigate scenarios with mobility \cite{mao2021mobility,sun2021mobility,matth2021mobility,huang2021transforming,Zegrar}, such as wireless systems in the presence of vehicles, trains, and UAVs.
In these scenarios, in particular, the RISs may even be mobile as they may be carried by UAVs or they may be deployed in trains.
In these scenarios, several challenging research problems emerge. For example, the estimation of rapidly time-varying channels becomes more challenging. Therefore, new channel estimation methods are required to
quickly and accurately track the channels, in order to enable the design of accurate phase shifts at the RIS. Also, since the channel is rapidly changing, the channel estimation and the RIS phase shift configuration need to be executed more frequently. As a result, it is crucial to design effective signal processing schemes to avoid a prohibitively high overhead in high-mobility scenarios.
Furthermore, the RIS has the potential to reduce the Doppler and delay spread \cite{matth2021mobility}, which deserves further research in more practical scenarios.

\subsection{Near-Field Channel}
To ensure a sufficiently large coverage, the aperture of the RIS planar array is typically large in order to compensate for the absence of power amplifiers and digital signal processing units for regenerating the signals.
Depending on the size of the RIS, the operating frequency, and its distance from the transmitter and the receiver, it may be necessary to utilize a near-field communication model
\cite{bjoson2020near,feng2021near,bjornson2021primer,wei2021codebook}.
Near-field communication models are, however, less understood than their far-field counterparts.
Therefore, it is  essential to consider accurate channel models to unveil the fundamental performance limits and the scaling laws of RISs in this context. It is well known, for example, that the scaling laws of the electromagnetic field radiated by the RIS is different in the near-field and far-field regions\cite{Danufane2021pathloss}. Also, new low-overhead channel estimation strategies are needed since the channel sparsity in the angle domain, which is based on the far-field planar wavefront assumption, may no longer hold when using the near-field spherical wavefront assumption \cite{cui2021channel}.

\subsection{Active RIS}
Quasi Passive RIS architectures have attracted significant attention due to their low hardware cost and energy consumption. Nevertheless, quasi passive architectures have their limitations. Since the signals reflected by the RIS is determined by the product of the distances from the transmitter to the RIS and from the RIS to the receiver in the far-field region, the received signal strength is relatively small compared with that from direct links, especially when the direct links are strong. The recently proposed active RIS architecture is a promising solution to overcome this issue \cite{zhang2021active,xu2021resource,you2021active,Khoshafa2021active}. After integrating active reflection-type amplifiers\cite{zhang2021active}, in addition to adjusting the phase shifts, active RISs can simultaneously amplify the magnitude of the reflected signals. However, the amplification at the active RIS requires an  additional power supply. Therefore, more dedicated beamforming at the BS and phase shifts at the RIS should be designed to balance the performance and energy consumption. Meanwhile, channel estimation with active RISs needs to incorporate the statistical properties of amplification-induced noise, which is more challenging and worth further investigation. Besides,  how to decide the optimal deployment of the active RISs is still an open question.

\subsection{Double/multi-RIS}
Most of the existing works have considered quasi passive beamforming designs and channel estimation schemes in systems with a single RIS.
In some scenarios, however, it may be convenient to enable the transmission of signals through reflections from multiple RISs in order to route the signals and bypass the blocking objects in a smart manner, directly at the electromagnetic level (electromagnetic routing)\cite{di2019smart}.
Thus, double/multiple RISs may be utilized to realize a blockage-free communication network via multiple signal reflections. Moreover, the proper design of the cooperative quasi passive beamforming can eliminate the inter-RIS interference, and achieve multiplicative beamforming gain from the inter-RIS reflection channel  \cite{double2020han,double2021bei,double2021mei,double2021nie}.
 However, these methods require highly accurate CSI, which is challenging to obtain due to the coupling of the different reflected signal links and more channel coefficients  to be estimated \cite{double2021you,double2021heng,double2021xiong}. In this context, channel estimation methods in double/multi-RIS aided communication scenarios in the presence of single/double/multi-reflection links need to be investigated.  In addition,  existing research works on double/multi-RIS aided systems  often ignore the impact of the secondary reflections among the RISs, which may be a reasonable approximation if the RISs are in the far-field of each other but it may not hold anymore if the RISs are closely located. Finally, the analysis and design of multi-RIS communications at high frequency bands is an open research issue as well.

  \subsection{Multifunction RISs }
In the existing literature, most of the research works concentrate on RISs that operate as anomalous reflectors or as reflecting lenses. An RIS, however, can realize multiple signal transformations depending on how the scattering matrix (or equivalently the surface impedance) is designed\cite{di2019smart}. Recently, notably, a few research attempts have been made to design RISs that operate as anomalous refracting mirrors or as anomalous refracting lenses\cite{cho2021mmwall} as well RISs that can simultaneously realize reflections and refractions in order to guarantee omni-coverage performance\cite{zhang2021omi,zhang2021intelligent,liu2021star,mu2021simultaneously}. Multifunction RISs are an emerging research topic, and the corresponding modeling, performance evaluation, and optimization are still at its infancy.

\section{Conclusions}\label{hrforh}

In this paper, we provided a comprehensive overview of state-of-the-art research for new and revolutionary RIS/IRS-aided wireless systems, with an emphasis on signal processing techniques for solving various channel estimation, transmission design and radio localization problems. Specifically, we first reviewed existing results on channel estimation under unstructured and structured channel models. Next, we provided a detailed overview of the research results from the perspective of different optimization techniques and  availability of CSI. In particular, several optimization techniques were described for optimizing RIS with discrete and continuous time shifts.
As far the availability of CSI is concerned, three main cases were considered, namely,  fully instantaneous CSI, two-timescale CSI, and fully long-term CSI. Simulation results demonstrated that the two-timescale CSI scheme constitutes a promising approach when the  pilot overhead is taken into account. Furthermore, radio localization is an important application of RISs, and it has been thoroughly reviewed by considering deployment scenarios and channel models that account for far-field and near-field propagation.  Finally,  several open research problems have been discussed.

\bibliographystyle{IEEEtran}
\bibliography{myre}

\end{document}